\documentclass[useAMS,usenatbib]{mn2e}
\usepackage{epsfig}

\newcommand{\bc}{\begin{center}}
\newcommand{\ec}{\end{center}}
\newcommand{\be}{\begin{equation}}
\newcommand{\ee}{\end{equation}}
\newcommand{\dd}{{\rm d}}
\newcommand{\kms}{\mbox{km s$^{-1}$}}
\newcommand{\msun}{\mbox{M$_{\odot}$}}
\newcommand{\msunyr}{\mbox{M$_{\odot}\,$yr$^{-1}$}}
\newcommand{\msunpc}{\mbox{M$_{\odot}\,$pc$^{-2}$}}
\newcommand{\msunpcub}{\mbox{M$_{\odot}\,$pc$^{-3}$}}
\newcommand{\tfb}{\tau_\mathrm{fb}}
\newcommand{\rt}{\rho_\mathrm{th}}
\newcommand{\nzl} {\textit {n0low}}  
\newcommand {\nzm} {\textit {n0med}}  
\newcommand {\nzh} {\textit {n0high}}  
\newcommand {\nol} {\textit {n1low}}  
\newcommand {\nom} {\textit {n1med}}  
\newcommand {\noh} {\textit {n1high}}  
\newcommand {\ntl} {\textit {n2low}}  
\newcommand {\ntm} {\textit {n2med}}  
\newcommand {\nth} {\textit {n2high}}  

\title[Feedback in simulations of disc-galaxy major mergers]
{Feedback in simulations of disc-galaxy major mergers}

\author[T.J. Cox, P. Jonsson, J.R. Primack, and R.S. Somerville]
{T. J. Cox$^{1, 2}$, Patrik Jonsson$^3$, Joel R. Primack$^1$, and Rachel S. Somerville$^4$ \\
        $^1$Department of Physics, University of California, Santa Cruz, 
                1156 High St., Santa Cruz, CA, 95064, USA\\
	$^2$Harvard-Smithsonian Center for Astrophysics, 
		60 Garden St., Cambridge, MA 02138, USA; \texttt{tcox@cfa.harvard.edu}\\
        $^3$Department of Astronomy and Astrophysics, University of California, 
                Santa Cruz, 1156 High St., Santa Cruz, CA, 95064, USA\\
        $^4$Max-Planck-Institut f\"ur Astronomie,
		K\"onigstuhl 17, D-69117 Heidelberg, Germany\\
        }

\begin{document}

\maketitle

\begin{abstract}
  Using hydrodynamic simulations of disk-galaxy major mergers we
  investigate the star-formation history and remnant properties when
  various parameterizations of a simple stellar feedback model are
  implemented.  The simulations include radiative cooling, a
  density-dependent star-formation recipe and a model for feedback
  from massive stars.  The feedback model stores supernova feedback
  energy within individual gas particles and dissipates this energy on
  a time-scale specified by two free parameters; $\tfb$, which sets
  the dissipative timescale, and $n$, which sets the effective equation
  of state in star-forming regions.  Via this model, feedback energy can
  provide pressure support to regions of gas that are thermally cold.
  Using a self-consistent disk galaxy, modeled after a local Sbc
  spiral, in both isolated and major-merger simulations, we investigate
  parameterizations of the feedback model that are selected with respect
  to the quiescent disk stability.  These models produce a range of
  star formation histories for disks evolved in isolation, or during a
  major merger, yet all are consistent with the star formation relation
  found by \citet{Kenn98}.  We suggest that this result is produced by
  the adopted recipe for star formation and is not a byproduct of the 
  feedback model.  All major mergers produce a population of
  new stars that is highly centrally concentrated, demonstrating a distinct
  break in the $r^{1/4}$ surface density profile, consistent with
  previous findings.  The half-mass radius and one-dimensional
  velocity dispersion are affected by the feedback model used.  In
  tests with up to an order of magnitude higher resolution, the star
  formation history is nearly identical, suggesting that we have
  achieved a numerically converged star formation history.  Finally,
  we compare our results to those of previous simulations of star
  formation in disk-galaxy major mergers, addressing the effects of
  star- formation normalization, the version of smoothed particle
  hydrodynamics (SPH) employed, and assumptions about the interstellar
  medium.  We conclude by suggesting several methods by which future
  studies may better constrain feedback models.
\end{abstract}
\begin{keywords}
galaxies: interactions -- galaxies: evolution -- galaxies: starburst -- 
galaxies: formation -- methods: numerical.
\end{keywords}

% --------------------------------
%   Introduction: tell me more
% --------------------------------
\section{Introduction}
\label{sec:intro}

It is commonly assumed that galaxy interactions trigger powerful
bursts of star formation.  This assumption is supported by a variety
of observational data which links morphologically disturbed or 
outright merging galaxies to enhanced levels of star formation 
\citep[e.g.,][]{LT78,JW85,Ken87,Bar03,Lam03,NCA04}.  In fact, the 
most vigorously star-forming 
galaxies in the local universe, the ultra-luminous infrared 
galaxies (ULIRGs), are nearly all mergers \citep{SM96,Bor00}.
Indeed, one of the fundamental 
assumptions of many galaxy-formation models is the efficient
conversion of cold gas into new stars whenever two galaxies merge.
These assumptions appear to be necessary in order to 
reproduce the abundance of Lyman-Break \citep*{SPF} and 
submillimeter \citep{Gui98,Bau05} galaxies.

The numerical simulations run to-date have supported the theory that
disk-galaxy mergers yield a burst of star formation.  The earliest
work established that galaxy mergers can concentrate a large amount of
gas in the central regions of the remnant \citep{NW83,H89,BH91,BH96}.
Later, by including simple recipes to capture star formation and
kinetic feedback from massive stars,
\citet[][hereafter MH96]{MH94majm,MH96} produced the first quantitative 
results for the star formation induced during a major merger.  These
authors found that mergers between two equal mass gas-rich galaxies, 
so-called major
mergers, trigger star formation at every close passage of the galactic
nuclei, culminating in a burst at the final merger.  In total,
these starbursts convert $\sim80$\% of the original gas mass into stars, an
efficiency apparently independent of galaxy orientation or the 
presence of a spheroidal bulge in the primary disk galaxy.

Subsequent work has become more ambitions.  
\citet[][hereafter S00]{Sp00} employed a slightly more elaborate 
treatment of the interstellar medium (ISM) by accurately treating gas
heating and cooling.  More recently, simulations
have included shock-induced star formation \citep{B04} and accreting
black holes \citep*{SdMH05}.  While all of these studies have supported
the fact that major mergers induce star formation, it has become clear
that the star-formation efficiency is highly dependent upon the 
star formation and feedback assumptions.
Because we lack a detailed theory of
star formation and the influence young stars have on their surroundings, 
there are no clear means to include these physical processes in numerical
simulations.  The typical approach is to formulate physically or 
observationally motivated recipes whose
impact is controlled by one or more free parameters.  This methodology
is also motivated by the finite numerical resolution achievable.  Current
state-of-the-art numerical simulations evolve
individual fluid elements which each represent $\sim10^5$~\msun\
and 100 pc.  As an example, star
formation on kpc scales is observationally found to trace the
gas density in what is known as the Schmidt law \citep{Sch59,Kenn98}.
Motivated by this relation, simulations can employ a recipe
that converts gas to stars based on the local gas density.
In this sense, a simulation can capture the sub-resolution
star formation averaged over scales that are numerically resolved.  
Unfortunately, there exists no clear analog to the Schmidt-law
for the effects of massive stellar winds and supernovae.

Effective parameterizations for stellar feedback have been included in
numerical simulations for the past decade.  The first attempts used
``thermal feedback,'' in which energy released by supernovae simply
increases the thermal energy of the gas \citep{Kz92}.  Subsequent
tests have yielded one of the most robust results regarding feedback:
thermal feedback is ineffective at regulating star formation due to
the short cooling times (and rapid thermal energy loss) in dense
star-forming regions \citep[S00;][]{NW93,SFTreeSPH}.  Extensions of
thermal feedback, in which a fraction of the supernova energy is
imparted as kinetic energy to nearby gas \citep{NW93,SFTreeSPH}, have
proved effective at producing stable, constantly star-forming disk
galaxies, yet have the negative property that their implementations
are resolution dependent (S00) and they have difficulty stabilizing disks
with a large gas fraction.  Beyond this kinetic feedback, most other
recipes have looked to prevent the efficient radiation of feedback
energy.  \citet{GI97} used UV radiation from young stars to offset
efficient cooling, S00 created a turbulent reservoir that retained
feedback energy for a set time-scale, and \citet{TC01} simply
inhibited cooling for 30 Myr subsequent to the injection of feedback
energy (see \citealt{TC00} and \citealt{Kay02} for a comparison of
various feedback methods).  More recently, methods have been developed
to model the multiphase nature of the ISM
\citep{Yep97,HP99,Sem02,Mar03,SH03} and the effects that supernovae
blastwaves may have on the ISM \citep{Stin06}.  While each of the above models
has advantages and disadvantages, many of these methods choose to
regulate star formation in an equivalent manner: by using feedback
energy to effectively pressurize the ISM in star-forming regions.

In this paper, we address the issue of supernova feedback in more
detail.  Using a generalization of the star-formation and feedback
model of S00, we explore a range of parameters and determine how the
resulting starburst is affected.  This is a slightly different
approach than previous work, in that instead of physically motivating
a specific feedback model, we attempt to understand how assumed
feedback parameters affect star formation.  Both the overall efficiency
of feedback and its dependence on gas density through an effective
equation of state are explored.
One of our long-term goals is to uniquely determine a single 
parameter set, or feedback model, which will be used for a
large series of major- and minor-merger simulations.  Unfortunately,
it is difficult to define a checklist against which each model can
be measured, and hence it is difficult for us to endorse one model
as significantly better than any other.  We can, however, identify
trends that are present in our various models and use these to guide
parameter exploration in future work.

An important aspect of the work presented here will be to detail how
assumptions made by MH96 and S00 affected their merger-induced starbursts.
Even though these papers arrived at similar conclusions, the galaxy
models, temperature and equation of state of the ISM,
implementation of
feedback, and N-body/smoothed particle hydrodynamics (SPH) code were
were all different.  This last point is particularly relevant as the 
conservation of energy and entropy depends on the formulation of SPH as
well as the numerical resolution \citep{H93sph,Tha00,SHEnt}.  In general,
all versions of SPH converge to an identical solution for high 
resolution, but can be quite different at low resolution.  This
result has recently motivated \citet{SHEnt} to develop a new 
formulation of SPH which manifestly conserves both energy and entropy,
when appropriate.
We will investigate what effects this new version of SPH has and how
this compares to older formulations of SPH.
We also address the numerical convergence of our simulations by
performing mergers with up to ten times the resolution.  In additional
tests, we vary the assumed gas metallicity and star-formation
efficiency and determine the resultant star formation.

In our study of the star-formation response to different parameterizations
of feedback, we find that the properties of the merger remnants are
also affected.  This is significant because mergers between spiral galaxies
are considered the most likely mechanism to form galactic spheroids.  This
``merger hypothesis'' \citep{TT72,T77} is generally supported by simulations
\citep[S00]{HRemI,HRemII,NB03}, and here we show that details of the feedback
model will affect the ability to test the merger hypothesis.  We reserve
an exhaustive study of the merger remnants to future work, but the
variation of feedback parameters alters the basic properties of the merger
remnant, and we document these effects here.

In short, we address the following issues in this work:
\begin{enumerate}
\item How does the star-formation rate and the global gas 
consumption depend upon the feedback model and feedback parameters?
\item Is the merger remnant affected by the choice of feedback
parameters?
\item Are our models consistent with previous simulations of disk-galaxy
interactions?  Does using a modified version of SPH that manifestly conserves
both energy and entropy, when appropriate, change the star-formation evolution?
\end{enumerate}

This paper is designed to be one in a series which compiles the
star-formation and remnant properties for a large number of galaxy
merger simulations.  Prior work has analyzed the kinematics of the
merger remnants \citep{Dek05}, the influence of dust on the
observed properties of the interacting system \citep{JonI}, the
shapes of the stellar and dark matter remnants \citep{Nov06}, and
the dissipative origin of elliptical galaxies \citep{DC06}.  Future
work will investigate the star formation induced during mergers 
between unequal mass galaxies (Cox et al., in preparation), propose
a model that predicts the properties of merger remnants
(Covington et al., in preparation), and compare the simulations
to observed samples of interacting and merging
galaxies using non-parametric morphological classification systems
(Lotz et al., in preparation).

We organize this paper as follows.
To begin, \S \ref{sec:sims} details the numerical techniques and the physical 
ingredients implemented into the simulations, and \S \ref{sec:isodisks}
introduces a disk galaxy which will be used to investigate our feedback
models.  This disk galaxy is simulated in isolation in \S \ref{ssec:isosfr}
and during a major merger in \S \ref{sec:majm}.  \S \ref{sec:majm} also
shows the star-formation history (\S \ref{ssec:mmsfr}) and takes a
brief look at the properties of the merger remnant (\S \ref{ssec:mrem}).
We discuss our results in \S \ref{sec:disc},
including a resolution study (\S \ref{ssec:res}), a comparison to
previous work (\S \ref{ssec:comp}), and the effects of varying the ISM
metallicity (\S \ref{ssec:cool}).  Finally, \S \ref{sec:concl} provides some
concluding remarks and a prospectus for future work using the techniques
we have developed in this paper.

%---------------------------------------
%    Describe Model and the Numerics
%---------------------------------------
\section{Numerical Simulations}
\label{sec:sims}

All numerical simulations performed in this work use the N-Body/SPH code 
GADGET \citep*{SpGad}.  Specifically, we use a variant of the publically
available version which has the ``conservative entropy'' version of SPH 
\citep{SHEnt}, and additional routines that track the radiative cooling
of gas and star formation.  This version of GADGET was obtained from
Volker Springel in 2001.  In addition, we implemented several new features
ourselves; stellar feedback, metallicity-dependent cooling, and the ability
of each gas particle to spawn multiple new stellar particles.  The 
following subsections describe in more detail the physical processes
included in our simulations.

% 
% -------------------
\subsection{Cooling}
\label{ssec:cooling}

Due to radiative cooling, baryons fall into the centers of dark-matter 
halos and achieve the cold, dense environment conducive to the formation
of stars.  The radiative cooling rate $\Lambda_{\rm net}$ is computed as
described in \citet*{Kz96}, where the gas is treated as a primordial 
plasma and the ionization states of H and He are explicitly tracked 
under the assumption of collisional ionization equilibrium.  In practice,
the radiative cooling rate $\Lambda_{\rm net}$ then simply a function of
the local gas density $\rho_{\rm gas}$ and internal energy per unit mass 
$u$.  However, the radiative cooling of astrophysical plasmas also highly
dependent upon the gas metallicity, we 
also ran merger simulations using the tabulated cooling curves of \cite{SD93},
which allow for a range of metallicities.  The results of these simulations are
presented in \S \ref{ssec:cool}.  In all simulations the spatial and temporal
changes in metallicity are ignored, although we keep track of these for use
in radiative transfer calculations through the dusty ISM \citep{JonI,JonSun}.  
Lastly, we remind the reader that, without 
adiabatic processes, the radiative cooling described here, in which molecular
cooling is neglected, effectively sets the minimum gas temperature to $10^4$K.

% 
% -------------------
\subsection{Star Formation}
\label{ssec:sf}

All numerical simulations presented here include star formation.  Star 
formation is assumed to be proportional to the local gas density and 
inversely proportional to the local dynamical time-scale 
\be
\frac{\dd \rho_{\star}}{\dd t} = c_\star \frac{\rho_{\rm gas}}{t_{\rm dyn}},
\label{sflaw}
\ee
where $c_\star$ is a free parameter determining the efficiency of star 
formation, and $t_{\rm dyn}$=$(4\pi G\rho_{\rm gas})^{-1/2}$.  This 
star-formation recipe mirrors the approach commonly implemented in numerical
simulations \citep[MH96; S00;][]{Kz92,SH03}, however we will briefly 
revisit this in \S~\ref{sssec:sfs}.

The star-formation prescription set forth by Equation~(\ref{sflaw}) 
was originally motivated by star-forming regions in our own galaxy
\citep{Sch59} but has been shown to hold in a wide range of environments.
The observational work of \citet{Kenn98} has shown that, on kpc
scales, the aperture-averaged star-formation rate per unit area is
correlated to the gas surface density.  This relation is empirically 
determined to be
\be \Sigma_{\rm SFR}= (2.5 \pm 0.7) \times 10^{-4}
\left( \frac{\Sigma_{\rm gas}}{ { \rm M_{\odot} pc^{-2}}}\right)^{1.4
\pm 0.15} \frac { {\rm M_{\odot}} }{ {\rm yr\,kpc^2}}
\label{KenLaw},
\ee 
and will be referred to as the ``Kennicutt law'' throughout the remainder 
of this paper.
While the index 1.4 resembles the $\dot{\rho}_{\star} \sim \rho_{\rm gas}^{1.5}$
implied by Equation~(\ref{sflaw}), we must admit that it is not immediately
clear that the three-dimensional Schmidt-law, based upon the gas volume density,
should be equivalent to the
empirical two-dimensional Kennicutt-law, based upon the gas surface density.
Nevertheless, the simulations do obey
(\ref{KenLaw}), as shown in \S\ref{ssec:isokenn}

\citet{Kenn98} found the existence of a surface density threshold
at $\sim 10$~\msunpc, below which star formation dramatically drops
\citep{Kenn98,MK01}.  To capture this effect in our numerical
simulations, we introduce a volume-density threshold, $\rt$.
Gas which has a density larger than $\rt$ is eligible
to form stars at a rate given by Equation~(\ref{sflaw}).  We adopt
$\rt = 0.0171$~\msunpcub, a value similar to that used
in S00 (MH96 did not include such a density threshold) and confirmed to
reproduce the observational suppression of star formation below 10 
\msunpc.  Interestingly, Equation~(\ref{KenLaw}) also appears to hold
for azimuthally-averaged gas densities and star-formation rates in
individual galaxies, although the scatter seems to be much larger and
the slope shallower \citep{MK01,WB02}.

Our implementation of star formation is fully determined by the one
free parameter in Equation~(\ref{sflaw}), $c_\star$.  To help us
select a numerical value for $c_\star$ we define the characteristic
time-scale for star formation as
\be t_\star = \frac{\rho_{\rm gas}}{\dot{\rho_\star}}= \frac{t_{\rm dyn}}{c_\star}.
\label{sfrtime}
\ee Equation~(\ref{sfrtime}) relates the star-formation time-scale to
the dynamical time by our one free parameter $c_\star$.  Both
time-scales become short at high densities.  Observations indicate
that the median gas consumption time-scale is much longer than the
dynamical time-scale, suggesting that $c_\star~\ll$~1 \citep{Kenn98}.
After experimenting with various values for $c_\star$ (see
\S\ref{ssec:comp}) we found that 0.03 provided a good fit to
Equation~(\ref{KenLaw}), although we caution that the observational
scatter is quite large - almost an order of magnitude higher or lower.
Hence there could be a range of allowed values for $c_\star$, all
consistent with the Kennicutt law.  Our value of $c_\star=0.03$ fixes
the time-scale for star formation to be 1.1 gigayear (Gyr) at
$\rt$.  Many previous studies also include star formation via
Equation~(\ref{sflaw}).  These works use a wide range
of star-formation efficiencies that are both higher
\citep{Kz92,NW93,K04,Gov04,KG05} and lower \citep*[MH96; S00;][]{SL03,SH03}
than ours.  In addition, several authors investigated a range of
values for $c_\star$ \citep{BS98,TC00} as we do in \S\ref{ssec:comp},
or use a hybrid model where $c_\star$ can change with redshift or
environment \citep{SL03,Oka05}.

Very little is known about the requisite conditions enabling star formation to occur.
Previous simulations \citep[see e.g.,][]{Kz92,Kv03,Gov04} have included star 
formation criteria such as gas convergence, Jeans instability, and 
%($\nabla\cdot {\bf v} < 0$), Jeans instability ($h/c<t_{\rm dyn}$), and 
%temperature thresholds (T $<$ T$_{\rm th}$).  With the resolution of current
temperature thresholds.  With the resolution of current
simulations, where individual particles represent $\geq 10^5$~\msun~of gas, it is
not clear that gas convergence should be required. The Jeans criterion is dependent
on the numerical smoothing length, and hence has the undesirable behavior that
it is a function of numerical parameters.  Finally, as mentioned above, our treatment
of cooling sets the minimum gas temperature at $10^4$~K, thus we do not expect gas
to approach physically reasonable temperatures for star formation ($\leq 50$K).  For 
these reasons we have not included any of the above criteria in the work presented 
here.

As gas is converted into stars at the rate dictated by Equation~(\ref{sflaw}), one must 
decide how to handle the newly formed stellar mass.  Since it is not computationally
feasible to form individual solar-mass particles, numerical simulations commonly 
treat SPH particles as ``hybrid'' particles, representing both gas and new stellar 
material \citep[e.g.][]{SFTreeSPH}.  While this procedure is relatively simple to 
implement, it has
the disadvantage that new stellar material is dynamically coupled to the gas.

An alternative approach, suggested by several authors
\citep[e.g.,][]{Kz92,SH03,Bot03}, is to treat star formation as a stochastic 
process.  Under this assumption each gas particle spawns a specified number $N_g$ of
equal-mass star particles.  Then, if each gas particle forms a new star
particle according to the probability
\be
P = N_g \left(1-e^{-dt/t_\star}\right),
\label{eq:stochsf}
\ee
the stochastic star-formation rate will equate to the designed continuous 
star-formation rate when averaged over the entire simulation time.  We follow
the above authors when implementing this procedure, and thus at each timestep actively
star-forming gas particles draw a random number between zero and one.  If this
random number is less than $P$, a new collisionless particle, of mass
$m_{\rm gas}/N_g$, is created.  We note that $m_{\rm gas}$ is the original gas
particle mass and that the SPH particle mass decreases as individual star 
particles are generated.
For the majority of the simulations used
in this work, we select $N_g=2$, although we have tried varying $N_g$ up
to 10 and found little difference in the results (see the end of \S\ref{ssec:mrem}
for details).

Throughout this work stars formed during the simulation will be called
``new'', as opposed to ``young'', stars.  We make this choice because at any
point during the simulation there exists a mixture of newly generated stars which
have a distribution of stellar ages.
While many of these particles are part of a young stellar population,
a small percentage will have ages of several gigayears by the time
the simulation is complete.

% 
% -------------------
\subsection{Metal Enrichment}
\label{ssec:metals}

Although it is not part of the analysis in the present paper, we have also 
included metal enrichment owing to stellar winds and supernovae ejecta.
The enrichment, which assumes a yield of 0.02 per solar mass of stars formed
and instantaneous recycling, is performed in a continuous fashion, based 
upon the instantaneous star-formation rate provided by Equation~(\ref{sflaw}).
The star-formation rate is calculated for each particle and metals are
recycled within the same particle, i.e., metals are not distributed among
neighboring particles nor is the diffusion of metals considered.
These metals are carried by each gas particle and subsequent generations
of new star particles adopt the metallicity of the parent gas particle at
the time when they are born.  Under these assumptions, the metallicity of
individual gas particles can only increase with time.  However, once the gas
particle is fully transformed into stars, these metals are forever locked in
the stellar component.  We intend to implement a more realistic approach 
where star particles return a fraction of their mass and metals back to the
interstellar medium in the future \citep[see, e.g.,][]{Mos01,Tor04,Oka05,
cSca06,cSca05}.  Finally,
as noted in \S~\ref{ssec:cooling}, even though we track the metallicity
of the gaseous component, this information does not affect the cooling rate.

% 
% -------------------
\subsection{Feedback}
\label{ssec:fb}

As discussed in the introduction, many of the methods by which 
feedback from young stars is incorporated into numerical simulations
involve artificially pressurizing gas in high-density, star-forming 
regions.  Within SPH, where particles represent fluid elements, this
pressurization is implemented on a particle by particle basis.  This
general approach is also used for the present study but, because there
exists no clear theoretical understanding of how feedback pressurization
should depend on the physical state of the star-forming regions, we
instead adopt a general formalism and investigate the star-formation
response to a variety of different parameter choices.

As the starting point for our exploration into feedback, we begin
by adopting the feedback model put forth by S00.   The relative
simplicity of this model and the ease with which it can be implemented
into GADGET motivate this choice.  Within this model, and adopting
the nomenclature of S00, gas is 
assumed to consist of two energy reservoirs.  One, the standard thermal energy 
$u$, can radiatively cool at a rate given by $\Lambda_{\rm net}$.  The second,
$q$, is an ``ad-hoc'' feedback 
reservoir.  The pressure of this medium is then calculated as 
\be P_{\rm eff} = (\gamma -1)(u+q)\rho, \label{eq:effPfull} \ee  
where $\rho$ is the gas density (the gas subscript is henceforth dropped),
and $\gamma$ is 5/3, consistent with the assumption of a monatomic ideal gas.
Equation~(\ref{eq:effPfull}) demonstrates that regions which are thermally cold, possibly
because the density is such that cooling is very efficient, can be pressure 
supported if the feedback reservoir $q\gg u$.

Feedback energy is assumed to come from core collapse supernovae and 
is calculated in a manner similar to \citet{NW93}.
New stars form with a power-law initial mass function of slope -1.5
between 0.1 and 40~\msun, and
every star above 8~\msun$ $ goes supernova releasing $10^{51}$ ergs of energy.  The
short lifetime of these high-mass stars is ignored and supernova energy is instantly
returned to the system.  
We have investigated the effects of including realistic delays in supernova 
feedback, but it made little difference to the results reported here.
We also investigated a limited range of excursions from the initial-mass function
and high-mass cutoff listed above.  In practice, though, these changes 
can be compensated with variations in our star formation or feedback
free parameters and we do not consider this further.
The supernova feedback energy per solar mass of newly formed stars 
$\epsilon_{\rm SN}$  is 
assumed to be first released into the feedback reservoir $q$, i.e.
\be \left( \frac{\dd q}{\dd t} \right)_{\rm feedback} = \epsilon_{\rm SN}
\frac{1}{\rho} \frac{\dd \rho_\star}{\dd t}. \label{eq:u} \ee
The feedback reservoir subsequently thermalizes its energy as
\be \left( \frac{\dd q}{\dd t} \right)_{\rm thermalize} = -
\left( \frac{\dd u}{\dd t}\right)_{\rm thermalize} = -f(\rho)q
\label{eq:q} \ee
where $f(\rho)$ parameterizes the efficiency of thermalization.

The original work of S00 chose $f(\rho)= \beta / \sqrt{\rho}$, i.e., the
thermalization of feedback energy is slower in high-density regions.
For star-forming gas whose pressure is determined by the feedback reservoir,
this choice results in a ``stiff'' equation of state $P \sim \rho^2$.  S00
provided two motivating factors for this model.  First, a self-gravitating
sheet of gas with this equation of state has a vertical scale height
that is independent of surface density, similar to observations.  Second,
if the star-formation law is given by Equation~(\ref{sflaw}), then this same 
self-gravitating sheet has a star formation rate per unit area $\Sigma_{\rm SFR}$
that scales as the gas surface mass density $\sigma$ to the 1.5 power, i.e.,
$\Sigma_{\rm SFR} \propto \sigma^{1.5}$.   In other words, the star formation
expected for an isolated disk manifestly scales in a similar fashion to the
observed Kennicutt law (Equation~\ref{KenLaw}).

While the formulation set forth by S00 was quite successful at regulating
star formation in both isolated
and merging galaxies and, as designed, reproduced the observed star formation
laws, there are several reasons to explore alternative formulations.  First,
it is not clear that all observations have converged on 
vertical scale heights of disk galaxies that are independent of radius.
Second, the observed star-formation law holds in physical environments, such
as the centers of spiral galaxies and in merging galaxies, that are very different from
isolated disks.  Thus, while the S00 equation of state naturally predicts isolated
disks that follow Kennicutt, this may be a necessary, but not a sufficient
condition.  Lastly, a quick survey of the literature returns a number of 
alternate formations of feedback which do not lead to the S00 equation of
state yet are still able to regulate star formation consistent with
the Kennicutt relation \citep[see e.g.,][]{SdMH05}.

For these reasons, we depart from S00 at this point and assume
$f(\rho)$ is a general power-law function of density, i.e.,
\be f(\rho) = \frac{1}{\tfb}
   \left\{ \begin{array}{l} 
	\left(\frac{\rho}{\rt}\right)^{(1-n)/2},
		{\rm for} \rho > \rho_{\rm th}  \\
	\left(\frac{\rho}{\rt}\right)^{-1/2}, 
		{\rm for} \rho < \rho_{\rm th}
	\end{array} \right.
\label{eq:fbf}
\ee
where $\rt$ is the critical density for star formation and we 
have introduced two free parameters: $\tfb$, the timescale with
which feedback energy is dissipated (and hence the efficiency
of the feedback), and $n$ which sets the polytropic index.  The
power-law below $\rt$ is fixed at $n=2$ to ensure that the 
feedback dissipation timescale gets shorter for low-density gas.
We note that setting $\tfb = \sqrt{\rt} / \beta$ and
$n=2$ transforms our model to the S00 model.

In order to gain some insight into our model, as well as facilitate the
selection of parameters to explore, we next cast our model in terms of
an effective temperature and pressure in star-forming regions.
Specifically, we consider a region of gas where
the adiabatic energy changes are negligible, but which is also at a
density greater than $\rt$ and hence is actively forming
stars.  Under these conditions the feedback reservoir will reach a
steady state in which the energy input due to supernovae (\ref{eq:u}) 
equals the
energy lost to thermalization (\ref{eq:q}), which is subsequently radiated
away, viz.,
\be q = \epsilon_{\rm SN} \frac{c_\star}{t_{\rm dyn}}
        \frac{1}{f(\rho)} 
      = \epsilon_{\rm SN} c_\star \left(
        \frac{\tfb}{t_{\rm dyn}} \right)
        \left(\frac{\rho}{\rt}\right) ^{(n-1)/2}.
        \label{eq:effq} \ee
This results is star-forming regions that have an effective
temperature and pressure given by
\be T_{\rm eff} \simeq \frac{(\gamma-1) \overline{\mu}}{k}
        \epsilon_{\rm SN} c_\star 
        \left(\frac{\tfb}{t_{\rm dyn}} \right)
	\left(\frac{\rho}{\rt}\right) ^{(n-1)/2},
	\label{eq:effT} \ee
and
\be P_{\rm eff} \simeq (\gamma-1) \epsilon_{\rm SN} c_\star
        \left(\frac{\tfb}{t_{\rm dyn}} \right)
	\left[\frac{\rho^{(n+1)/2}}{\rt^{(n-1)/2}} \right],
        \label{eq:effP} \ee
where $\overline{\mu}$ is the mean molecular weight,
and $k$ is Boltzmann's constant.

As mentioned previously, and shown explicitly by Equation~(\ref{eq:effP}),
$n$ fixes the polytropic index, i.e., the power-law dependence of
pressure on gas density $P \propto \rho^{1+n/2}$ (remember that
$t_{\rm dyn} \propto \rho^{-1/2}$).
When $n=0$, the pressure increases linearly
with the density, as in an isothermal gas.  When $n>0$, the
pressure increases faster than linear, and in this case the gas has a
``stiff'' equation of state.  In this work we will investigate three
choices of $n$: 0, 1 and 2, thus from isothermal to the same $n=2$
as investigated in S00.  While the present study assumes that $n$
is a constant, there is no reason why $n$
could not be a function of gas density.  In fact, this is
the case in a recent study by \citet{SdMH05} where their
``multi-phase'' model results in $n=2.8$ near $\rt$ and then
gradually softens to $n=0.7$ at higher densities.

To minimize the effects of the feedback reservoir $q$ in
low-density regions, i.e. below the critical threshold for star 
formation, $f(\rho)$
is fixed to $\tfb \rho^{-1/2}$
(i.e., $n=2$) for $\rho < \rt$.  The coefficient
$\tfb$ ensures that $f(\rho)$ is continuous at $\rt$.

The numerical implementation of these star formation and feedback
processes is handled in a continuous fashion on a particle-by-particle
basis. At each timestep, active gas particles can form
stars and instantaneously deposit energy into the ISM.
Thus, Equations~(\ref{eq:u}) and (\ref{eq:q}) become 
source terms in the standard hydrodynamic equations that are
evolved within GADGET.

% 
% ------------------------------
\subsection{Parameter Sets}
\label{ssec:psets}

% ---------------------------
%  Table 1 - Parameter Sets
% ---------------------------
\begin{table}
\bc
\caption{Star-formation and feedback parameter sets
used in either isolated galaxy or major-merger simulations,
or both.  $c_\star$ is the star-formation efficiency and $n$ the
density dependence of the feedback energy thermalization timescale.
$\tfb $ is the thermalization timescale, and $T_{\rm eff}$
the effective temperature, in Kelvin, at the threshold density
for star formation $\rt$. S00 used the same feedback parameters
as our $n=2$ ``low-feedback'' model, but a lower star-formation
efficiency which is below the Kennicutt value.}
\begin{tabular}{lcccc}
\hline
Model & $c_\star$ & $n$ & $\tfb$ (Myr) & $T_{\rm eff}$ at
$\rt $ (K)  \\
\hline
\hline
S00 & 0.004 & 2 & 0.83 & $3.5\times 10^3 $  \\
\textit{low} & 0.03 & 0,1,2 & 0.83 & $2.6\times 10^4 $ \\ 
\textit{medium}   & 0.03 & 0,1,2 & 8.3 & $2.6\times 10^5 $  \\
\textit{high}  & 0.03 & 0,1,2 & 82.7 & $2.6\times 10^6 $  \\
\hline
\end{tabular}
\label{tab:sffbps}
\ec
\end{table}

The star-formation and feedback model presented in the previous sections
has three free parameters: $c_\star$, controlling the efficiency of star formation, 
$\tfb$, determining the timescale of feedback thermalization, 
and $n$, specifying the feedback
density dependence, or, as Equation~(\ref{eq:effP}) demonstrates, the equation of state.  
While the star-formation efficiency $c_\star$ was fixed in
\S~\ref{ssec:sf}, $\tfb$ and $n$ are essentially free parameters.  

The approach of the work presented here is to survey a range of feedback
parameters which sample, but not exhaust, viable possibilities.  Unfortunately,
it is unclear how to judge the viability of feedback parameters.  One 
possibility that has been frequently used in the literature is to require that
the feedback be sufficient to stabilize the gaseous disk of a quiescent spiral
galaxy against dynamic instabilities.  This appears to be a zeroth-order 
requirement because observed spiral galaxies {\it are} dynamically stable.
While this procedure is well defined, it does not
generally yield a unique parameter set, but rather sets a minimum amount of 
feedback necessary to stabilize a specific galaxy model. 
Galaxy models that have high gas fractions, high baryon
fractions, and slowly rising rotation curves will be more susceptible to disk
instabilities and thus require more feedback.

In the current paper, we will also use the quiescent disk galaxy stability as
a tool to investigate parameter choices.  However, the model disk galaxy we use
is designed to represent a Sbc Hubble-type galaxy (this will be introduced in
the following section), and thus is a large, gas-rich spiral that represents one
of the most gravitationally unstable galaxies observed.  Since Sbc spirals {\it are}
stable, it seems reasonable to require that any plausible feedback model will
need to stabilize our
Sbc model.  In practice, the feedback model may be more stable, and thus we will 
investigate a range of parameter sets and attempt to understand their consequences.

In order to define a reference feedback timescale $\tfb$, we look at the 
stability of our isolated galaxy model gas disk as measured by the Toomre
$Q$ parameter \cite[eq. 6-49]{BT}, i.e.,
\be Q=
\frac{c_s \kappa}{\pi G \Sigma_{\rm gas}} ,
\label{eq:tq}
\ee
where $c_s^2$=$\gamma$$P\rho^{-1}$ is the sound speed, $\kappa$ the
epicyclic frequency, and $\Sigma_{\rm gas}$ the gas surface density.
For our model, the pressure of star-forming gas, and hence its
sound speed, is approximated by Equation~(\ref{eq:effP}).  We also note that
observations suggest that most disk galaxies reside very close to 
$Q\sim$1 \citep{MK01,WB02}.  Thus, by inserting (\ref{eq:effP}) into
(\ref{eq:tq}), and setting $Q=1$ we arrive at an expression for the
feedback efficiency
\be
\tfb =  \frac{t_{\rm dyn}}{\gamma (\gamma -1) \epsilon_{\rm SN} c_\star}
        \left(\frac{\Sigma_{\rm gas} \pi G}{\kappa} \right)^2
        \left(\frac{\rho}{\rt} \right)^{(1-n)/2} .
        \label{eq:b} \ee
Thus, via Equation~(\ref{eq:b}), $\tfb$ can be determined from the
star-formation efficiency $c_\star$, the feedback density dependence
$n$, the dynamical time at a specified density and a disk-galaxy 
model which specifies $\kappa$ and $\Sigma$.  Equation~(\ref{eq:b})
has the nice property that if it is evaluated at $\rho=\rt$, as we
will do here, then the dependence on $n$ is removed.
In practice, both $\kappa$ and $\Sigma$ are functions of disk radius,
but their ratio reaches a minimum at approximately twice the
exponential disk scale length and we adopt this value for use in
Equation~(\ref{eq:b}).

We will label the $Q=1$ case as the ``medium'' parameter set and use this
as a reference for explorations to a super-stable ``high'' parameter set
and an unstable ``low'' parameter set, which are ten times higher and
lower, respectively, and should maintain the disk at $Q\simeq3$ and
$Q\simeq0.3$.  These values of $\tfb $ serve to sample the range of
plausible feedback efficiencies, and our model disk galaxy will be
simulated in isolation and during a major merger for each parameter set.
Table~\ref{tab:sffbps} lists the three values of $\tfb$ as well as
the labels that will be used to reference each parameter set from this
point forward.

%---------------------
%    Isolated Disks
%---------------------

\section{Isolated Disk Galaxies}
\label{sec:isodisks}

\subsection{The Basics of Model Disk Galaxies}
\label{ssec:basics}

% -----------
%  Table 2
% -----------
\begin{table}
\begin{center}
\caption{Properties of the Sbc disk-galaxy model used in this
work.  M$_{\rm vir}$ is the virial mass, including dark matter
and baryons.  The dark-matter halo has a NFW profile with a
concentration $c$ and spin parameter $\lambda$, before this
is contracted due to the presence of the baryons.  The baryonic
mass is distributed into one of three components: an exponential
stellar disk, an exponential gas disk, or a stellar bulge.  The
mass fraction of each baryonic component is given by m$_d$, m$_g$, and
m$_b$.  R$_d$ is the
stellar disk exponential scale length and z$_0$ is the vertical
scale height.  The gas disk has a scale length $\alpha$ times
R$_d$.  Both gas and stellar disks are rotationally supported
with angular momentum j$_d$ and j$_g$ given in terms of the
total angular momentum.  Finally, the bulge has a spherical
exponential distribution with scale radius R$_b$.
}
\begin{tabular}{lr}
\hline 
 &  Sbc  \\
\hline
\hline
Virial Mass, M$_{\rm vir}$  & 8.12 $\times 10^{11}$ \msun\\
Concentration, c=R$_{\rm vir}/r_s$  & 11 \\
Spin Parameter, $\lambda$  &  0.05 \\
Disk Mass Fraction, m$_d$  &  0.048 \\
Radial Disk Scale Length, R$_d$  & 5.5 kpc\\
Vertical Scale Height, z$_0$  & 1.0 kpc\\
Stellar Disk Spin Fraction, j$_d$  & 0.058 \\
Gas Mass Fraction, m$_g$  & 0.065 \\
Gas Scale Length Multiplier, $\alpha$  & 3.0 \\
Gas Spin Fraction, j$_g$  & 0.219 \\
Bulge Mass Fraction, m$_b$  & 0.012 \\
Bulge 3D Scale Length, R$_b$  & 0.45 kpc\\
\hline
N$_{\rm halo}$  & 100,000 \\
N$_{\rm disk}$  & 30,000 \\
N$_{\rm gas}$   & 30,000 \\
N$_{\rm bulge}$ & 10,000 \\
\hline 
\end{tabular}
\label{tab:ics}
\end{center}
\end{table}

In this section we describe the basic ingredients of our model disk
galaxy.  In general, these models are similar to those described in
\citet{H93}, \citet{MH96}, \citet{SW99}, and S00, as they contain a
stellar and gaseous disk, which is rotationally supported, a spheroidal
stellar bulge, all surrounded by a massive dark-matter halo.  In the 
following subsection we will describe the parameter choices for our fiducial 
model, which is designed to approximate a large, gas-rich Sbc disk
galaxy.  In \S~\ref{sssec:comp} we will also build an exact
replica of the model used by S00, which will aid in our comparison;
however, we postpone discussion of this model until then.

The initial disk is rotationally supported and is assumed to have the
following distribution
\be
\rho_d(R,z) = \frac{M_d}{4\pi z_0 R_d^2}
              {\rm sech}^2\left(\frac{z}{2 z_0}\right)
              {\rm exp}\left(-\frac{R}{R_d}\right),
              \label{eq:expdisk} \ee
where $M_d$ is the disk mass, $R_d$ is the radial disk scale length,
and $z_0$ is the vertical scale height.  The disk is in general
composed of both gas and stars, with a gas mass fraction
\be
 f_d = \frac{{\rm M}_{\rm d, gas}}{{\rm M}_{\rm d}}.
\ee  We also
allow for the possibility that the gas and stellar disks may have
different radial distributions.  While we continue to assume that
both distributions follow Equation~(\ref{eq:expdisk}), we will 
set the gas radial disk scale length $R_{d, gas} = \alpha R_d$.

We also define $m_d$ and $m_g$, the stellar and gaseous
disk mass fractions, as 
\be
m_d = \frac{{\rm M}_{\rm d, stars}}{{\rm M}_{\rm vir}}
\ee
and
\be
m_g = \frac{{\rm M}_{\rm d, gas}}{{\rm M}_{\rm vir}},
\ee
where M$_{\rm vir}$ is the total virial mass of the galaxy, including all
components, and M$_{\rm d}= {\rm M}_{\rm d, stars} + {\rm M}_{\rm d, gas}$.

The vertical structure of the stellar disk is fixed by requiring that
the disk be Toomre stable, i.e., $Q\geq1$.  In practice, this results
in a $z_0$ that is set to 20\% of the radial disk scale length $R_d$,
and a vertical velocity dispersion of $\sim40$~\kms at $R_d$, both
consistent with observations \citep[see e.g.,][]{Bot93}.

The vertical structure of the gaseous disk is more complicated as the
temperature is fixed by the equation of state, rather than the velocity
dispersion (which is, in turn, fixed by our choice of $z_0$).
This formalism has recently been updated by \citet{SdMH05}
to account for pressure gradients in the gas disk and hence, they claim, can
be used to construct purely gas disks.  While we have not thoroughly
investigated the effects of their new method, nor are we using it here, we note 
that after an initial equilibration period our disk, with a relatively high gas
fraction, is quite stable.  This will be shown in the following section.

We also allow for the inclusion of a centrally concentrated stellar bulge.
The bulge, when included, is non-rotating and has a spherically symmetric
exponential distribution
\be
\rho_b (r) = \frac{{\rm M}_b}{8 \pi R_b^3} {\rm exp} \left(-\frac{r}{R_b}\right).
\label{eq:expbulge}
\ee
The exponential distribution is supported by observations which show
that late-type spirals typically have bulge profiles more consistent with 
exponential than the commonly assumed $r^{1/4}$ profile
\citep{dJ96,Bal03}.  In order to allow for this second possibility,
we also allow our model bulges to follow the \citet{H90} distribution
\be
\rho_b (r) = \frac{M_b}{2 \pi} \frac{a}{r(r + a)^3}.
\label{eq:Hbulge}
\ee
We found very little difference in the star formation between the two 
bulge distributions, and thus all galaxies in this work contain a bulge
described by Equation~(\ref{eq:expbulge}).

The disk-bulge system is surrounded by a massive dark-matter halo.  The
dark matter profile is assumed to follow the Navarro-Frenk-White (NFW)
fitting formula \citep*{NFW97}, characterized by two quantities, M$_{vir}$,
the total mass, and the halo concentration R$_{\rm vir}$/$r_s$, where 
R$_{\rm vir}$ is the virial radius and $r_s$ is a scale radius.  In all
of the galaxy models studies in this paper, the dark halo is contracted
due to the adiabatic growth of the central baryons \citep{Blu84,Flo93,
MMW98,Gne04}.

The angular momentum of the dark halo is set by the spin parameter $\lambda$.
In the \citet{MMW98} model for disk formation, the specific angular momentum
of the baryonic component is conserved, and thus fixing $j_d$, the disk
angular momentum fraction, sets the size of the corresponding disk.  We 
keep this formalism for the stellar disk, but treat the gaseous disk
separately.  The gas disk, whose size is a specific fraction ($\alpha$) of 
the stellar disk, has a much larger angular momentum fraction, $j_g$.

Once the above parameters are selected, the compound galaxy is constructed 
in a manner suggested by \citet{H93}, \citet{SW99}, and S00.  In short, this
method fixes the mass distribution according to the above fomula, and then
assumes the velocities can be approximated by the Jeans equations.  We refer
the interested reader to these works for more details.

Each disk galaxy is represented by N particles.  In \S \ref{ssec:res} we
investigate the choice of N by running some higher resolution simulations.
Motivated by the criteria of \citet{Pow03}, all
simulations presented here adopt a gravitational softening length $h=400$~pc
for the dark matter particles and 100~pc for the stellar and gas particles.
We remind the reader that, in GADGET, forces between neighboring particle
become non-Newtonian for separations $<2.3\,h$.

Now that we have outlined the basic ingredients of our disk galaxy model,
we next introduce the specific model used for our investigation of star formation
and feedback.

\subsubsection{Sbc Galaxy Model}
\label{sssec:sbc}

Since our primary goal in this work is to investigate star formation and 
feedback parameters, we select a galaxy model which is gas-rich and thus likely
to be gravitationally unstable.  To this end we choose a large gas-rich disk galaxy
whose properties are motivated by local Sbc galaxies.  This type of initial
condition has the added benefit that the large gas supply provides a significant
fuel supply and is thus likely to generate a large burst of star formation when
participation in a galaxy merger.  Furthermore, large, gas-rich disks such as
this may be similar to disks at high redshift where gas fractions are presumably
higher.

Because we choose the parameters of this model from observations of local
Sbc's, we will need to make some inferences to translate the observed 
quantities to our model inputs.  To begin, we note that the optical radius
R$_{\rm opt}$, dynamical mass M$_{\rm dyn}$, ratio of HI to M$_{\rm dyn}$ 
within the optical radius, and absolute blue luminosity L$_B$ are neatly 
compiled as a function of Hubble type in \citet{RH94}.  Thus, for instance,
the radial disk scale length is determined from the optical radius by assuming
the relation R$_{\rm opt}$ = 3.2 $\times$ R$_d$, which is valid for a Freeman
disk.

Next, the bulge-to-disk ratio and bulge size are constrained to be
consistent with the Sbc galaxies as found in \citet{dJ96}.  Because we
have selected a late-type spiral, this necessarily results in a small
compact bulge.  We can determine the total stellar mass by using the
average Sbc B-band luminosity L$_B$ and the B-band mass-to-light ratio
of 1.0 found for late-type spirals \citep{BdJ01}.  The total stellar
mass and the bulge-to-disk ratio can be used to determine the exact
mass in both the stellar disk and bulge components.  We note that with
these assumptions the bulge makes up slightly less than 10\% of the
baryonic mass.

Sbc galaxies are gas-rich.  Observationally, there is a wide range of
gas properties associated with Sbc galaxies, but in general, the gas
component is roughly exponentially distributed and more extended than the
stellar disk \citep{BvW94}.  Taking this into consideration, we assume
the gas disk to have an exponential profile with a gas scale radius
R$_g$ which is a factor $\alpha$ times the stellar disk scale radius,
R$_d$.  Based on the most extended distribution for Sbc galaxies
reported in \citet{BvW94}, we choose $\alpha=3$.  
The mass of the gaseous components
can be found from the ratio of HI to M$_{\rm dyn}$ within the optical
radius, with our assumed profiles, corrected for helium.

At this point, the mass and distribution of the baryonic components
are completely specified.  However, the dark matter distribution is
not.  In order to get some handle on the dark mass and profile we can
use the dynamical mass.  In practice, we assume an adiabatically
contracted NFW halo with a concentration of 11 and then iteratively
search for dark matter mass which gives us the correct dynamical mass
at R$_{\rm opt}$.  This procedure results in a dark halo which is
$8.1\times10^{11}$~\msun.  The resulting rotation curve has a maximum
speed of approximately 204 \kms~which is larger than the
$\sim190$~\kms$ $ suggested by the baryonic Tully-Fisher relation
\citep{BdJ01}.  The large rotation velocity of this model is a typical
byproduct of adiabatically contracting an NFW halo \citep[see,
e.g.,][]{Cole00}.

The above procedure does not use the halo spin parameter to set the
size of the stellar disk.  Arguably, the actual scale lengths of
galaxies are more well constrained than halo spin parameters.  This
also means that the angular momentum of the disk components is
independent of the halo angular momentum.  The halo spin parameter was
set to the median value from \citet{Bul01J}. Decoupling of the
halo and disk angular momenta is also motivated by the discrepancy
between the specific angular momentum distribution of a typical dark
matter halo and that of an exponential disk \citep{Bul01J}.

Each disk galaxy is represented by 170,000 N-body particles, 100,000 of which
represent the dark-matter halo, 30,000 each for the stellar and gas disks,
and 10,000 for the bulge.  In \S\ref{ssec:res} we investigate our choice
of particle number.  Table~\ref{tab:ics} lists the
properties for our Sbc disk-galaxy model.

%
% ----------------------------------------------
\subsection{Star Formation}
\label{ssec:isosfr}

% -------------------------------------
%  Figure 1 : Isolated Star Formation
% -------------------------------------
\begin{figure*}
\begin{center}
\resizebox{5.5cm}{!}{\includegraphics{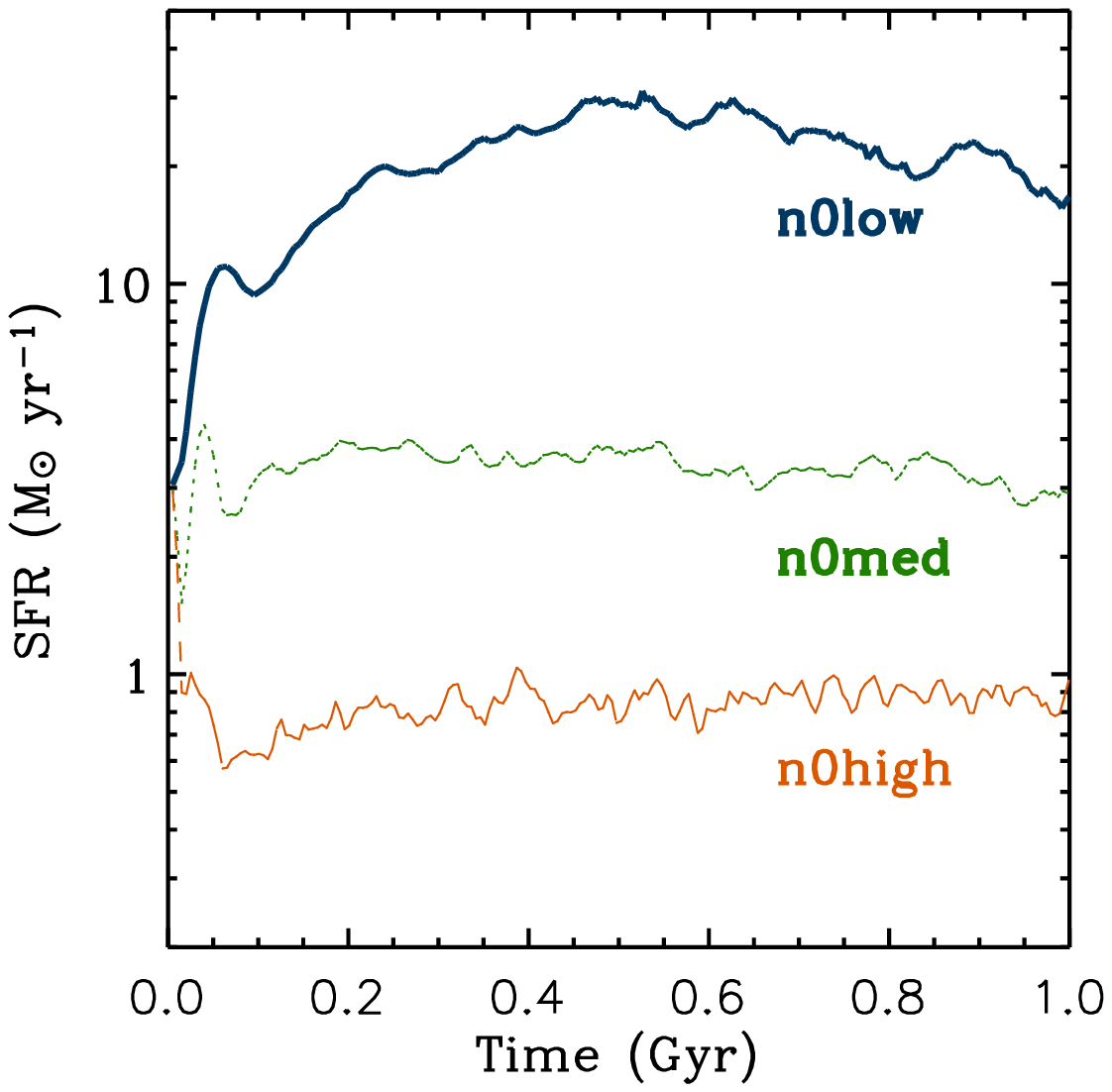}}
\resizebox{5.5cm}{!}{\includegraphics{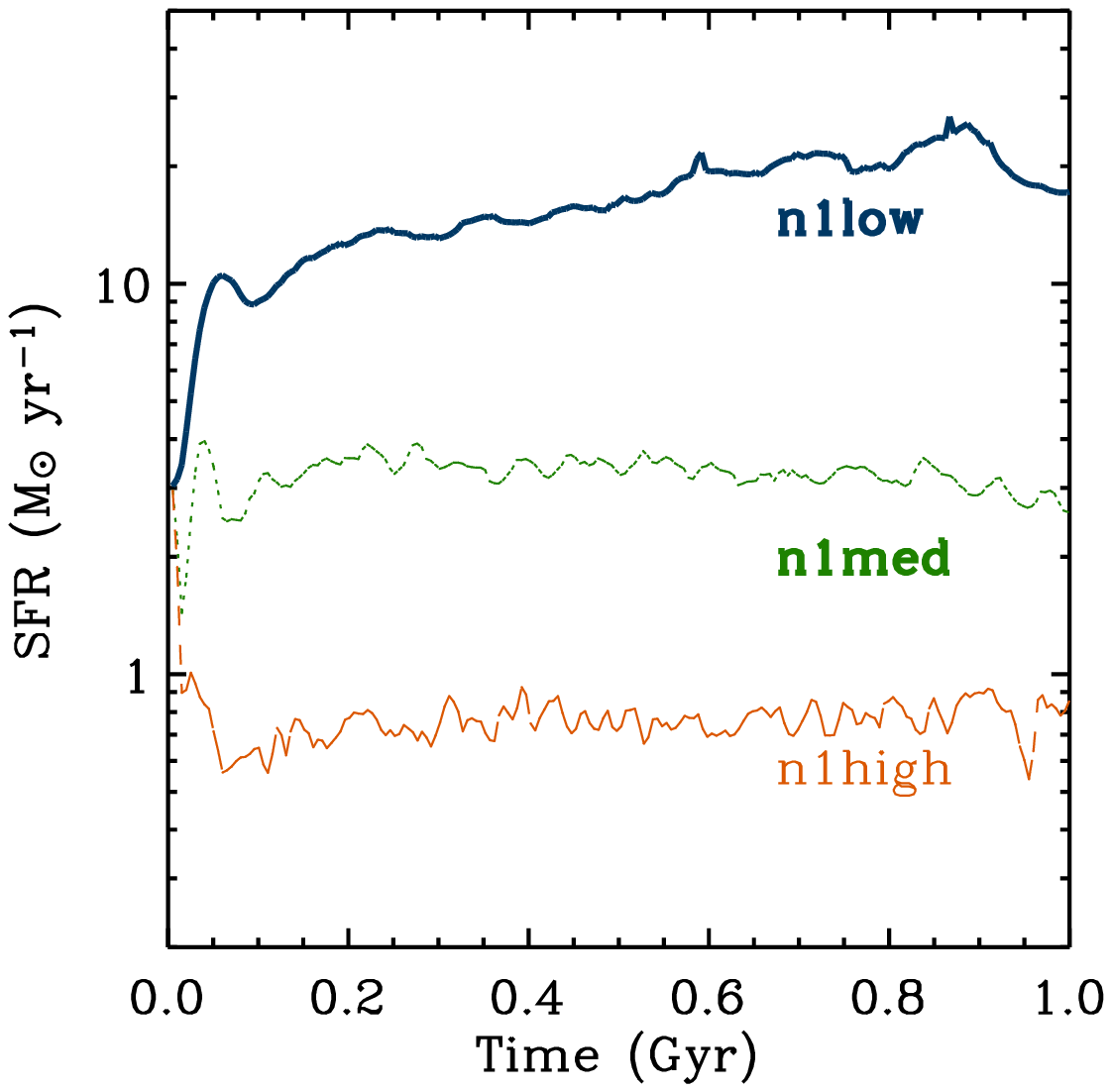}}
\resizebox{5.5cm}{!}{\includegraphics{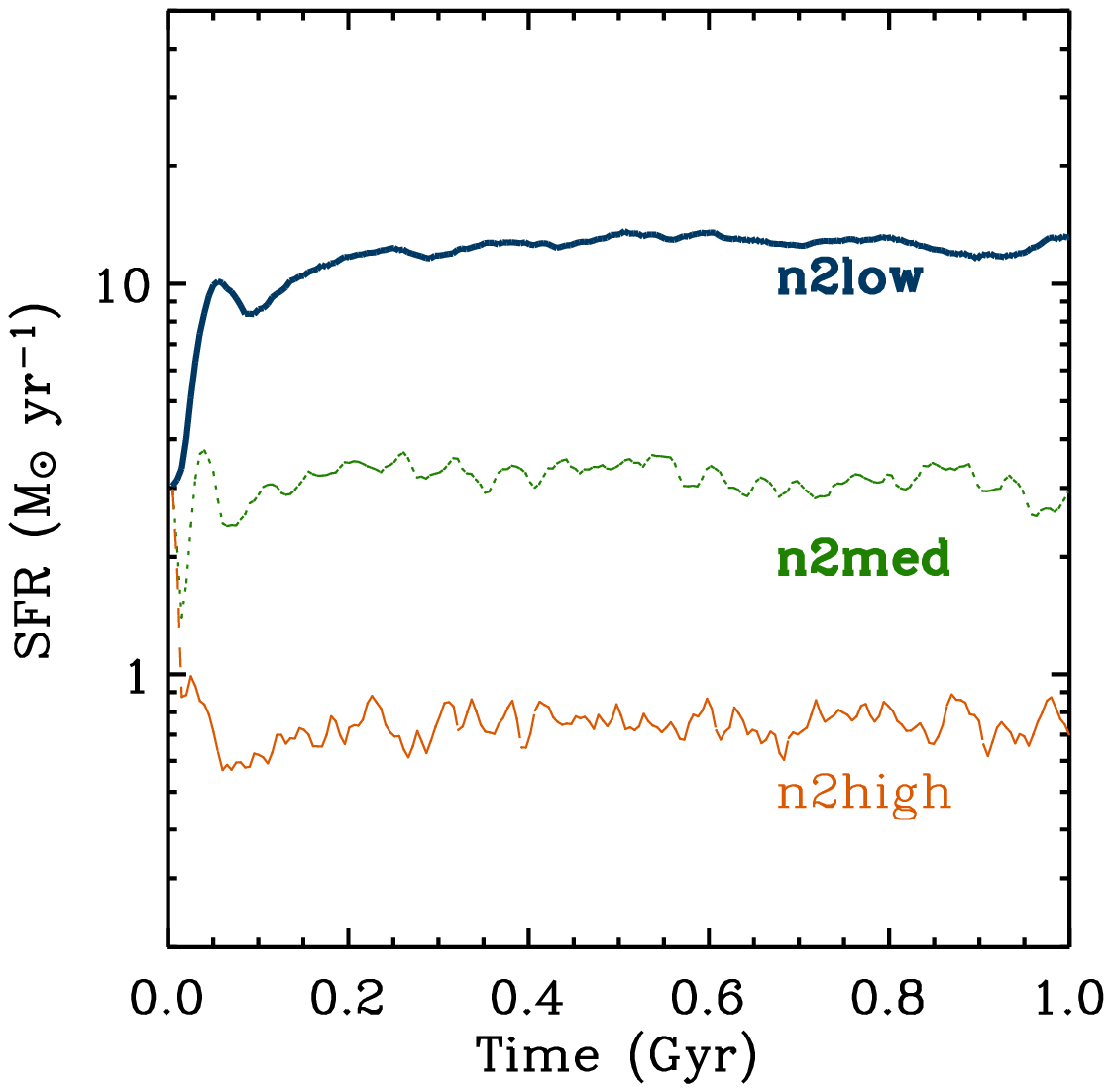}}\\
\caption{Star formation for an isolated Sbc disk galaxy.
The left, middle and right show the runs for the 
$n=$0, 1 and 2, feedback models, respectively.  The 
``low'' feedback model has the highest star-formation rate
and the ``high'' feedback model has the lowest rate.
For a given  $n $, the
star-formation rate scales roughly as $\tfb^{-1/2} $.
\label{fig:isosfr}}
\end{center}
\end{figure*}

% -----------
%  Figure 2
% -----------
\begin{figure*}
\begin{center}
\resizebox{16.5cm}{!}{\includegraphics{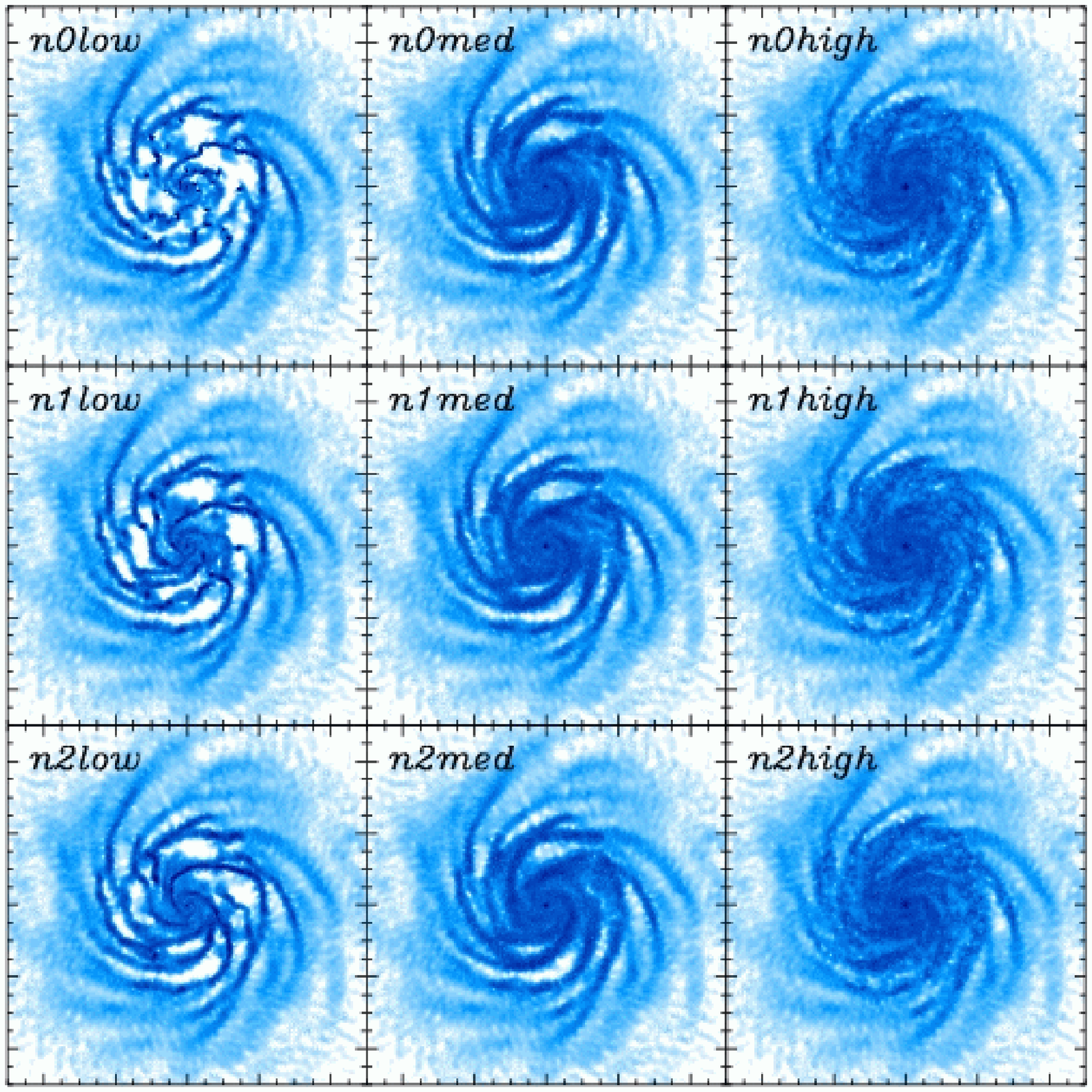}}
\caption{Projected gas density for the Sbc disk galaxy
model for each parameter set shown after 500~Myr, or
$\sim3$ rotation periods at the half-mass radius, of 
evolution. 
The ``low'' parameter sets have led to the break up of the
gaseous disk and hence accelerated star formation as shown in
Fig.~\ref{fig:isosfr}.
\label{fig:isogsd}}
\end{center}
\end{figure*}

The feedback model outlined in \S\ref{sec:sims} was designed to
stabilize the gas disk and prevent excessive star formation due to
instabilities in the gas disk.  To test our model, the Sbc galaxy
model was simulated in isolation for one Gyr, with the last nine
parameter sets in Table~\ref{tab:sffbps}.  The resulting
star-formation rates are shown in Fig.~\ref{fig:isosfr}.  Except for
\nzl\ and \nol, all models have continuous star
formation at a roughly constant rate.  The spikes of large star
formation in the first $\sim100$~Myr of the simulation result from the
initial model not being exactly in equilibrium.  In particular, the 
feedback reservoir $q$ is empty at the beginning of the simulation and 
thus provides no pressure support in the initial disk.
The low-feedback parameter sets
have star-formation rates of over 10 \msunyr, the medium-feedback
parameter sets 2 -- 3 \msunyr, and the high-feedback parameter sets
$<1$ \msunyr.
The equation of state index, $n$, has a relatively minor effect on the
average star-formation rate of the
isolated galaxies.  For instance, run \nzm\ has an average star-formation rate of
3.3 \msunyr, while the average of \ntm\ is 3.0 \msunyr.

Figure~\ref{fig:isogsd} displays the gas surface density for the
models.  All models with constant star formation are qualitatively
similar.  In the \nzl\ and \nol\ models, the gas disk has fragmented
due to the growth of instabilities in the inner disk.  The \ntl\ model 
also looks fragmented, but its star formation rate is still regulated
at a constant rate of 10 \msunyr.  This evolution is consistent with 
other studies that have evolved isolated disks with insufficient
feedback \citep[S00;][]{Bot03,Rob04,SdMH05,Li05,Li06}.  We also note that these
instabilities were likely seeded by noise in the potential \citep{T81,
H93} and because the initial disk is not in perfect equilibrium.  In 
the stable disks, this noise is then swing amplified to produce a 
spiral pattern.  Representing the halo with a fixed potential or 
increasing the number of particles used for the halo both delayed the 
onset of gravitational instabilities or spiral structure.

The gravitational instabilities are also manifested in the star-formation
rates which are much larger than the stable disks' and are not maintained at
a constant value.  This is especially apparent in the $\nzl$ and $\nol$
models in which run-away star formation progresses until a large fraction 
($\sim60$\%) of the original gas is consumed.  In general, however, relatively
little of the original gas is converted to stars during the 1~Gyr evolution
we follow; only $\sim$~5\%  for the $high$ feedback models and $\sim20$\%
for the $medium$ feedback models.   One reason for these low gas consumptions
is the very extended distribution of gas.  Half of the original gas mass is
at radii greater than 27.6 kpc, whereas star formation predominantly takes
place at the galaxy center and along transient spiral structures,
always at radii less than 25 kpc.  A more subtle feature is the transient
spike of star formation at $T\approx0.05$~Gyr, present in all models, that
results because the initial system in not in perfect equilibrium.  One
reason for this is that the initial feedback reservoir is initially empty
and requires a small amount of star formation to fill it sufficiently enough
to provide pressure support.  We tested one simulation that started with 
a filled feedback reservoir and while the initial transient star-formation
rate was suppressed, the evolution beyond $T=0.1$~Gyr was unchanged.

Within our model, the time-averaged star-formation rate in a quiescent
spiral disk is directly set by $\tfb$.  Figure~\ref{fig:isosfr} demonstrates
that each $low$, $med$, and $high$ feedback model forms stars at a nearly
constant rate, regardless of $n$.  We can understand the origin of this 
scaling by considering that the gas disk is in hydrostatic equilibrium 
within the galaxy potential.  Since the gas is a small fraction of the total
mass, the potential and thus the gas pressure (if it is in hydrostatic equilibrium)
are expected to be fixed by the galaxy model.  In other words, the gas
pressure is constant regardless of the specific feedback model.  For
star-forming gas, where the effective pressure is provided by 
Equation~(\ref{eq:effP}), this yields a simple relation between the gas density
and the feedback timescale
\be
\rho \propto \tfb^{- 1/(1+n/2)}. 
\ee
Since the integrated star formation rate is 
\be
 SFR \propto \dot\rho_\star V \propto \dot\rho_\star/\rho
\propto \rho^{0.5}, 
\ee
where the third proportionality comes from Equation~(\ref{sflaw}),
it follows that 
\be\label {eq:sfrn} 
SFR \propto \tfb^{-
  1/(2+n)}.  
\label{eq:sfscal}
\ee 
In the simplest, $n=0$, case the star-forming gas has an isothermal
equation of state and Equation~(\ref{eq:sfscal}) predicts 
the star formation to scale as $SFR \propto \tfb^{-1/2}$.  In our
range of models, $\tfb$ was varied by factors of ten so that we expect
the star formation rates to vary by a factor of $\sqrt{10} \sim 3$.
This appears to match the simulated star-formation rates quite well, 
where the values for the $high$, $med$, and $low$ models are
$\sim 1$, 3, and 10.

These star-formation scalings are complicated by the distinct density
threshold for star formation as well as the self-gravity of the gas.
In particular, the $n=2$ model appears to follow the $\tfb^{-1/2}$
scaling the closest, yet its expected scaling is a much softer
$\tfb^{-1/4}$.  The reason for this discrepancy, as we will see 
more clearly in \S~\ref{ssec:isophase}, is that most of the
gas within these models piles up at the threshold density for star formation
$\rt$ and thus has the effective temperature at this density
($T_{\rm eff}$ in Table~\ref{tab:sffbps}).  Thus, the gas scales 
as an isothermal gas rather than $n=2$.

%
% --------------------------------------
\subsection{The Kennicutt Law}
\label{ssec:isokenn}

% -----------
%  Figure 3
% -----------
\begin{figure}
\begin{center}
\resizebox{8.0cm}{!}{\includegraphics{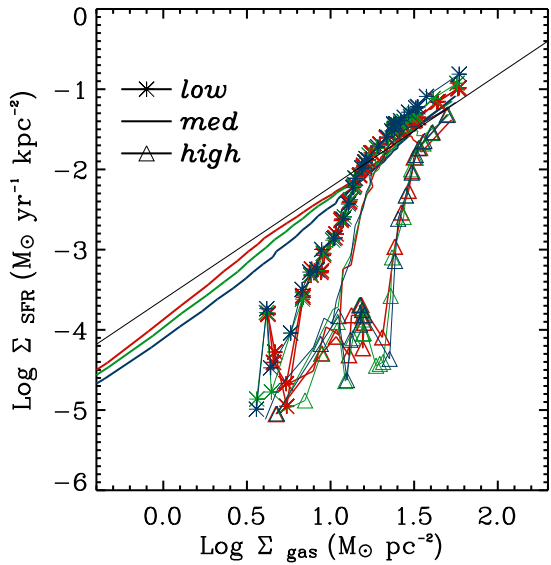}}\\
\caption{Star-formation rate per unit area and gas surface density averaged
within azimuthal annuli for the fiducial isolated Sbc disk galaxy at 
$T=0.1$~Gyr.  Shown with different symbols are the $low$ (asterisks), 
$med$ (continuous line), and $high$ (open triangles) parameter sets for
each value of $n$ ($0=$~blue, $1=$~green, $2=$~red).  Also shown are three
$med$ feedback runs that did not include an explicit gas density 
threshold for star formation.
The straight (black) line is the empirical Kennicutt law.  This plot
can be directly compared to Figure 3 from \citet{Kenn98}.
\label{fig:isokenn}}
\end{center}
\end{figure}

Star formation in the simulations was designed 
to be consistent with the empirical Kennicutt law
\citep[Equation~\ref{KenLaw}]{Kenn98}.  The success of this is
verified in Figure~\ref{fig:isokenn}, where we show the star-formation
rate and gas mass averaged in azimuthal annuli for the isolated
Sbc galaxy simulations.  The Kennicutt law is shown as a straight black
line, which the simulations closely track.

While Figure~\ref{fig:isokenn} demonstrates that all simulated 
disks track the Kennicutt law at high gas surface densities, the 
simulations that included a threshold density $\rt$ fall below the
Kennicutt law at lower gas surface densities.  As mentioned
previously, a similar feature is observed by \citet{Kenn98} near
$\sim 10$~kpc~pc$^{-2}$.  Figure~\ref{fig:isokenn} also shows that
higher feedback models fall below the Kennicutt law at higher gas
surface densities.  This is a result of the increased vertical scale
height owing to the high feedback pressurization.

To ensure that it is $\rt$ that sets the departure from the observed
Kennicutt law we reran the three medium feedback models ($n0med, n1med$,
and $n2med$) with $\rt=0$.  Figure~\ref{fig:isokenn} shows that these
simulations closely track the Kennicutt law across all gas densities.

%
% --------------------------------------
\subsection{Disk Scale Heights}
\label{ssec:diskz}

The differences in star-formation rates between the low-, medium-, and
high-feedback parameter sets is accompanied by corresponding
differences in the scale height of the gas.  The medium-feedback
simulations have scale heights of $\sim$~300 parsec at a radius of 5
kpc, increasing to 450 parsec at 25 kpc where star formation
ceases.  The high-feedback simulations, in contrast, have scale
heights of 450 parsec at 5 kpc radius, increasing to 500 parsec at 25
kpc.  The $n=0$~and~1 low-feedback simulations do not have a well
defined scale height because the unstable disk has fragmented.
The $n=2$ low-feedback simulation has a scale height of $\sim$~200 
parsec, near the numerical resolution of our simulations.

For a given value of $\tfb$, the vertical structure is 
independent of the value of $n$.  This fact, which may be surprising
at first, arises because for all these models the feedback is
efficient enough to limit the maximum gas density to slightly above
the threshold density while most of the gas is at densities slightly
below the threshold, even in the star-forming regions of the disk.  Since
$n$ describes the scaling of effective pressure with density above
the threshold, and the star-forming gas is confined to a narrow range
of densities above the threshold, it follows that $n$ has only a very
small influence on the structure of the isolated galaxies.  (As
mentioned earlier, for gas below the threshold density a value of $n =
2 $ is used in all cases.)

The fact that the star formation comes from gas at the threshold
density also explains how a $\sim$~50\% difference in scale 
heights between the med- and high-feedback sets can explain a
difference in star formation rate of more than a factor of 3.  Without
a density threshold, Equation~(\ref{sflaw}) predicts that the
integrated star-formation rate
should scale as $\rho^{1/2} $, but at the threshold density the
effective star-formation law is much steeper than indicated by 
Equation~(\ref{sflaw}).

%
% --------------------------------------
\subsection{Phase Diagram}
\label{ssec:isophase}

% -------------
%  Figure 4
% -------------
\begin{figure*}
\resizebox{16.5cm}{!}{\includegraphics{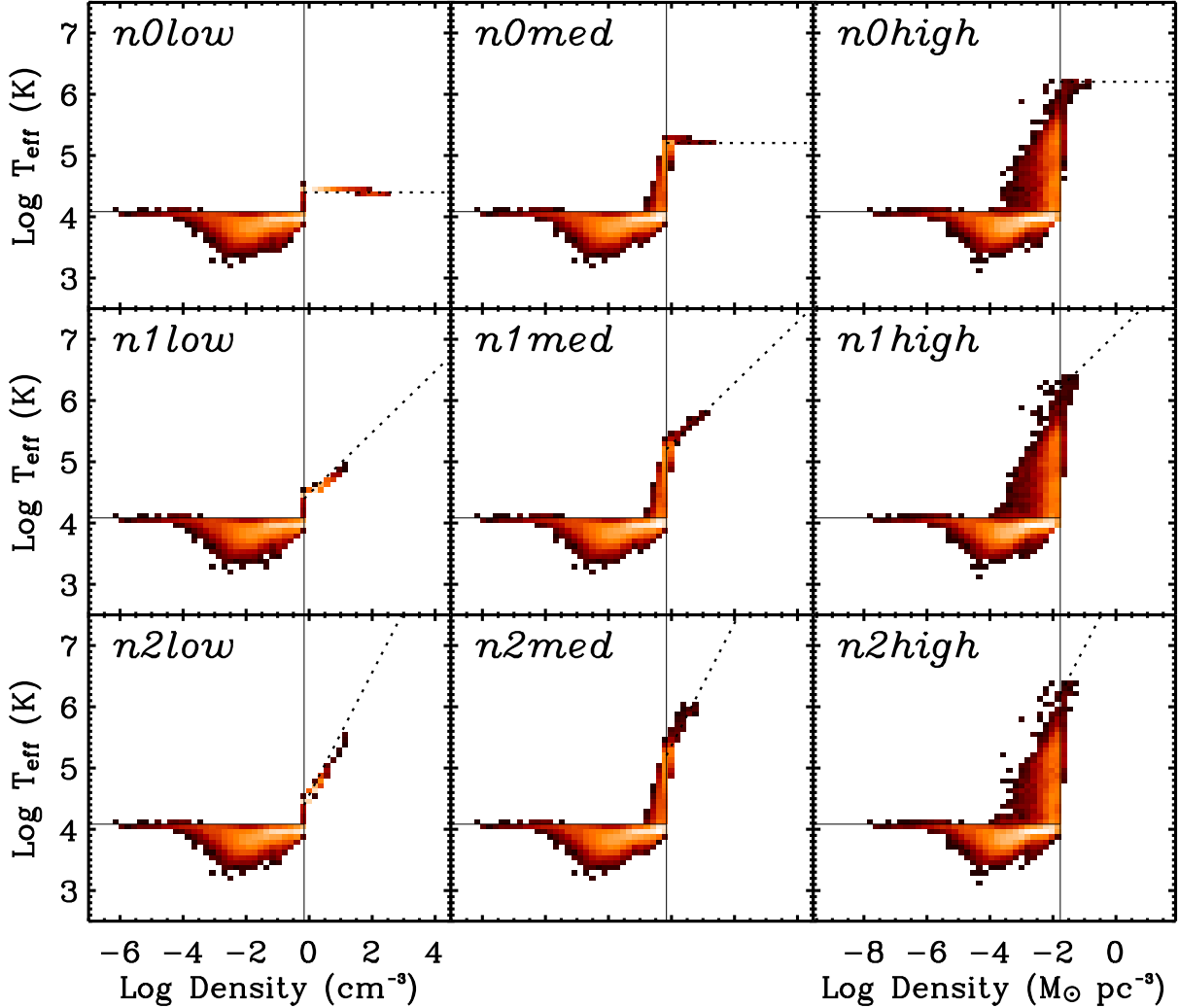}}% 
\caption{Each panel shows the effective temperature -- gas density
phase diagram for the isolated Sbc disk at $T=0.25$~Gyr for the entire
set of feedback models considered.  The effective temperature is 
calculated as
T$_{\rm eff}$=$\bar{\mu}$P$_{\rm eff}$/$k\rho$, where P$_{\rm eff}$
is given by Equation~(\ref{eq:effPfull}).  The vertical line
shows the threshold gas density for star formation $\rt$, and the
horizontal line at 1.2$\times 10^4$~K is a guide to the eye that
demarcates the floor of the cooling function.  The expected effective
temperature determined by our feedback model, Equation~(\ref{eq:effT}),
is shown by a dotted line at densities above $\rt$.
\label{fig:isophase}}
\end{figure*}

To illustrate the effects of our pressurizing feedback, and its dependence
on the free parameters $\tfb$ and $n$, we present temperature--gas
density phase diagrams in Figure~\ref{fig:isophase}.  On the vertical axis of
the phase diagram is the effective temperature, which includes a contribution,
if any, from the feedback reservoir.
There are several lines plotted in Figure~\ref{fig:isophase} that delineate
distinct phases of gas.  The horizontal line at $10^4$~K shows the initial gas
temperature as well as the minimum of the cooling function.  Most low-density
gas resides near this temperature.

The feedback reservoir becomes significant for gas which is above the
threshold density for star formation $\rt$, shown by the vertical solid
line in Figure~\ref{fig:isophase}.  At high gas densities
($\rho_{\rm gas} \geq \rt$) star formation acts to maintain the gas
at a temperature and pressure specified by Equations~(\ref{eq:effT})
and (\ref{eq:effP}), which
is shown by a dotted line in Figure~\ref{fig:isophase}.  From these
relations it is apparent that the slope of the dotted line is specified
by $n$, and the normalization by the combination $c_\star \tfb$.
For example,
when $n=0$, star-forming gas is isothermal with an effective temperature
of 2.5$\times10^4$~K, 2.5$\times10^5$~K, and 2.5$\times10^6$~K (see
Table~\ref{tab:sffbps}) for the $low$, $med$, and $high$ models, 
respectively.

Figure~\ref{fig:isophase} demonstrates the effects of the equation of state
in star-forming regions set by $n$.  For instance, the $n0low$ model is
nearly identical to no feedback at all.  Gas is always close to $10^4$~K, and
the disk is insufficiently pressurized and subsequently fragments
(see Figure~\ref{fig:isogsd}).  In the case of $n2low$, however, the steep
equation of state above $\rt$ means that high-density gas eventually becomes
pressurized and limits the amount of fragmentation.

In general, the gas phases below the threshold density $\rt$ are very
similar for each $high$, $med$, and $low$ model, regardless of $n$.
For the medium- and high-feedback models, the high temperature and
significant pressurization cause a large amount of the gas to pile up
at $\rt$.  This effect is a byproduct of the long timescale for energy
dissipation, set by $\tfb$, in the $high$ and $med$ models.  In this
scenario, because of recent star-formation, the feedback
reservoir has been filled and the increased pressure support has pushed
the particles to densities below $\rt$.

For the quiescent spiral galaxy we simulate here, we have not included any 
diffuse hot gaseous halo.  Thus, gas only becomes hot ($> 10^{4}$~K)
because of star formation.  We also expect hot diffuse gas to be efficiently
produced by shocks \citep{Cox04} and active galactic nuclei \citep{Cox06x} 
that may attend the galaxy merger.  This will be explicitly shown in 
\S~\ref{ssec:mmphase}.

%---------------------------------------
%   Major Mergers: pics, sf, rems
%---------------------------------------
\section{Major Mergers}
\label{sec:majm}

% -------------
%  Figure 5
% -------------
\begin{figure*}
\resizebox{16.5cm}{!}{\includegraphics{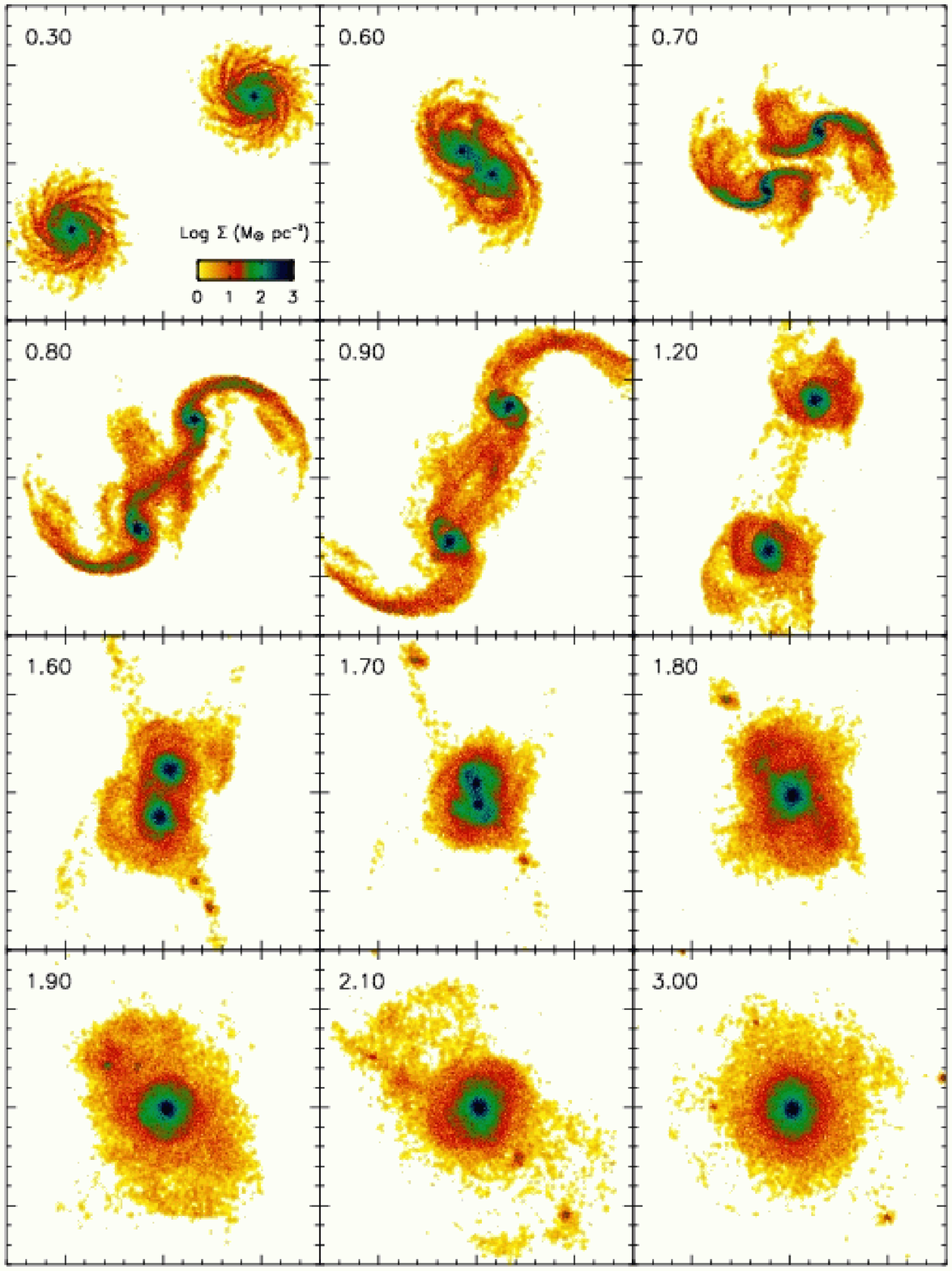}}% 
\caption{Projected stellar (disk plus bulge) density during the
\ntm\ major merger.   Each panel measures 160 kpc on a side, and 
the number in the upper left of each frame is the simulation time 
measured in Gyr.  The disk galaxy initially on the lower left is
coplanar with the orbital plane while the upper right disk is tilted
by 30$^\circ$.  The logarithmic density scale, in units of \msunpc,
is given by a colorbar in the upper-left plot.
\label{fig:mmoldstarmorph}}
\end{figure*}

% -------------
%  Figure 6
% -------------
\begin{figure*}
\resizebox{16.5cm}{!}{\includegraphics{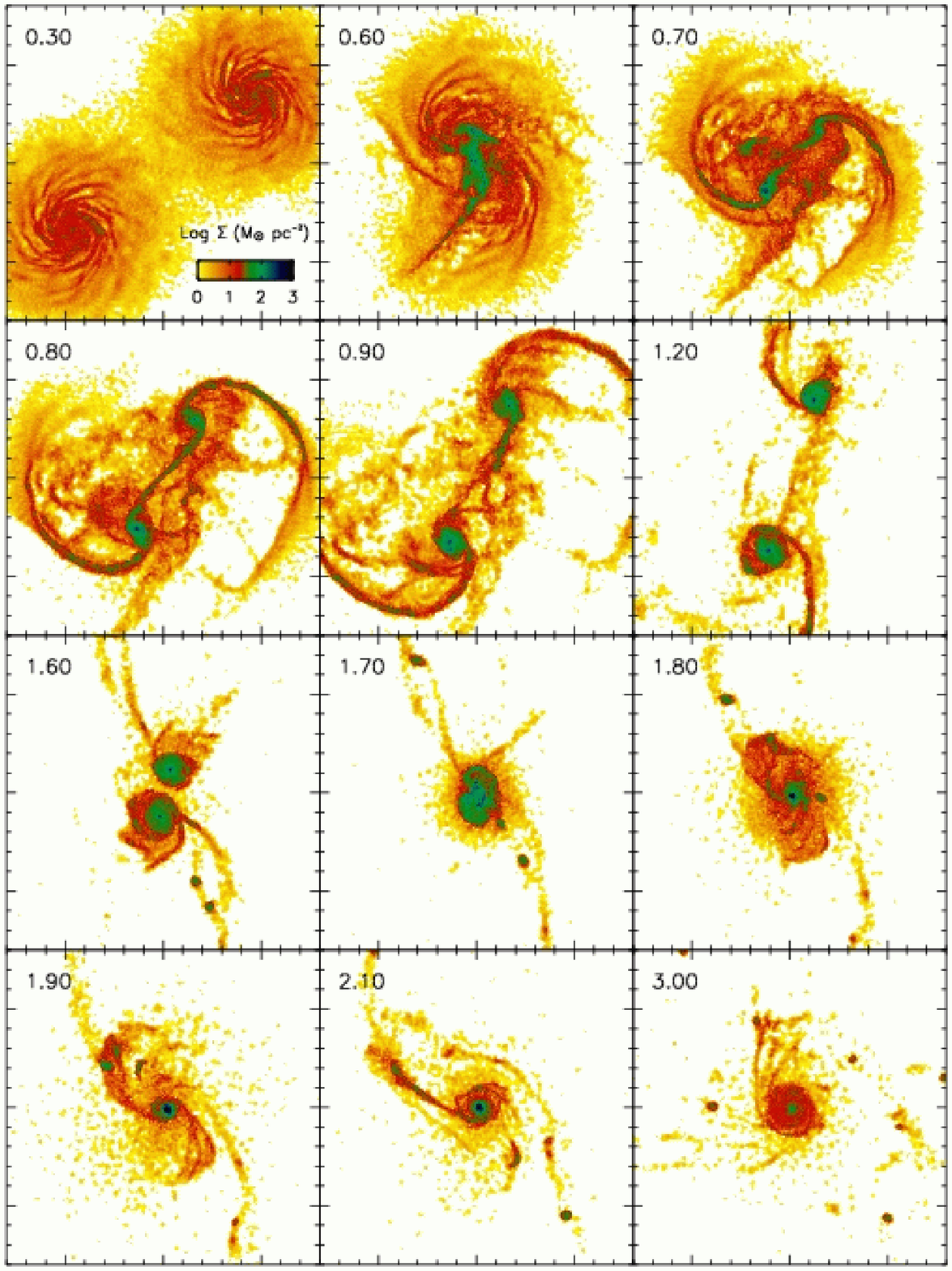}}% 
\caption{Same as Fig.~\ref{fig:mmoldstarmorph}, except here the 
projected gas density is shown. The density scale is identical to
Fig.~\ref{fig:mmoldstarmorph} and is given in the
top-left plot.
\label{fig:mmgasmorph}}
\end{figure*}

To generate a galaxy merger, identical copies of the isolated galaxy model 
are initialized on a prograde parabolic orbit with pericentric distance 
$R_{peri}=11$ kpc ($\sim$~2~R$_d$ or $\sim$~0.05~R$_{\rm vir}$) and 
an initial separation of 200 kpc.  One disk is in the orbital plane while the
other is inclined by $30^o$.  This generates a fast, nearly radial, collision 
consistent with orbits found for dark-matter halos in a cosmological 
simulation \citep{KB04}.

Figure~\ref{fig:mmoldstarmorph} displays the stellar (disk plus bulge)
density projected onto the orbital plane and Figure~\ref{fig:mmgasmorph}
shows the projected gas density from an equivalent viewing angle.  
The simulation begins with
a wide separation between the two galaxies, and they are essentially
unaffected by each other's presence.

As the galaxies reach pericenter
($T\approx$~0.6 Gyr) each disk becomes tidally distorted.  Long tidal
tails carry out loosely bound material while the central regions form
transient bar-like structures (see $T=0.7-0.9$~Gyr).  At this point
dynamical friction has increased the spin of the dark-matter halos and
extracted orbital energy from the originally unbound orbit.  As the
galaxies separate, a bridge forms that connects the two galactic
centers, and at $T=1.1$~Gyr the galaxies reach apocenter, with a
separation of 80 kpc.

The final merger is a messy process starting at
$T=1.6$~Gyr and lasting for 200 Myr.  The disks are effectively
destroyed and we are left with a spheroidal-looking object
\citep[MH96; S00;][]{HRemI}.  Owing to gas shocking, kinetic energy is
converted to thermal energy and subsequently radiated away.  This results
in an offset between the stellar and gaseous components that torques the
gas into the dense central regions of each galaxy (see $T=0.7-0.9$~Gyr).
Since we have tied the formation
of stars to the gas density through Equation~(\ref{sflaw}), a burst of
star formation ensues \citep[MH96; ][]{MH94majm}.  From $T=1.2-1.6$~Gyr,
the extended distribution of the gaseous component produces long tidal
tails while the stellar tails have already fallen back to the central
galaxy.  These gaseous tails
are sharper and thinner than their stellar companions, and continually
rain back on the central disks.

The final merger is particularly
violent for the collisional gas as large amounts of gas shocking occur
\citep{Cox04}.  These shocks provide a significant source of heating
to the gas, and its evolution depends on its density.  High-density
gas is able to radiatively cool on a short time-scale, and provides
fuel to continue the starburst.  Low-density gas, on the other hand,
expands from the central region and contributes to a hot gaseous halo
in the merger remnant.  At $T=3.0$~Gyr, 1.2 Gyr after the merger is
complete, $\sim18$\% of the current gas content has cooled and has formed a
nuclear ($<10$~kpc) disk, similar to what was found by
\citet{B02} and \citet{SH05}.  The merger remnant also contains
several tidal condensations, formed from loosely bound tidal material
and composed entirely of gas and new stars
\citep[c.f.,][]{BH96,Duc04}.

%
% -----------------------
\subsection{Star Formation}
\label{ssec:mmsfr}

% -------------
%  Figure 7
% -------------
\begin{figure*}
\begin{center}
\resizebox{5.5cm}{!}{\includegraphics{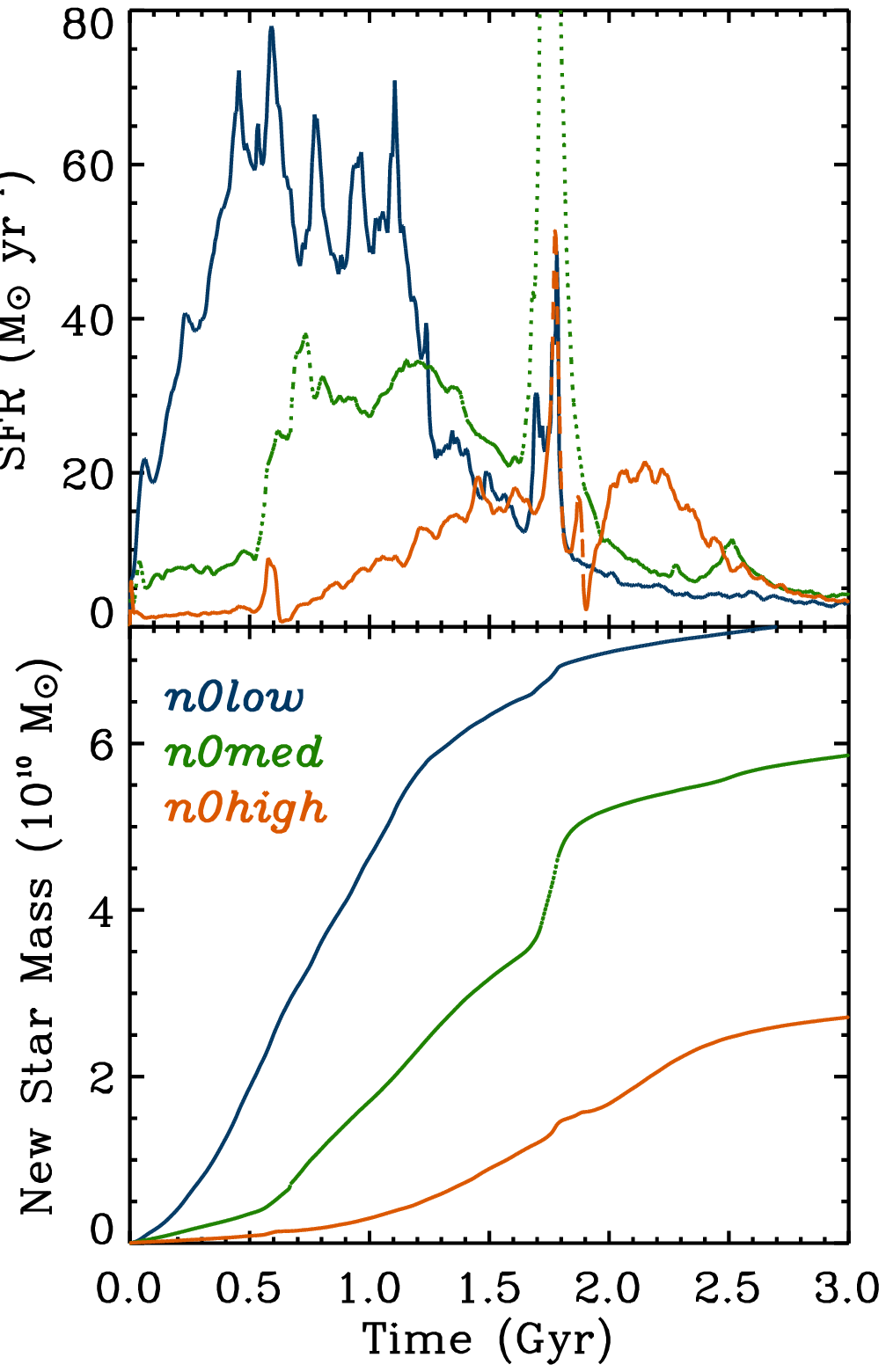}}
\resizebox{5.5cm}{!}{\includegraphics{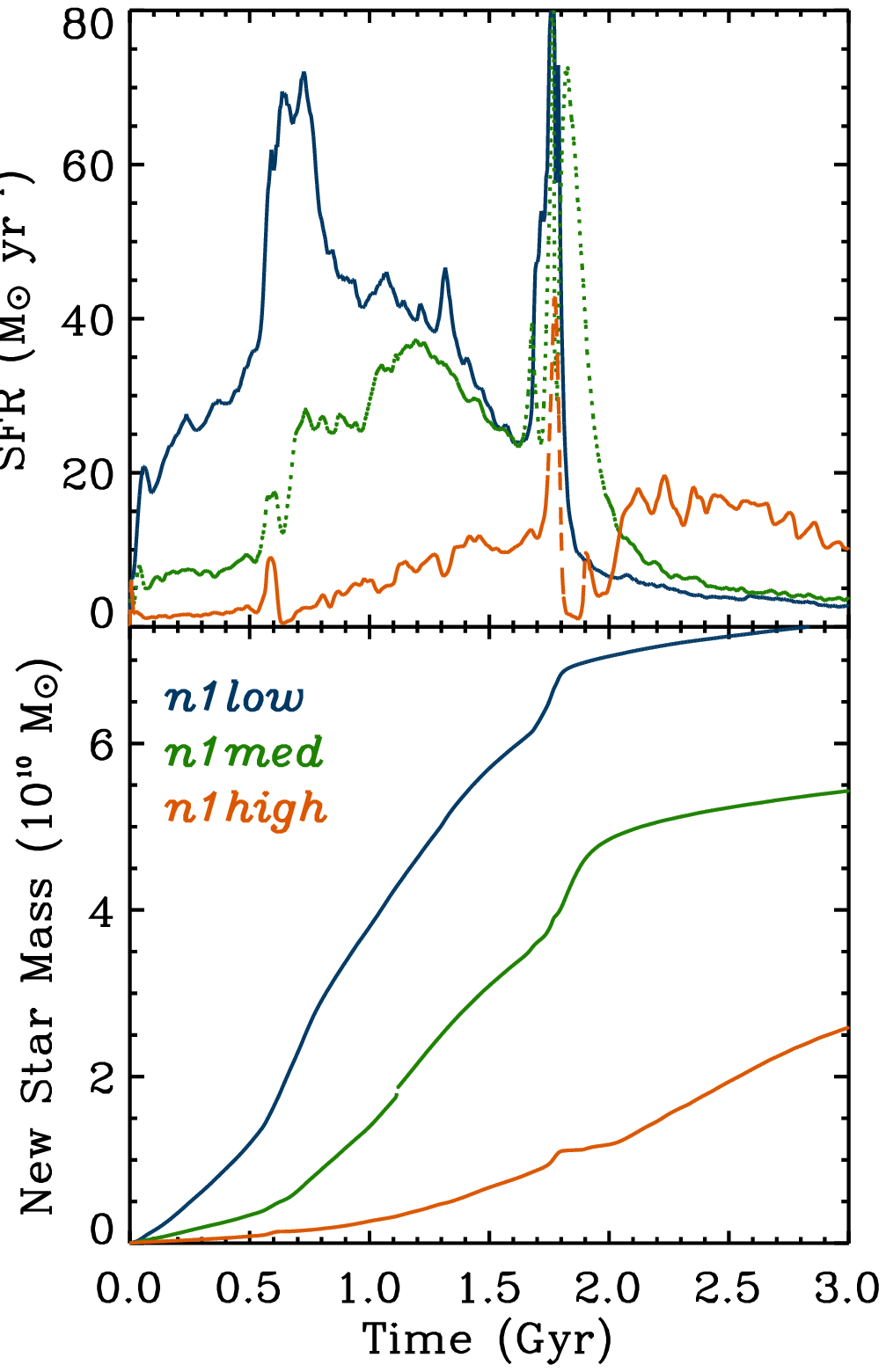}}
\resizebox{5.5cm}{!}{\includegraphics{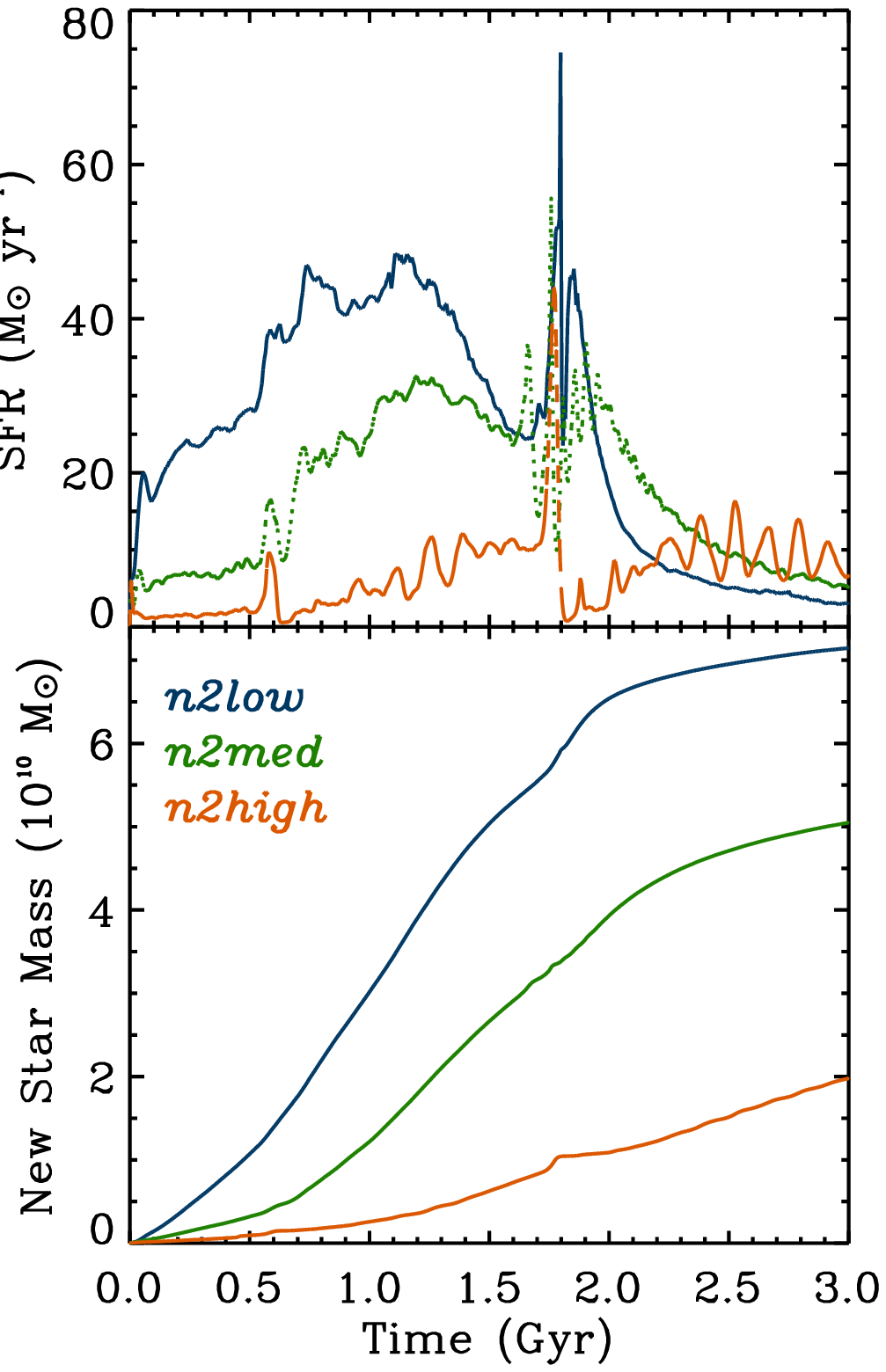}}\\
\caption{Star formation for a major merger between two identical
Sbc disk galaxies.  The $n=0,1$, and 2 feedback models are shown
in the top, middle and bottom panels, respectively.  In every instance
the low feedback model has a higher star-formation rate than the high
feedback model between $T\la2$~Gyr.  The final merger occurs at
$T=1.8$~Gyr and this is coincident with the highest peak of star
formation in all models with sufficient feedback to stabilize the
isolated disks, i.e. the medium- and high-feedback parameter sets.
\label{fig:mmsfr}}
\end{center}
\end{figure*}

Figure~\ref{fig:mmsfr} displays the star-formation rate during the
simulated major merger for all feedback parameter sets.  The
star-formation history is punctuated by several significant dynamical
events.  Prior to the first close passage ($t<0.5$) the star-formation rate
is that of two quiescent disks.  The first encounter of the disks, at
$T=0.58$~Gyr, produces a strong shock between the gas disks and a
ridge of high density gas stretches across both disks.  This feature
generates an enhancement of the global star-formation rate, especially
in the low and medium feedback models.

Subsequent to the first passage, starting at $T~\sim~0.65$~Gyr,
tidally ejected gas is steadily stripped of its angular momentum and
continuously falls back on the original disk galaxy.  This process
lasts for $\sim1$~Gyr, but the resulting star-formation rate is
strongly influenced by the feedback model.  The low-feedback models
provide gas with minimal pressure support.  Gas quickly reaches high
densities in the galaxy center, fueling a prolonged burst of star
formation. High-feedback models provide much stronger pressure
support, which limits the amount of gas at high density and hence the
star-formation rate.

At first glance, the prolonged star formation during and immediately
after the first disk encounter seems at odds with the results of MH96
and S00, whose progenitor disks, which contained a bulge as ours do
here, demonstrated very minor star formation after the first
encounter.  In fact, the galaxy models employed by MH96 and S00
contained bulges more massive (one-third the disk mass) than that used 
here (one-tenth the disk mass), and a much smaller
fraction of gas.  As noted by MH96, this suggests that the suppression
of gaseous inflow due to the presence of a bulge is very dependent
upon the bulge mass and the disk composition.  More work will be
needed to quantify the relationship between bulge mass, galaxy
composition, and early gas inflow leading to bursts of star formation.

The final merger is preceded by an increase in the star formation rate
at $T\sim1.6$~Gyr.  At this time, gas begins to see the nuclei as a
single potential, while it is also being stripped of angular momentum
due to the abundant shocks that occur.  At $T=1.8$~Gyr, the final
coalescence occurs.  The strongest burst of star formation precedes the
merger by $\sim50$~Myr in every model.  The strength and duration of
this burst is a strong function of the feedback model.  Models with
isothermal equations of state have a strong, prolonged burst (except
the \nzl\ and \nol\ models, which by this time
have consumed a large portion of the available gas), while
the models with stiffer equations of state suppress large bursts of
star formation due to the efficient pressure support provided by the
feedback reservoir.

% ---------
%  Table 3
% ---------
\begin{table}
\begin{center}
\caption{Star formation properties of major merger simulations.
Here $e$ is the fraction
of the original gas consumed by star formation during the entire
3 Gyr simulation.  SFR$_{\rm max}$ is the maximum star formation
rate during the merger.  T$_{\rm LIRG}$ is the duration, in Gyr, at which
the star formation rate is greater than 20 \msunyr, corresponding
to when this galaxy pair would be classied as a luminous
infrared galaxy, LIRG.}
\begin{tabular}{lccc}
\hline
Model & $e$ & SFR$_{\rm max}$ & T$_{\rm LIRG}$ \\
  & & (\msunyr) & (Gyr) \\
\hline
\hline
% S00  & 0.0 & 0 & 0.0 \\
\nzl  & 0.70 & 78 & 1.55 \\
\nzm  & 0.55 & 159 & 1.31 \\
\nzh & 0.25 & 51 & 0.25 \\
\hline 
\nol  & 0.69 & 84 & 1.72 \\
\nom  & 0.51 & 105 & 1.29 \\
\noh & 0.24 & 43 & 0.08 \\
\hline 
\ntl  & 0.67 & 75 & 1.85 \\
\ntm  & 0.47 & 56 & 1.34 \\
\nth & 0.18 & 44 & 0.04 \\
\hline 
\end{tabular}
\label{tab:sfprops}
\end{center}
\end{table}

To provide a more quantitative view of the star formation induced by
the galaxy merger, we calculate the fraction $e$ of the original gas
converted to stars during the simulation.  Table~\ref{tab:sfprops}
lists $e$ for each merger.  Two trends are
noticeable for $e$, both consistent with the isolated disk
galaxy simulations: first, high-feedback models consume less than half
as much of the original gas as the low-feedback models do, for
equivalent $n$.  Second, models with a stiffer equation of state (i.e.,
higher $n$) suppress gas consumption, although this is a much smaller
effect.

Another relevant measure of the global star formation is the peak
star-formation rate SFR$_{\rm max}$, which is also listed in
Table~\ref{tab:sfprops}.  The equation of state $n$ has a much larger
influence on SFR$_{\rm max}$ than on $e$.  For example \nzm\ has a
SFR$_{\rm max}$ roughly three times larger than \ntm, yet the
overall gas consumption is larger by only 8\%.  Similarly, the peak
star-formation rate of \nth\ is only 20\% lower than \ntm, while
the gas consumption is less than half as much.

The different dependence of $e$ and SFR$_{\rm max}$ on the feedback
parameters results from the varying effects of $\tfb$ and $n$.
The effective pressure is proportional to $\tfb$ (see 
Equation~\ref{eq:effP}), thus large values produce significant 
pressure support and restrict the amount of gas that can reach 
star-forming densities.  Thus the overall gas consumption $e$ is
primarily a result the value of $\tfb$.  On the other hand, once
gas reaches star-forming densities, $n$ determines the ability of this
gas to reach even higher densities, and because the star-formation 
rate scales non-linearly with density (see Equation~\ref{sflaw})
the maximum star-formation rate SFR$_{\rm max}$ is a strong function
of $n$.

It is interesting to note that the simple star-formation rate scaling,
derived as a function of $\tfb$ in Equation~(\ref {eq:sfrn}), that only
moderately described the behaviour of the isolated galaxies, performs
much better for the merging galaxies star-formation rates.
Recall that the
expected scaling was $SFR \propto \tfb^{-1/(2+n)}$.  Applying
this to the maximum star-formation rate, we would expect a ratio of
3.2, 2.2, and 1.8 for the $n = 0 $, 1, and 2, respectively.  The
ratios between the $med$ and $high$ runs from Table~\ref {tab:sfprops}
 are 3.1, 2.4, and 1.3.  Because of the unregulated star formation in
the $low$ models, the scaling appears to break down for these.

One interest in studying star formation induced by galaxy mergers is
their possible relation to luminous infrared galaxies \citep[LIRGs;
][]{SM96}.  LIRGs are classified as galaxies which have an infrared
luminosity greater than $10^{11}$ L$_{\odot}$ and hence are forming
stars at $\geq20$~\msunyr \citep[using the L$_{IR}$-SFR conversion of
][]{Kenn98}.   Listed in Table~\ref{tab:sfprops} is T$_{\rm LIRG}$, the
total time during the merger in which this system would be detectable
as a LIRG.  The choice of feedback has a significant effect on the
length of time this merger would classify as a LIRG.  We note that we
have not distinguished between the two progenitors, and thus do not
account for viewing angles where the two galaxies would be classified
as separate.  Nor have we included a full treatment of the obscuring
effects of dust and its re-radiation in the infrared \citep[but see
][]{Patrik,JonI,JonSun}.

It is clear that these three quantities do not describe the full
star-formation history of each model.  Instead, they are gross
measures, useful to characterize the effects $n$ and $\tfb$ have
on the star formation induced by a galaxy major merger.  Future work
will compile quantities such as these for a wide range of initial
conditions, assumed merger orbits, and merger mass ratios.  These
relations will be useful in analytic or semi-analytic models of galaxy
formation, and in using observations to clarify which feedback models
are realistic.

As a final remark related to the star-forming histories shown in
Figure~\ref{fig:mmsfr}, we draw attention to the oscillatory
star formation that occurs in nearly every model, but is most 
pronounced in the high-feedback models.  These oscillations occur
primarily after peaks in the star formation rate and are thus
clearly visible after the first passage and the final burst.
However, even the isolated disk high-feedback models contains some
oscillatory star formation.  The periods of oscillation are almost
exactly $2 \times \tfb$, the thermalization timescale, indicating
that the oscillations result from the interplay between pressurizing
feedback efficiently quenching star formation followed by the 
dissipation of this pressure support, gravitational collapse, and an
increase in the star-formation rate.

%
% ----------------------------
\subsection{The Kennicutt Law}
\label{ssec:mmkenn}

% -------------
%  Figure 8
% -------------
\begin{figure}
\begin{center}
\resizebox{8.0cm}{!}{\includegraphics{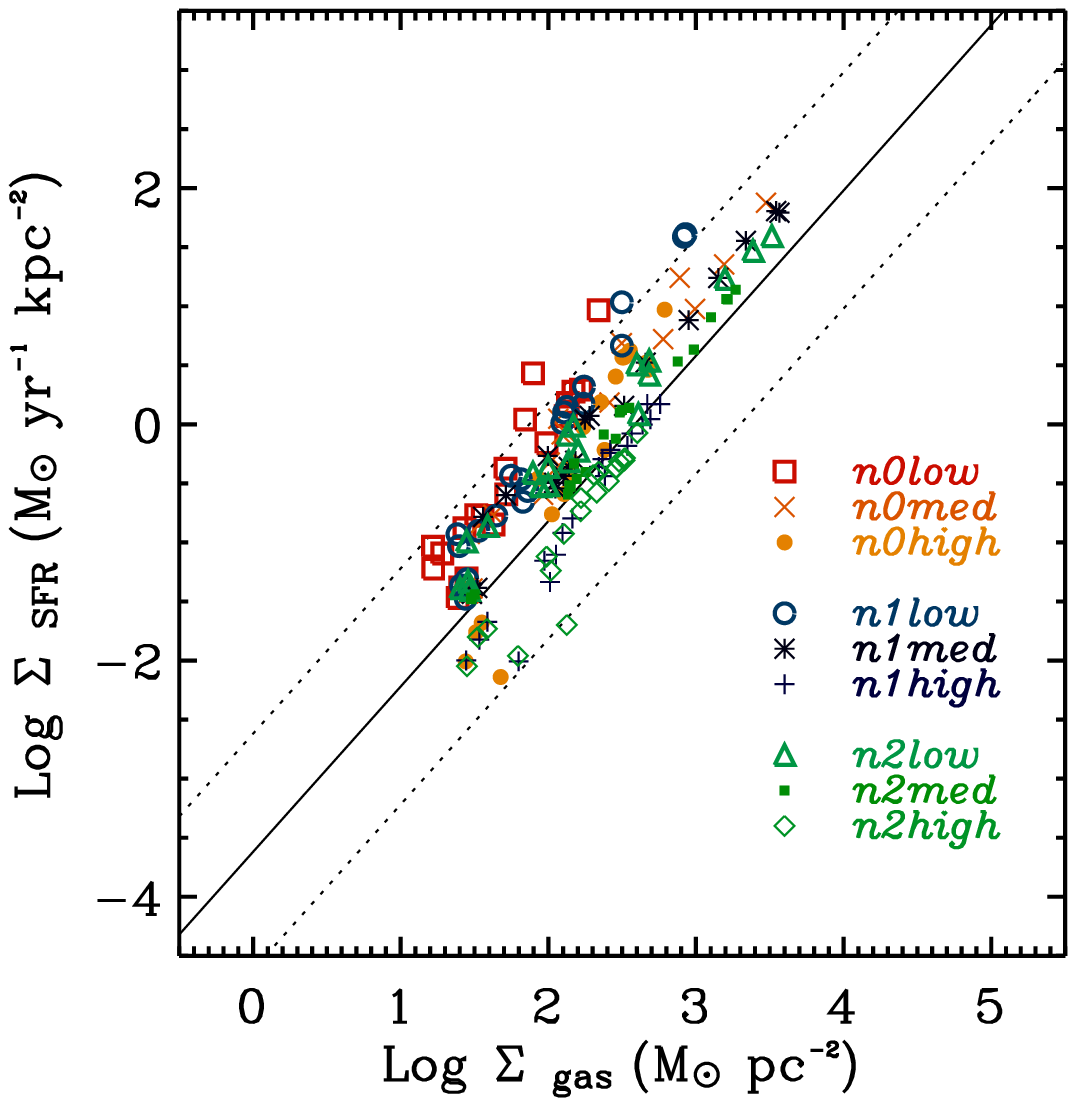}}\\
\caption{Star-formation rate per unit area and gas surface density averaged
within an azimuthal aperture of radius 2.0 kpc, one point type for each
simulation.  For each run points are plotted every 200~Myr during the interaction, and
every 50~Myr during the final merger.
The solid line is the empirical Kennicutt law, and
the dotted lines
represent the envelope in which {\bf all} observation reside.  This plot
can be directly compared to Figure 6 from \citet{Kenn98}.
\label{fig:mkenn}}
\end{center}
\end{figure}

In \S~\ref{ssec:isokenn} we demonstrated the the individual spiral
galaxies matched the empirial Kennicutt law \citep[Equation~\ref{KenLaw}]{Kenn98},
as they were designed to do.  In Figure~\ref{fig:mkenn}, we show the
star-formation rate and gas mass averaged in a 2 kpc aperture plotted at 
representative times during each merger.  As was done previously, we overplot the
Kennicutt law with a solid black line, and the observed envelope for which
{\it all} observations are contained is plotted with a dotted line.
Similar to the isolated galaxies, the merger simulations closely track the 
empirical star-formation law.  There are a few points that are above the upper
envelope, however these are from the unstable $\nzl$ and $\nol$ runs.
The fact that all of the stable models agree with the
Kennicutt law, regardless of $n$ and $\tfb$, suggests that it is our
formulation of star formation according to Equation~(\ref{sflaw}) and the
value of the free parameter $c_\star$ (which are identical for all runs)
that determines the agreement with the Kennicutt law.

%
% -------------------------
\subsection{Phase Diagram}
\label{ssec:mmphase}

% -------------
%  Figure 9
% -------------
\begin{figure*}
\resizebox{16.5cm}{!}{\includegraphics{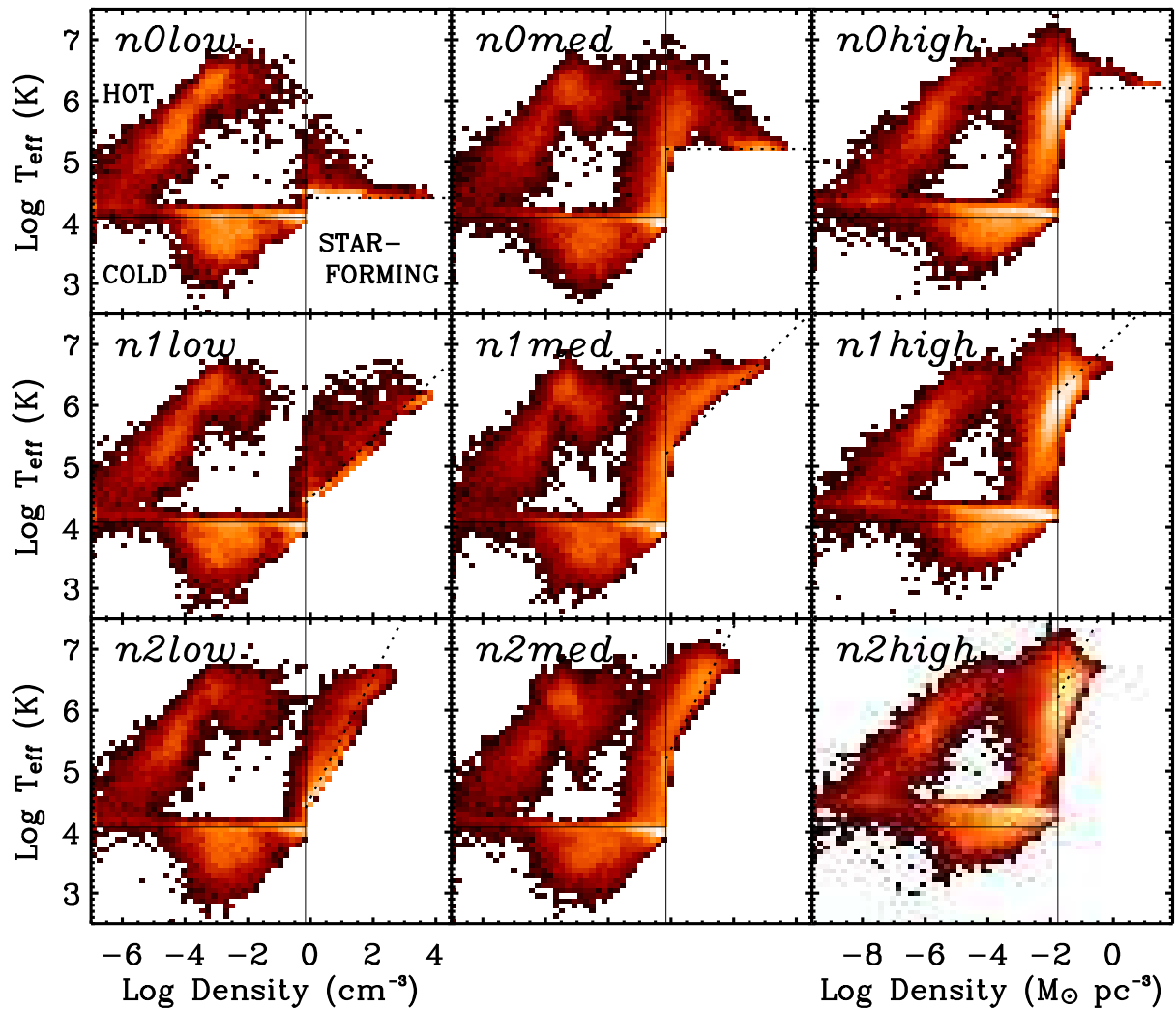}}% 
\caption{The effective phase diagram for the galaxy merger at
$T=1.75$~Gyr for all models.  The calculation of T$_{\rm eff}$ and all
of the lines are identical to Figure~\ref{fig:isophase}.  The 
top-left panel ($\nzl$) denotes the three phases of gas, hot, cold,
and star-forming as described in the text.
\label{fig:mmphase}}
\end{figure*}

At this point we revisit the phase diagrams originally introduced in
\S~\ref{ssec:isophase} in the context of isolated Sbc disk galaxies.
We expect two differences when we look at the phase diagrams of the
interacting galaxies in comparison to the isolated disks.  First, the
collisional shocks that attend the interaction generate a significant 
amount of hot low-density gas.  Second, because of the interaction,
much more gas will reach high densities.

Figure~\ref{fig:mmphase} shows the phase diagram for the galaxy merger
of each parameter set at a point in time 50 Myr prior to the final
coalescence.  At
this point, most models are near their peak of star-formation. 
All gas resides in one of three distinct regions in the phase diagram:
hot, cold, or star-forming.

Gas which is below the critical density threshold for star formation $\rt$
is designated as either in a hot or cold phase, depending on whether its temperature 
is above or below $1.2\times10^4$~K, denoted by a horizontal line in 
Figure~\ref{fig:mmphase}.  When the simulation begins, as well as
in the isolated disks, all non-star forming
gas is cold owing to efficient cooling.  This gas remains
cold until the first passage, when gas contained in the tidal tails is
ejected and adiabatically cools to $\sim10^3$~K.  Gas that is not part of
the tidal material collides with gas from the other disk.  These collisions
generate shocks that heat the gas to temperatures well above $10^4$~K
where it is deemed hot.  Because the cooling time of this hot, low-density
gas is greater than the merger time-scale, most of this gas expands out of
the central region and it remains in this phase until the merger is complete.

During the final merger, the significant collision results in shock heating,
and therefore a lot of hot gas \citep{Cox04}.  While most of the gas
remains hot after the final coalescence ($T\ga2.0$), some of the shock-heated gas
cools and flows to the galaxy center.  This gas, as well as the continual
infall of tidal material, fuels the steadily decreasing star formation.

Gas which is at, or above, the threshold density for star formation is strongly affected
by the feedback model.  In this sense, the merger is similar to the isolated disk.
The main difference is that gas in the merger reaches much high densities than 
the in the isolated 
case.  Figure~\ref{fig:mmphase} shows that, like the isolated disks, the effective
temperature of star-forming gas is close to Equation~(\ref{eq:effT}), which is 
delineated with a dotted line in the figure.

Figure~\ref{fig:mmphase} also shows a large spread in the effective temperature
of star-forming gas.  This spread is considerably larger than in the isolated
disks and is a direct result of the varied history of shock-heating, star
formation, and feedback that attends the violent nature of the merger.
Figure~\ref{fig:mmphase} also demonstrates that the med- and high-feedback
models have a significant amount of gas that piles up at $\rt$ owing to the
pressure support provided by these models.  This feature also relates to the
oscillatory star formation discussed at the end of the previous section as
gas continually cycles into and out of star-forming regions based upon their
recent star-formation history, current dynamics, and dissipation of
feedback reservoirs $q$.

% -------------------------
\subsection{Merger Remnants}
\label{ssec:mrem}

Most of the merger simulations run to-date have focused on the role disk-galaxy
major mergers may play in forming elliptical galaxies, as first
suggested by \citet{TT72} and \citet{T77}, and commonly referred to 
as the ``merger hypothesis''.
To this end, studies have addressed whether galaxy mergers can reproduce the 
$r^{1/4}$ surface brightness profile which many elliptical galaxies appear to have
\citep{B92,HRemI,HRemII,HRemIII,MH94dsc,BH96,Sp00,NT06}, the kinematic structure
of merger remnants \citep{NB03,GG05b,GG05c,Cox06rot},
or if the merger remnants fall on the fundamental plane \citep{BS97,Bek98fp,
NB03,Dan03,NLC03,BK05,BK06,Rob06fp}, 
a tight correlation between the velocity dispersion, effective radius,
and luminosity, on which all ellipticals reside \citep{DD87,Dre87,BBF92,Bern03fp,Pad04}.
However, very few of these simulations have included recipes for star formation
and feedback or explored how these parameters affect the remnant.  In light of this, 
this section looks at the properties of the merger remnants to determine if studies
of this sort might be sensitive to the prescriptions used to include star formation
and feedback.  We 
reserve a complete analysis of merger remnants for future work because this will
require a much larger set of runs spanning a range of progenitor galaxy masses.
Here, we simply present
the stellar mass profile, the remnant size, as measured by R$_{\rm e}$, the half-mass 
radius, and the one-dimensional central velocity dispersion
for each merger remnant.  We show that the size and velocity dispersion depend strongly 
on the feedback model while the stellar mass profile is only weakly dependent.

%
% -------------------------
\subsubsection{Stellar Mass Profiles}

% --------------
%  Figure 10
% --------------
\begin{figure*}
\begin{center}
\resizebox{5.5cm}{!}{\includegraphics{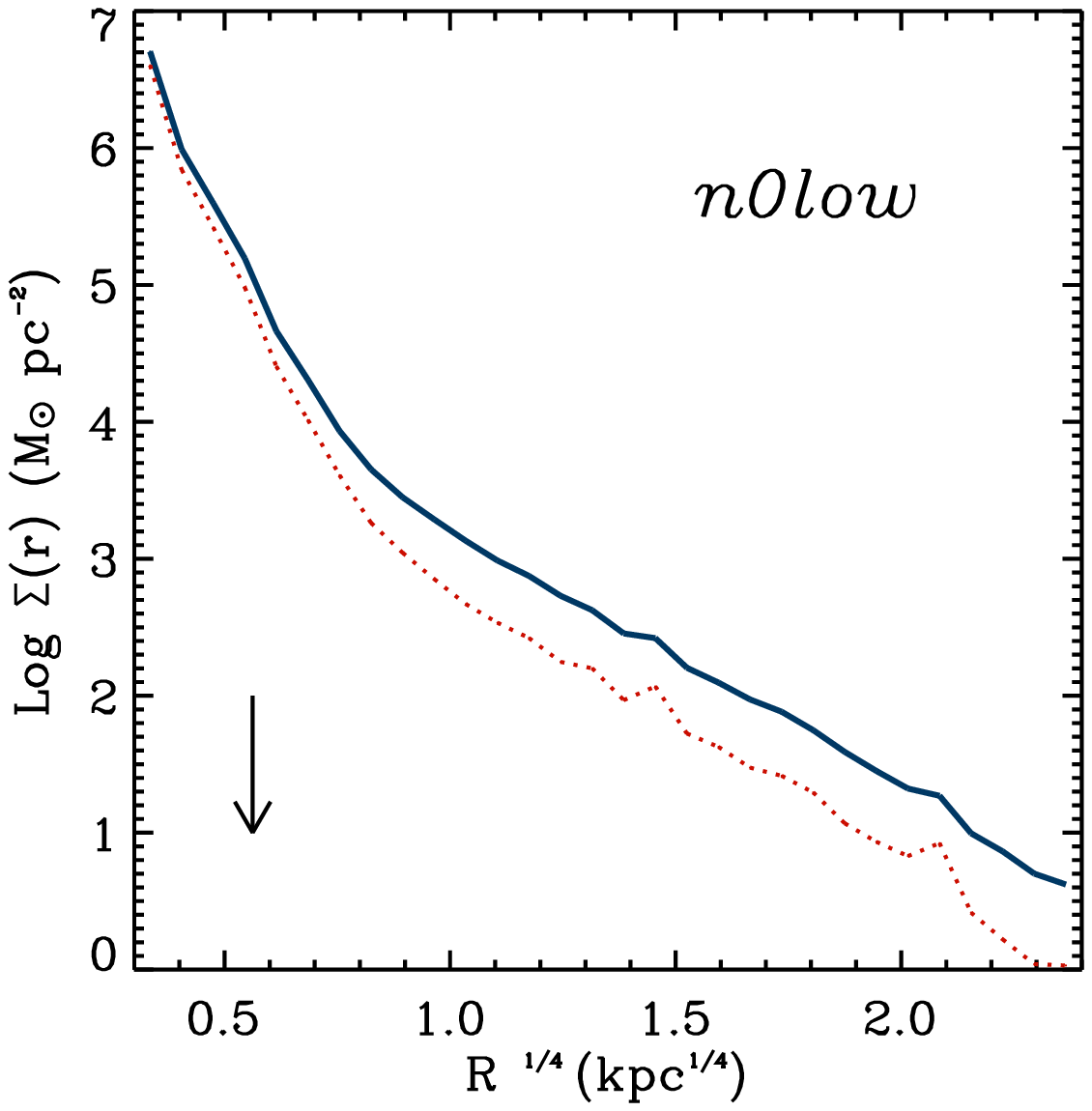}}% 
\resizebox{5.5cm}{!}{\includegraphics{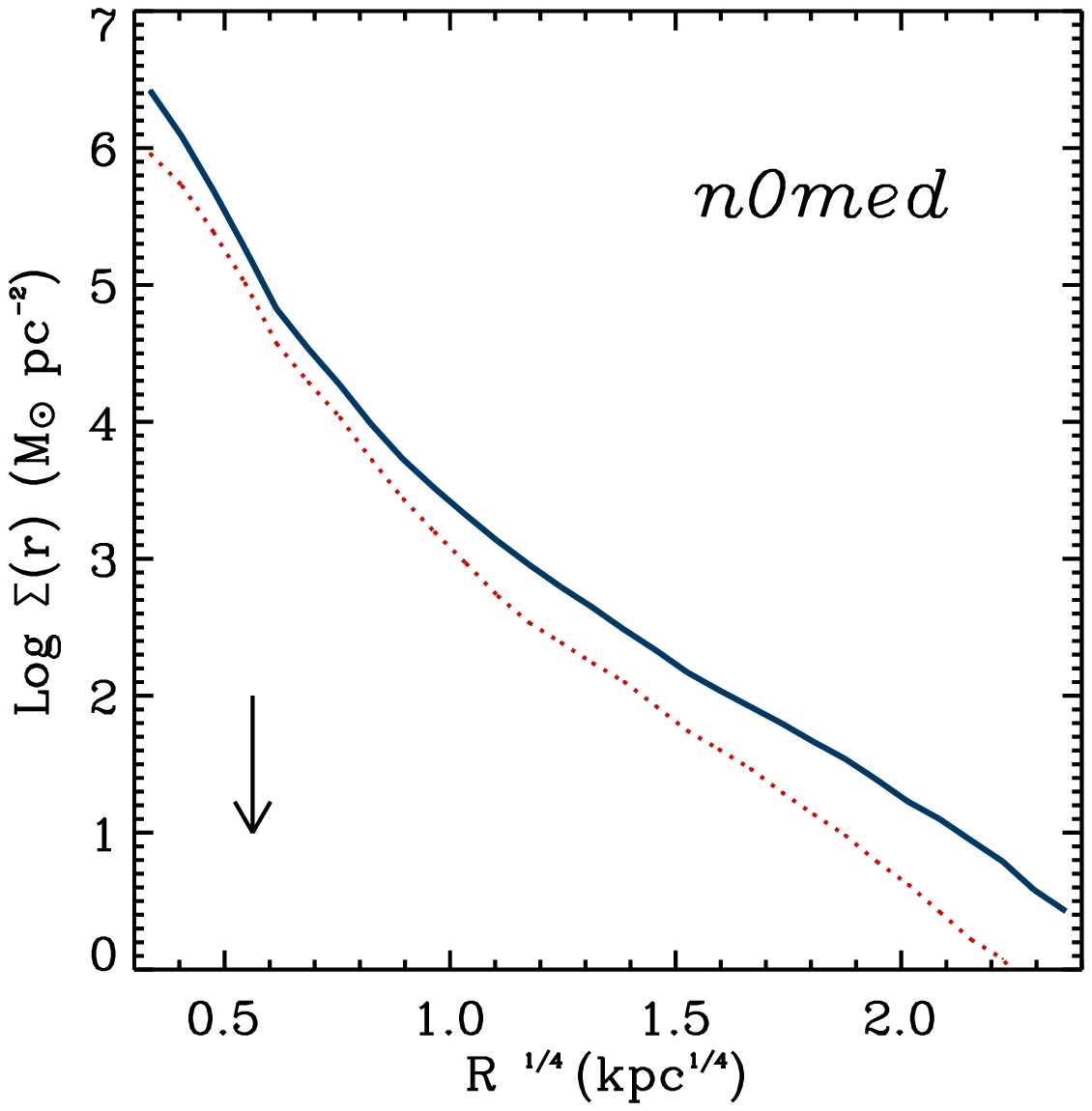}}%
\resizebox{5.5cm}{!}{\includegraphics{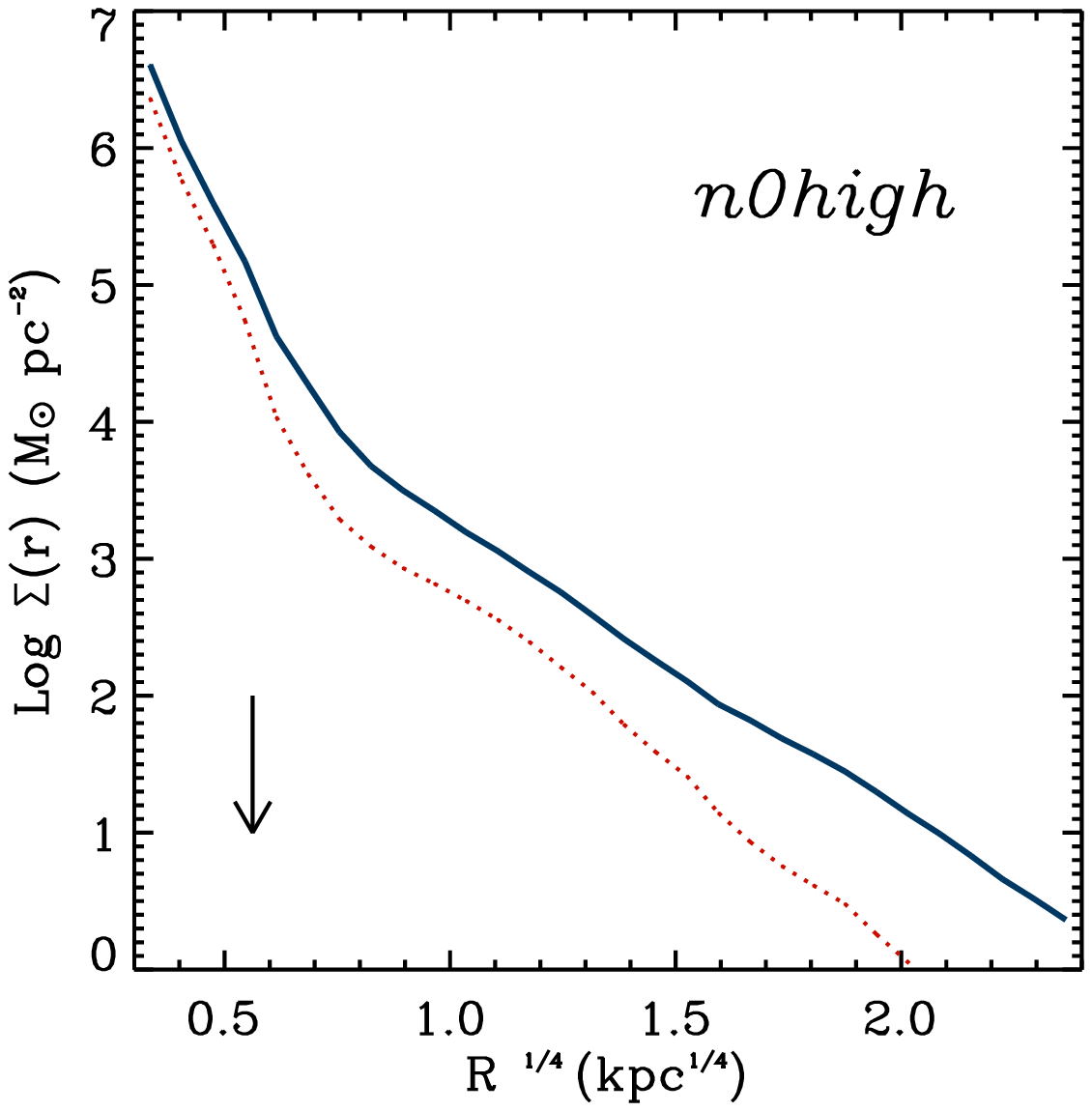}}\\
\resizebox{5.5cm}{!}{\includegraphics{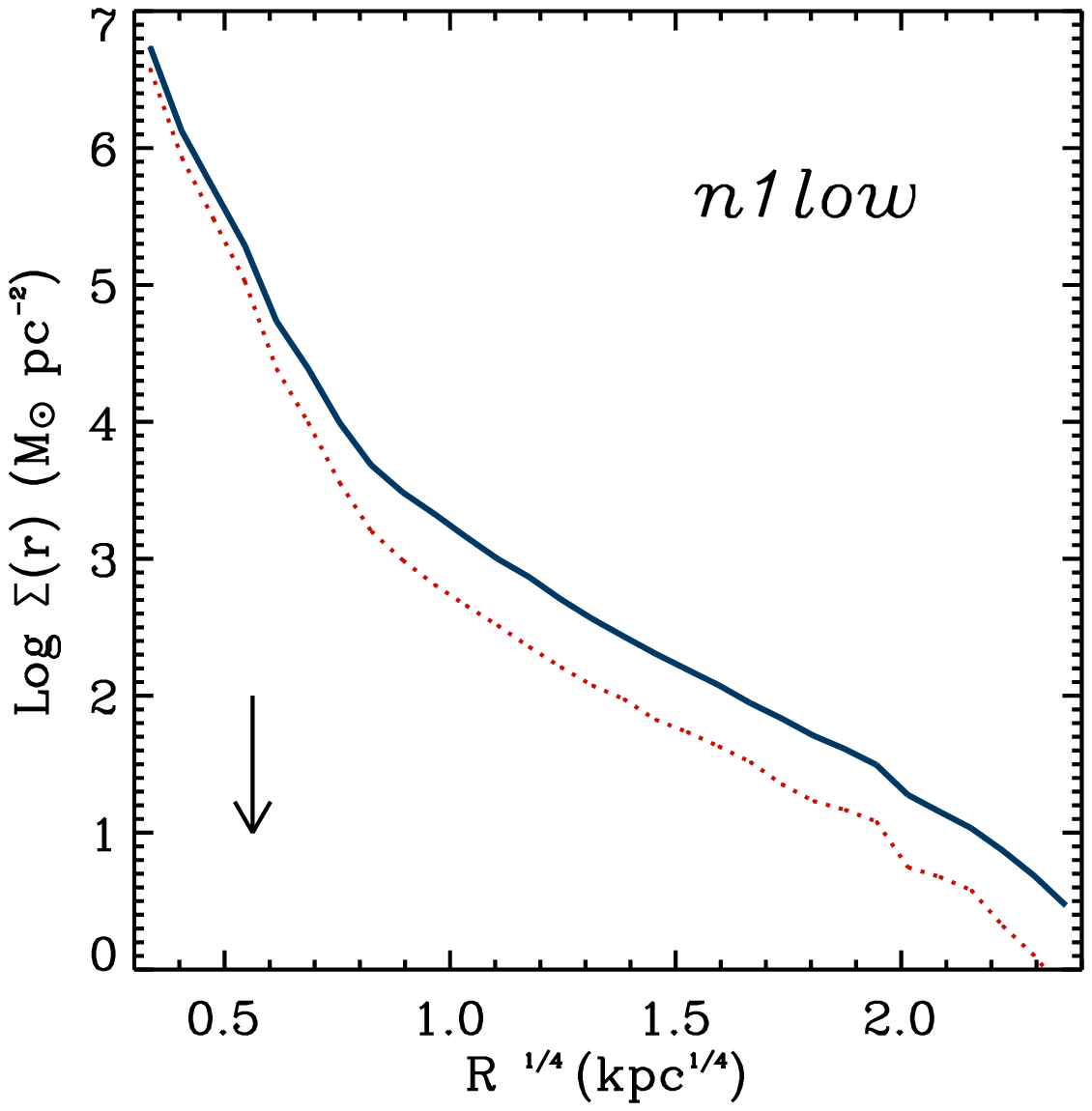}}% 
\resizebox{5.5cm}{!}{\includegraphics{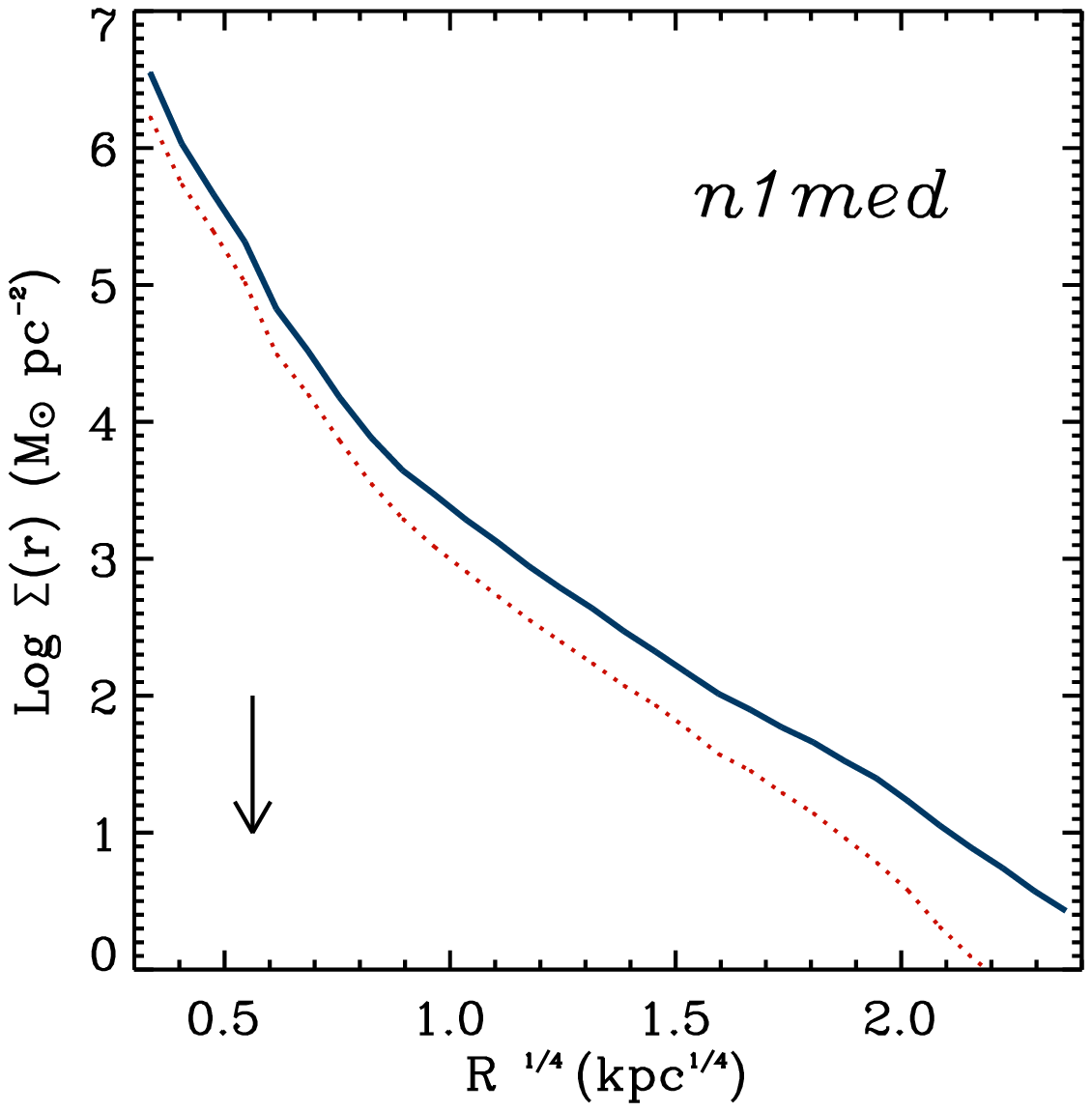}}%
\resizebox{5.5cm}{!}{\includegraphics{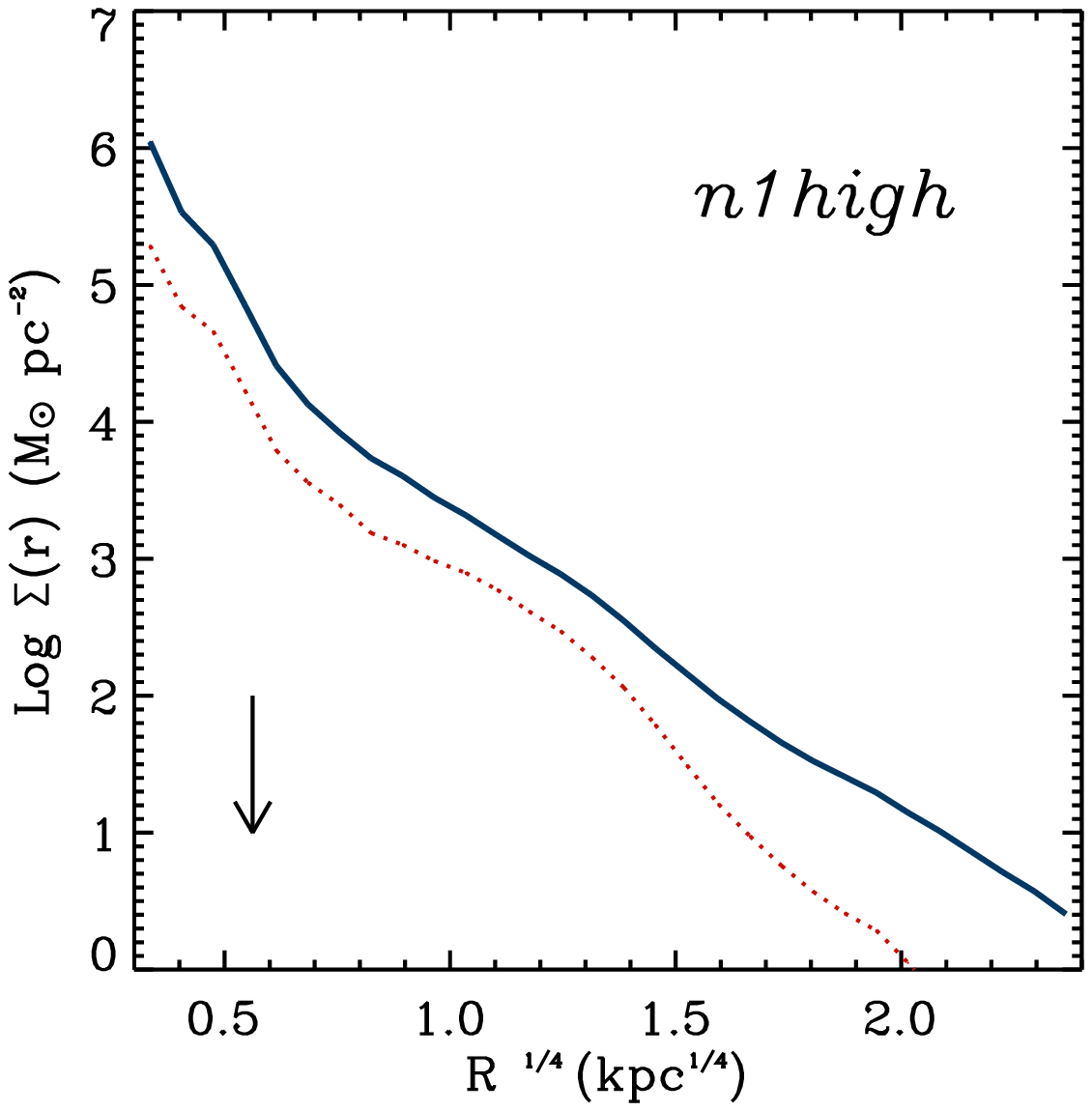}}\\
\resizebox{5.5cm}{!}{\includegraphics{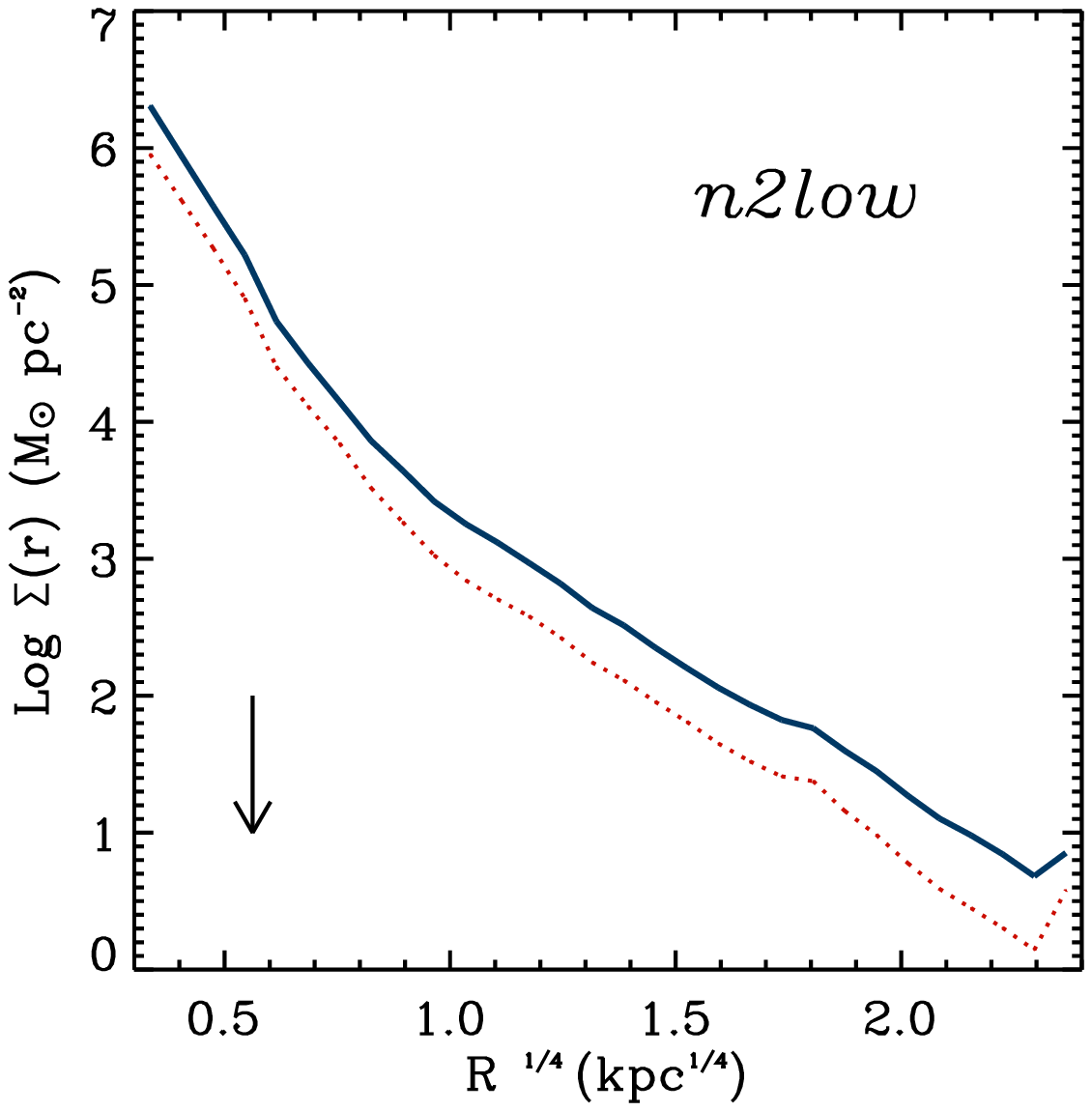}}% 
\resizebox{5.5cm}{!}{\includegraphics{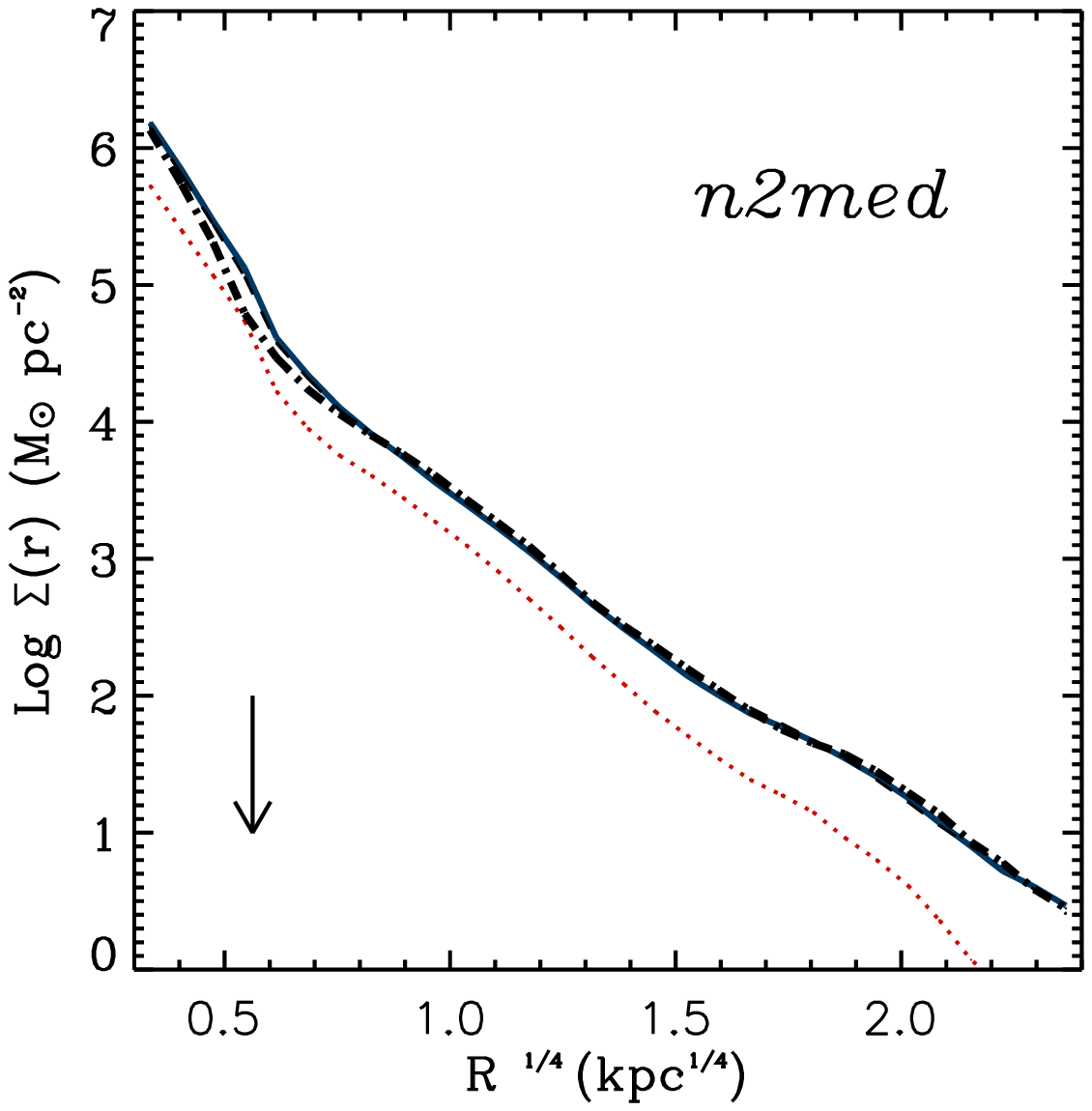}}%
\resizebox{5.5cm}{!}{\includegraphics{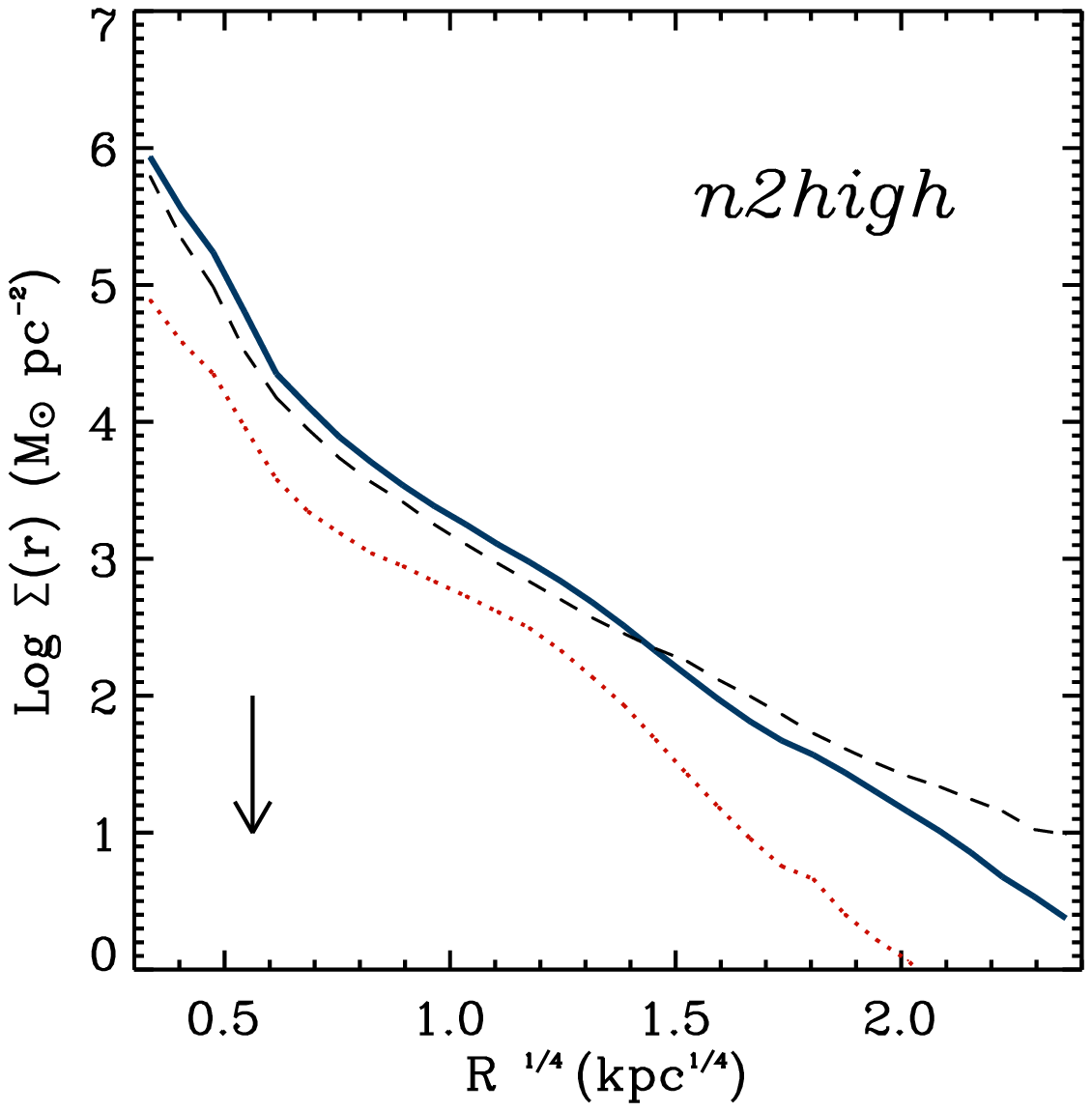}}\\
\caption{Surface density of each merger remnant projected onto the orbital
plane versus the fourth root of the radius.  Other projections are qualitatively
similar.  The solid line is for all stellar mass
and the thin dotted line is stellar material formed during the major-merger
simulation.  The arrow in each plot points to the gravitational softening length 
of 0.1 kpc.  The \ntm\ plot, in the lowest middle panel, also shows the profile
of an identical simulation with ten times the resolution and another where 
gas particles can spawn 10, instead of the standard 2, stellar particles.  These
variant simulations produce nearly identical profiles.  The \nth\ plot, in the 
lower right-hand panel, contains a dashed 
line showing the total stellar profile of a purely collisionless run.
\label{fig:devac}}
\end{center}
\end{figure*}

Figure~\ref{fig:devac} shows the azimuthally averaged surface density profile
versus the fourth root of the projected radius for all merger simulations.
A pure $r^{1/4}$-type profile would be a straight line in this plot.
Similar to the work of \citet{MH94dsc} and S00, the merger remnants show
central cusps that are disjoint from the outer density profiles.
The central-density excess begins at R~$\sim200$~pc, near our resolution
limit, and is comprised of both stars that are newly formed during the merger
(shown as a dotted line), as well as the pre-existing bulge.  In higher
resolution simulations (see \S\ref{ssec:res}) the profile changed very little,
as indicated by the overplotted profile in the $\ntm$ panel, suggesting
that it is not just poor resolution which generates the central profile cusp.
Further, because all models shown in Figure~\ref{fig:devac} display a cuspy
profile, it does not appear to be a result of the strength or density-dependence
of the feedback model, although there is a trend for the high feedback models
to form fewer stars and thus the surface density is predominantly bulge-dominated
toward the center.

The ubiquitous nature of the central cusp indicates that it is a byproduct
of our initial galaxy model.  As noted by S00, the remnant has not forgotten
about the dense stellar bulge in the progenitor disk.  Even a collisionless
run, shown as a dashed line in the lower-right plot, demonstrates a break 
in the surface density.  We also performed one simulation without a central
bulge which produced a profile that exhibited a more subtle upturn toward
the center.  Not surprisingly, the central density excess in this case was
solely composed of stars formed during in the simulation.  These results
are consistent with \citet{MH94dsc} and S00.

It is not yet clear if the surface densities plotted in Figure~\ref{fig:devac}
are consistent with observations, or even if there exists a definitive sample
with which to compare these profiles.  In a recent study of the K-band
surface-brightness profiles of 51 merger remnants by \citet{RJ04}, 16 (30\%)
had profiles steeper than $r^{1/4}$.  Similar studies of ellipticals 
\citep{Byu96} and ULIRGs \citep{Vil02} have found a much smaller percentage
that deviate from $r^{1/4}$ indicating that the discrepancy between the
observations and simulations is far from obvious.  In fact, during the one
Gyr evolution subsequent to the final galaxy merger, the central excess
slowly increases due to the remnant gas slowly being converted to stars.  This trend 
might have been abated, or even reversed, if a more violent injection of
energy would have dispersed, or removed, the remaining gas thereby causing
the density profile to remain constant or slightly decrease.  While no
source of energy is inherent to our feedback model, recent work by 
\citet{SdMH05red} has shown that an active galactic nucleus (AGN) can
indeed expel large amounts of gas and completely shut off star formation.
Correlations between mergers and AGN \citep{Vil95,WL03} and mergers and
outflows \citep{M99,M05,Ru05II,Ru02} provide evidence for this scenario.
Moreover, merging black holes at the galactic center could also influence
the central stellar distribution \citep[see, e.g.,][]{Mil01}.  In a more
pessimistic vein, \citet{SpGad2} has recently shown that time integration
methods commonly employed in N-body simulation can lead to significant
errors when using adaptive timesteps and highly eccentric orbits (see his
Fig.~5).  Thus, the central densities may be the result of numerical
errors which could be remedied by using new codes such as Springel's GADGET-2.
A much more thorough study of the time evolution of the surface density
profile, the robustness to numerical integration parameters, and the dependence
on various initial conditions will be required to determine if merger remnants
evolve into ``regular'' elliptical galaxies, and if so, which types of ellipticals.

As a final comment, we remark that the centrally concentrated new stellar 
population present in all models has affected even the collisionless component.
The contraction of the collisionless component is clearly demonstrated in the
lower-right plot of Figure~\ref{fig:devac}.  The total
surface density for \nth\ intersects the collisionless profile at 
R~$\approx4.5$~kpc, signaling that the gaseous dissipation and star formation
included in \nth\ has
contracted the mass profile in comparison to the collisionless case.

%
% -----------------------------------
\subsubsection{The Fundamental Plane}

% -----------
%  Table 4
% -----------
\begin{table}
\begin{center}
\caption{Projected remnant properties averaged over 1000 random lines
of sight.  R$_e$ is the half-mass radius of all stars, while
R$_{e,{\rm os}}$ is the half-mass radius of the old stellar
component, both disk and bulge, and R$_{e,{\rm ns}}$ is the
half-mass radius of stars generated during the course of the merger
simulation.   $\sigma$ is the velocity dispersion of all the
stars measured within half R$_e$, and $\sigma_5$ is the velocity 
dispersion measured inside an aperture of 5 kpc.  The total and new star
effective radii are not calculated for the collisionless simulation
because there was no star formation included and hence no new
stars.}
\begin{tabular}{lccccc}
\hline
Model & R$_e$ & R$_{e,{\rm os}}$ & R$_{e,{\rm ns}}$ & $\sigma$ & $\sigma_5$ \\
      &  (kpc) & (kpc) & (kpc) & (\kms) & (\kms) \\
\hline
\hline
%S00 & x.x & x.x & x.x & xxx & xxx \\
\nzl  & 7.4 & 8.1 & 6.1 & 204 & 212 \\
\nzm  & 3.8 & 5.7 & 1.8 & 263 & 243 \\
\nzh & 4.0 & 6.1 & 0.8 & 226 & 213 \\
\hline
\nol  & 6.8 & 6.8 & 6.9 & 224 & 231 \\
\nom  & 3.8 & 5.8 & 1.8 & 263 & 242 \\
\noh & 4.2 & 5.9 & 1.9 & 196 & 185 \\
\hline
\ntl  & 5.9 & 6.0 & 5.8 & 220 & 224 \\
\ntm  & 4.0 & 6.0 & 2.0 & 232 & 205 \\
\nth & 5.0 & 6.2 & 2.3 & 184 & 176 \\
\hline
collisionless & -  & 7.2 & - & 155 & 155 \\
\hline
\end{tabular}
\label{tab:rems}
\end{center}
\end{table}

To begin, we note that the previous section demonstrated that the 
collisionless stellar distribution responds to gaseous 
dissipation and star formation in
the galaxy center by becoming more concentrated.
To quantify the contraction of stellar mass, we measure the half-mass radius R$_e$ of
the total, the new, and the old stellar (bulge plus disk) mass and list these in 
Table~\ref{tab:rems}.  The half-mass radius is determined as the radius which 
encloses half the projected stellar mass.  We do not attempt to compute a luminosity
profile, nor do we try to fit a $r^{1/4}$-type luminosity profile since it does
not appear to accurately characterize the surface density.  Hence, R$_e$ is not
identical to the effective radius, but it serves as a useful proxy.
The half-mass radii of the new stars R$_{e,{\rm ns}}$ shows that
half of all stars formed during the medium and high feedback merger 
simulations end up within $\sim2$~kpc of the center of the merger remnant.
Further, R$_{e,{\rm ns}}$ depends on $n$, the feedback reservoir equation
of state index, such that increasing $n$ increases the half-mass radius of the newly 
formed stellar population.
The dissipation, gas infall and star formation has affected the old stars
as well.  The half-mass radius of the old stars R$_{e,{\rm os}}$ is $\sim14-20$\% smaller
in the runs including gas, cooling, star formation, and feedback 
compared to the purely collisionless simulation.

The central concentrations of newly formed stars affect the kinematics of the entire
stellar population.  To quantify the stellar kinematics, we calculate the one-dimensional 
velocity dispersion within an aperture centered on the potential minimum. 
Table~\ref{tab:rems} lists the velocity dispersion within apertures of radius one-half 
R$_e$ and 5 kpc, averaged over a thousand projections. The typical scatter due to projection
effects is $\sim10$\%.
Because the one-dimensional velocity dispersion is a measure of the depth of the 
potential well, the remnants with a smaller effective radius naturally lead to a 
larger velocity dispersion, and in all cases the velocity dispersions are much larger
than the collisionless simulation -- the largest by over 50\%.  The dispersion profile
also changes.  In the collisionless case, the velocity dispersion is flat, i.e., the
dispersion is identical when measured within an aperture of 3.6 or 5 kpc.  In contrast,
the medium and high feedback simulations have a velocity dispersion
that increases toward the remnant center.

While we reserve a more extensive study of these merger remnants to future work, it
is interesting that the effective radii and velocity dispersions presented in 
Table~\ref{tab:rems} are consistent with the early-type $R_e$-$\sigma$ scaling relation 
found in the Sloan Digital Sky Survey \citep[see their Fig. 4 and 5]{Bern03corr}.
Further, the differences generated by the feedback models 
scatter the merger remnants perpendicular to this scaling relation, suggesting that it is
the interplay between star formation and feedback which fixes the position of a merger
remnant on the fundamental plane of hot stellar systems.  The sensitivity of merger
remnants to feedback models is potentially a useful discriminator for selecting the
best model or parameters.

% -------------
%  Figure 11
% -------------
\begin{figure}
\begin{center}
\resizebox{8.0cm}{!}{\includegraphics{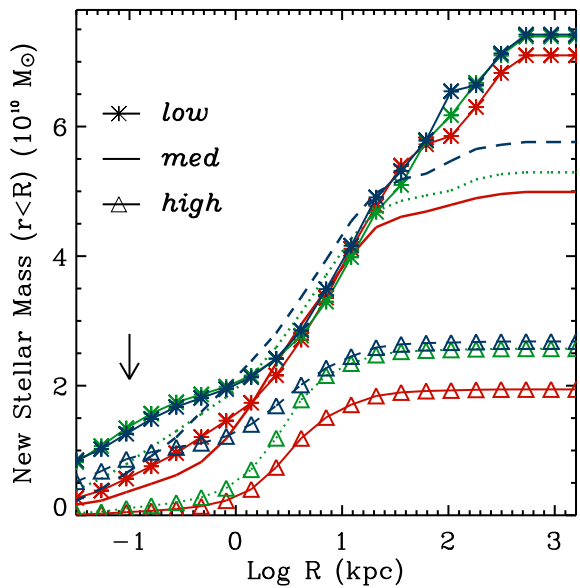}}\\
\caption{Cumulative new (formed during the merger simulation) stellar mass in the
  merger remnant for each model.  Shown as solid, dotted and dashed
  thick lines that represent the $n=2$, $n=1$, and $n=0$ feedback models,
  respectively.  Then, the degree of feedback, i.e., $high$, $med$, and 
  $low$, is denoted by the symbols as shown in the key.  As in
  Figure~\ref{fig:devac}, the arrow indicates the gravitational
  softening length of 0.1~kpc.
\label{fig:massinr}}
\end{center}
\end{figure}

Figure~\ref{fig:devac} and R$_{e,{\rm ns}}$ both demonstrate that the majority of new stars reside
within the inner regions of the remnant.  This is confirmed in Figure~\ref{fig:massinr},
which plots the cumulative mass distribution of new stars in each merger
remnant.  In all models approximately 80\% of {\it all} new stellar mass is
within 10 kpc of the remnant center.  In contrast, the simulations of MH96 led to
merger remnants in which all of the new stellar material was within 1~kpc of the center.
This suggests that our 
feedback model restricted the gas inflow and dispersed the central star-forming region. 
The new star mass cumulative profile is highly correlated with $n$.  The $n=2$ models have the 
lowest cumulative mass distribution for both the high- and low-feedback series.  Apparently, the 
stiffer equation of state has provided sufficient pressure support for stars to be
formed over a larger range of radii, thus decreasing the fraction of stars which 
reside within the inner $\sim1$~kpc of the remnant.

Because stars are stochastically spawned from collisional gas (see the last paragraphs in
\S~\ref{ssec:sf}), it is possible that the new stellar component spuriously loses energy 
and angular momentum due to its artificial coupling to the gas.  This was
explicitly checked by rerunning the \ntm\ simulation with $N_g=10$, i.e., each gas 
particle now spawns 10 individual stellar particles as opposed to the fiducial value of 2.
The resulting surface-density profile was very similar to \ntm.  The new
stars were slightly more extended, with a half-mass radius R$_{e,{\rm ns}}=2.2$~kpc, about 10\%
larger than \ntm.  The more extended new stars also affected the velocity field and the 
dispersion was lower at half R$_e$ and higher at 5~kpc, but both of these effects were only
$\sim3$\% of the values listed in Table~\ref{tab:rems}.

% ------------------------------------------
%   Discussion: various parameter studies
% ------------------------------------------
\section{Discussion}
\label{sec:disc}

%
% ---------------------------
\subsection{Convergence Study}
\label{ssec:res}

% -------------
%  Figure 12
% -------------
\begin{figure}
\begin{center}
\resizebox{8.0cm}{!}{\includegraphics{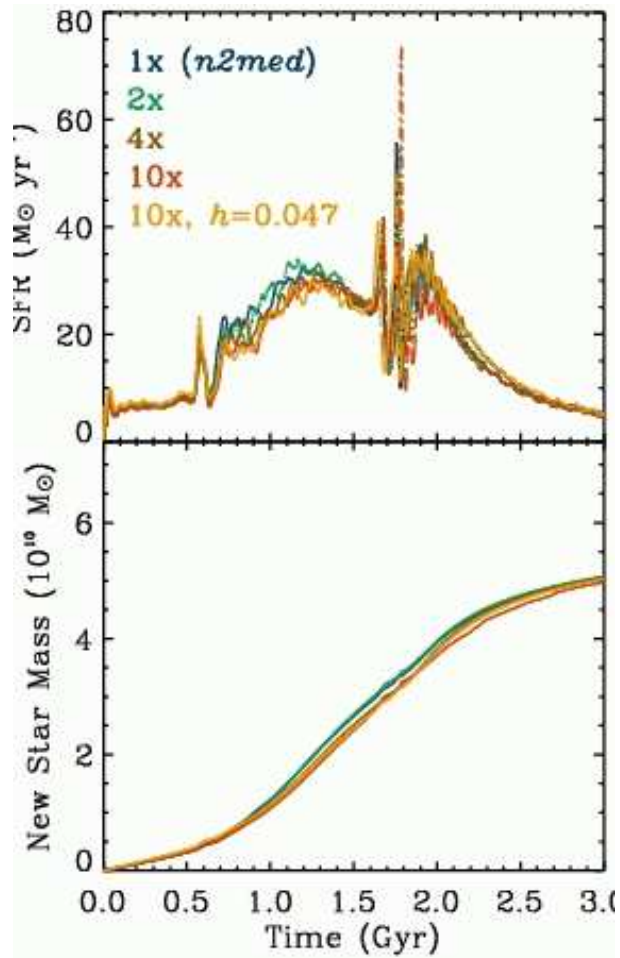}}
\caption{Star-formation rate for our fiducial major merger
with runs of varying resolution.
\label{fig:sfrres}}
\end{center}
\end{figure}

The resolution of N-body simulations such as those performed here is
determined by the particle number N and the gravitational softening
length $h$.  Increasing N and decreasing $h$ both result in higher
resolution.  Of course, this higher resolution comes at the expense 
of computational time.  In practice, the selection of appropriate
values for N and $h$ is a balance between the resolution necessary 
to address the relevant processes and the available resources.
In this section we explore variations in
particle number and smoothing length for our fiducial merger simulation.
S00 performed a similar study for the isolated disk.  To
isolate the effect of particle number, we select only one (\ntm) of
the feedback parameters to resimulate 
with the total particle number increased by factors of 2,
4 and 10.  The ratio of dark-matter to collisionless disk to
collisional particle number remained fixed.  The highest resolution
simulation contained 1,000,000 dark matter particles, 300,000 each of
stellar disk and collisional gas, and 100,000 bulge particles for each
galaxy, making the total number of particles used in this simulation
3.4 million.  The individual particle masses were
$7.1\times10^5$~\msun$ $ for the dark matter, $1.8\times10^5$~\msun$ $
for the collisional gas, $1.3\times10^5$~\msun$ $ for the stellar disk
and $1.0\times10^5$~\msun$ $ for the stellar bulge.

In general, increasing N allows the gravitational softening length
to be decreased as roughly N$^{-1/3}$.  However, because of the increased
computation time resulting from this scaling, we have chosen to use a smaller
softening length for only one simulation.  We chose our highest resolution
simulation that had ten times the particle
number.  In this case, the softening length was $10^{1/3}$ (0.46) times
smaller than the fiducial 0.1~kpc.

Figure~\ref{fig:sfrres} shows the star-formation rate for simulations
of varied resolution.  As the resolution is increased, either via N or
$h$, we find a slight trend of lower star-formation rates during the
initial passage, but the effects are less than 5\% and, overall, there
is excellent agreement between the global gas consumption.  The
consistency of star formation for all resolution runs suggests that our
fiducial particle number is sufficient to resolve the star-formation
history.

The merger remnants are also insensitive to the increased resolution.
The surface density, half-mass radii, and velocity dispersions are all
within 2\% of the \ntm\ simulation, with no trend as a function of
higher resolution.  However, there are subtle differences between the
high- and low-resolution simulations.  In particular, the number of
small condensations orbiting the central gas disk in the merger
remnant (see bottom-right panel in Fig.~\ref{fig:mmgasmorph}) is
increased by $\sim30$\% in the higher resolution simulation.

%
% ----------------------
\subsection{Comparison to Previous Work}
\label{ssec:comp}

The simulations presented here extend prior studies of star formation 
and feedback in equal-mass gas-rich disk-galaxy major mergers.  As
in the work of MH96 and S00, we find that mergers drive gas inflows
which efficiently form stars and lead to a central concentration of
newly formed stellar mass.  Furthermore, we have quantified 
how these results depend on the assumed feedback efficiency and density
dependence.  However, the comparison to previous work is complicated
by the many differences between the methodology used here and that
employed in previous work.  For clarity and definiteness, we find it
useful to enumerate the differences between the 
simulations presented here and those of MH96 and S00:
\begin{enumerate}
\item To represent stellar feedback, MH96 imparted momentum ``kicks'' while
S00 and this work chose to pressurize the ISM.  In a recent paper, \citet{SdMH05}
showed these ``kicks'' to be a weak form of feedback that was dependent upon the
numerical resolution.  We do not consider this further.
\item Each simulation used a different disk-galaxy model.
\item The star-formation recipe employed is identical in all simulations, but
the normalization, i.e. the choice of $c_\star$, is different.  The fiducial
value used in all simulations here was 0.03, while S00 and MH96 both used 0.004.
For completeness, we will investigate a range of values both an order of
magnitude smaller and larger than ours.
\item MH96 represented the ISM as an isothermal gas, at 10$^4$ K.  They argued
that gas in disk galaxies efficiently cools and thus fixing the
gas to this temperature scarcely affects the gas dynamics.  S00 included
adiabatic gas processes and shock heating as we do here.
\item Our simulations use a version of SPH which integrates the entropy, while
MH96 and S00 used a version of SPH which integrated the energy.
\end{enumerate}

In order to better understand these differences, we will use this section to
describe two types of tests.  First, we will build the exact initial conditions of S00 
and attempt to duplicate his star-formation history.  In addition, we will run 
his initial conditions with one of our models.  In the second series, we use
our fiducial model in a number tests which go through the above items, in
order, and resimulate our fiducial \ntm\ model with slight
modifications aimed at illustrating the outcomes of each assumption.
As mentioned above, we do not fully address item (i) because the 
comparison to the ``kicks'' feedback used in MH96 has been 
recently performed by other authors \citep[see][]{SdMH05}.  We also
do not perform a full exploration of the second difference listed above, 
the disk galaxy model.  A complete understanding of the dependencies
associated with the initial conditions is outside the scope of this work
(the beginning of such a project was broached in \citet{Thesis}).
However, our first type of test, the explicit comparison using the models
of S00, will suffice to provide a basic understanding of item (ii)
above.

\subsubsection{Comparison to S00}
\label{sssec:comp}

As a first task we'll verify that our model can reproduce one of
the quiescent star-formation histories and one of the merging 
disk-galaxy star-formation histories previously shown in S00.  In
order to perform this test, we must contruct identical initial 
disk models to those used in S00.  As a byproduct of this test, we
will gain some insight as to how the disk galaxy model
determines the resulting star-formation history.

% -------------------------------------------
%  Figure 13 : S00 Isolated Star Formation
% -------------------------------------------
\begin{figure}
\begin{center}
\resizebox{8.0cm}{!}{\includegraphics{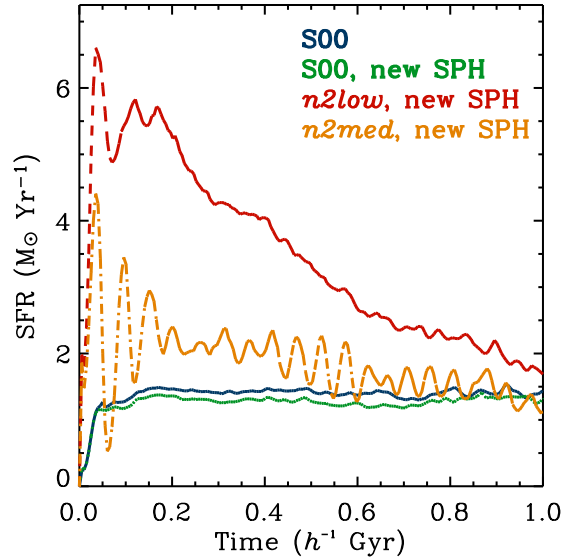}}
\caption{Star formation for model I1 of S00.  Shown are four
runs, two that have the identical parameters as in S00, which we
have labeled accordingly, and the final two runs use the 
$\ntl$ and $\ntm$ parameters used throughout this work.
Because the original work of S00 used the
``energy, standard'' formulation of SPH, our primary 
comparison run uses this as well.  The other two runs use
the newer version which is the standard for the main body
of this paper.
\label{fig:S00iso}}
\end{center}
\end{figure}

For the isolated galaxy comparison, we choose the I1 galaxy model
outlined in \S~4.1 of S00.  This model has $V_{200}
=160$~\kms, and thus a virial mass of 1.4 $\times10^{12}$~\msun.  The
halo has a concentration $c=5$ and a spin parameter $\lambda=0.05$.  All
of the baryons are in a disk, i.e., there is no bulge component, and
20\% of the baryons are in a gaseous form.  This results in mass fractions
of $m_d=0.04$, $m_g=0.01$, and $f_d=0.2$. 
Further, the gas and stellar disks are assumed to
follow the same exponential distribution (i.e., $\alpha=1$) with a
radial disk scale length of 6.4~kpc.  The model is realized
with 30,000 dark matter, 20,000 gas and 20,000 stellar disk particles
for the same resolution used in S00.

When we introduced the parameter sets in \S~\ref{ssec:psets} we
provided the conversion to the exact model used in S00.  This parameter
set had a lower star formation normalization $c_\star=
0.004$, but was otherwise very similar to our $\ntl$.  Although this
wasn't used for any of the models thus far, we now use it to simulate
the galaxy model I1 in isolation.  It is also significant to
point out that the original S00 simulations used the ``energy, standard''
formulation of SPH where we have used a newer version introduced by
\citet{SHEnt}.  We will simulate the S00 parameter set with both here.
As a reference to the parameter sets we've developed in this work, we
will also run the I1 model with the $\ntl$ and $\ntm$ parameter sets.

Figure~\ref{fig:S00iso} shows the resulting star formation for the I1 
galaxy model evolved 1.4~Gyr.  As a confirmation of our ability to 
replicate the initial disk model, we note that both of the simulations
that used the S00 parameter set have star-formation histories that are
nearly identical to that shown in S00 (see his Fig. 7).  Since the only
difference between these two runs is the version of SPH, we conclude
that the version of SPH makes little difference for evolving disk 
galaxies in isolation.

The two simulations which evolved I1 with the $\ntl$ or $\ntm$ parameter
set are quite different.  The $\ntl$ model has a much higher star 
formation rate because it is nearly identical to S00, only it has a
larger value of $c_\star$.  Because of this, nearly three times 
as much gas is converted to stars during the simulation and this rapid
star formation quickly depletes the disk of high-density gas
resulting in an subsequential decline of the star-formation rate.
The average star-formation rate of the simulation that uses the $\ntm$ 
parameter set is much closer to
that of the S00 runs, however there are strong oscillations present.
These oscillations were seen in the med- and high-feedback runs during the Sbc
major merger and were discussed briefly at the end of \S~\ref{ssec:mmsfr}

An interesting comparison can be made between the star formation
reported in Figure~\ref{fig:S00iso} for the isolated I1 disk, and that
shown earlier in Figure~\ref{fig:isosfr} for the isolated Sbc disk.
First, with the $\ntl$ parameter set, we note that the Sbc disk has a 
steady star-formation rate of $\sim0.75$~\msunyr.  This same parameter set
for the I1 disk begins with a much larger star-formation rate ($\sim6$
\msunyr) that quickly declines as the high-density gas is consumed.
Not surprisingly, one concludes that the star-formation history is a 
direct consequence of the amount and distribution of high-density gas,
both of which are a byproduct of the initial disk model \citep[see also][]
{Li05,Li06,SdMH05}.  This leads us
to suggest a natural extension to our work: extend our tests to a large
set of initial disk models with properties drawn from observations, as 
with our Sbc-disk model.  By comparing these tests to observed relations
between Hubble-type and star formation, one may provide further 
constraints on the star formation and feedback model.

% -------------------------------------------
%  Figure 14 : S00 A1 Merger Star Formation
% -------------------------------------------
\begin{figure}
\begin{center}
\resizebox{8.0cm}{!}{\includegraphics{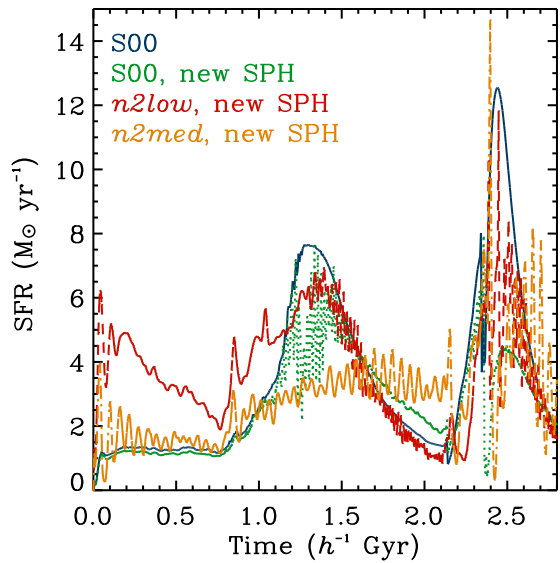}}
\caption{Star formation for the A1 bulgeless major merger of
S00.  As with Figure~\ref{fig:S00iso}, we show one run that
is designed to reproduce the star-formation history of S00,
and three other runs which have small variants in either the
star formation and feedback parameters, or the version of SPH.
\label{fig:S00mm}}
\end{center}
\end{figure}

After assuring ourselves that we are able to evolve the exact I1 model
of S00 in a similar fashion and investigating the differences induced
by either the version of SPH, or the alternate star formation and 
feedback model, we now move on to test one of his major mergers.  For
this test, we select the bulgeless A1 major merger.  These initial 
disk models are slightly smaller than the I1 disk, with $V_{200}=
120$~\kms, and thus their mass is 6.0 $\times10^{11}$~\msun.
The halo has a concentration $c=5$ and a spin parameter $\lambda=0.05$.
As with the I1 model, all of the baryons are in a disk, i.e., there 
is no bulge component, and the mass fractions are $m_d=0.04$ and $m_g=0.01$,
where the gaseous and stellar disks are similarly distributed.
This model is realized with an identical particle number as I1.

Figure~\ref{fig:S00mm} shows the star formation rate for the A1 major
merger.  As with the I1 test, we run four simulations: one identical
to S00, one with the same star formation and feedback parameters as
S00 but with the newer version of SPH, and two with our parameter 
sets $\ntl$ and $\ntm$.  Once again, when we use the S00 parameter set
and the identical version of SPH, we can reproduce the A1 merger
star-formation rate exactly (see Fig. 14 of S00).

In contrast to the isolated case, when we use the S00 parameter set
with the newer version of SPH,
the star-formation history shows marked changes.  The star-formation rate
in this case is oscillatory and suppressed relative to the S00 result.
These changes occur
primarily during the bursts of star formation that result after the first
passage and the final merger and are likely a result of the improved
entropy conservation in the newer version of SPH.  We will perform more
tests along this line, and discuss this further in \S~\ref{sssec:sphv}

The other two runs, which use the $\ntl$ and $\ntm$ parameter sets,
also show distinct differences from S00.  As in the case of the isolated disk,
$\ntl$ shows much larger star formation during the quiescent stages of
evolution.  However, the bursts are relatively similar in amplitude, 
especially to the S00 model with the same SPH version.  The $\ntm$ run
shows the oscillations that were also present in I1 and Sbc, and there
is almost no burst of star formation subsequent to the first passage
at $T=1.0 - 1.7$~$h^{-1}$~Gyr.  The final merger, however, contains
the highest star-formation rate of any of the four models.

From the tests performed in this section, we arrive at three primary 
conclusions.  First and foremost, our model reduces to that of S00, 
and as such, if we use his initial conditions, we can reproduce his
results.  Second, we note that the increased value for the star-formation 
normalization $c_\star$, unsursprisingly, produces an increased 
star-formtion rate.  This increase occurs only during the 
quiescent evolution.  During the merger-induced bursts, 
the star-formation rate is nearly equivalent, as demonstrated in 
Figure~\ref{fig:S00mm}.  The third conclusion pertains to the version
of SPH.  While this made very little difference while evolving the quiescent
disk, there are substantial differences that occur 
during the more dynamic merger.  We will perform a more thorough
exploration of these last two dependencies, $c_\star$ and the version
of SPH, using our Sbc disk-galaxy model in the sections that follow.

%
% ------------------------------------------
\subsubsection{Star-Formation Recipe: Normalization}
\label{sssec:sfn}

% -------------
%  Figure 15
% -------------
\begin{figure}
\begin{center}
\resizebox{8.0cm}{!}{\includegraphics{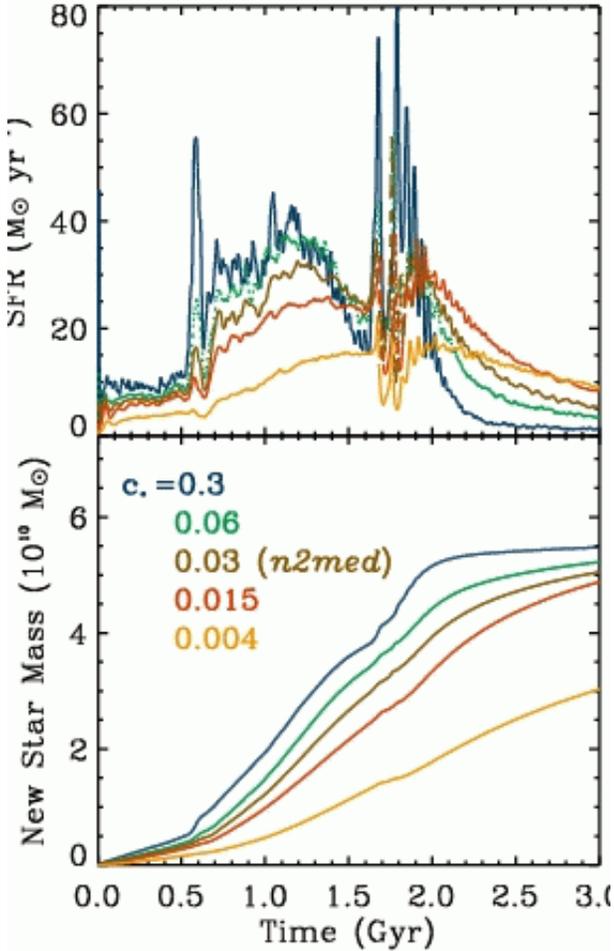}}\\
\caption{Star-formation rate during the fiducial merger
for different values of $c_\star$, the parameter
that sets the star formation efficiency.
\label{fig:sfr_cstar}}
\end{center}
\end{figure}

% ------------
%  Table 6
% ------------
\begin{table}
\bc
\caption{Star-formation properties of major merger simulations
as a function of $c_\star$.  Quantities are defined in
Table~\ref{tab:sfprops}.}
\begin{tabular}{lccc}
\hline
$c_\star$ & $e$ & SFR$_{\rm max}$ & T$_{\rm LIRG}$  \\
 &  & (\msunyr) & (Gyr) \\
\hline
\hline
0.3  & 0.51 & 82 & 1.30 \\
0.06 & 0.49 & 57 & 1.36 \\
0.03 (\ntm\ )  & 0.47 & 56 & 1.34 \\
0.015 & 0.45 & 38 & 1.21 \\
0.004 & 0.28 & 22 & 0.01 \\
\hline 
\end{tabular}
\label{tab:sfprops_cstar}
\ec
\end{table}

The parameter $c_\star$ normalizes the course-grained recipe to
convert gas to stars via Equation~(\ref{sflaw}).  While
\S\ref{ssec:mmsfr} showed that $c_\star=0.03$ reproduced the
observational Kennicutt law (\ref{KenLaw}) over two orders of
magnitude in gas surface density, the observational scatter appears
large enough to allow for a range of star-formation efficiencies.  In
this section, we resimulate the fiducial \ntm\ model with $c_\star$=
0.004, 0.015, 0.06, and 0.3.  These values of $c_\star$ represent
excursions from our fiducial value (0.03) of an order of magnitude
(0.004 and 0.3) and a factor of two (0.015 and 0.06).  Both MH96 and
S00 used $c_\star\approx0.004$.

Figure~\ref{fig:sfr_cstar} shows the star-formation rate for the
fiducial \ntm\ simulation and the four mergers with alternate
star-formation normalizations.  Not surprisingly, the star formation
prior to the final merger correlates with the star formation
efficiency $c_\star$, but the situation reverses after the merger.  In
fact, the low efficiency of star formation means very little gas is
consumed prior to the final merger, and by the end of the simulation,
at $T=3.0$~Gyr, the $c_\star=0.004$ simulation has the highest rate of
star formation.  In a manner similar to \S \ref{ssec:mmsfr}, the star
formation can be quantified by the gas consumption, the maximum
star-formation rate, and the duration in which the merger would
qualify as a LIRG, which are listed in Table~\ref{tab:sfprops_cstar}.

Varying the star-formation efficiency highlights the differing effects
of star formation versus feedback.  For example, changing the fiducial
\ntm\ model by decreasing $c_\star$ to 0.004 (as done above) or
increasing $\tfb$ to 2.0 (\nth), reduces SFR$_{\rm max}$ and $e$.
Increasing the feedback, as in \nth, restricts the ability of gas to
reach high densities, but gas which does so will be rapidly converted
to stars.  Hence, the gas consumption $e$ is strongly affected while
the peak star-formation rate is only slightly less than \ntm.  In
contrast, decreasing $c_\star$ only slightly alters the gas density
but strongly reduces the star-formation efficiency of the high-density
gas.  In this case, the maximum star-formation rate is less than half
\ntm, but the gas consumption $e$ is higher than when increasing the
feedback.

% ------------
%  Figure 16
% ------------
\begin{figure}
\begin{center}
\resizebox{8.0cm}{!}{\includegraphics{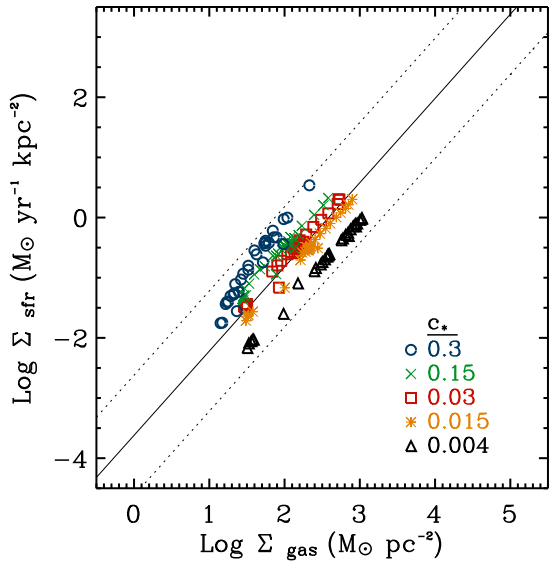}}\\
\caption{Identical to Figures~\ref{fig:mkenn} except shown here are
simulations with varying star-formation normalization $c_\star$.  The
(red) squares are the fiducial \ntm\ simulation and were
similarly shown in Figure~\ref{fig:mkenn}.
The (blue) circle, (orange) asterisk, (yellow) cross, and (black)
triangle symbols are for $c_\star=0.3, 0.06, 0.015,$~and $0.004$, 
respectively.
\label{fig:kenn_cstar}}
\end{center}
\end{figure}

Because the star-formation rate was affected by different values of the star-formation
efficiency $c_\star$, it's reasonable to ask if the agreement with the
Kennicutt law has been destroyed.  Figure~\ref{fig:kenn_cstar} answers this question
by showing the gas and star-formation rate surface densities in a 2 kpc aperture for
each of the simulations with different $c_\star$.  As expected, due to our coupling of 
star formation to the gas density via Equation~(\ref{sflaw}), the slope of each 
simulation tracks the observed Kennicutt law.  However, the various normalizations
have shifted the amplitude of star formation so that the values 0.3, and 0.004 are
on the edges of the observed range.

%
% ------------------------------------------
\subsubsection{Star-Formation Recipe: Slope}
\label{sssec:sfs}

% -----------
%  Figure 17
% -----------
\begin{figure}
\begin{center}
\resizebox{8.0cm}{!}{\includegraphics{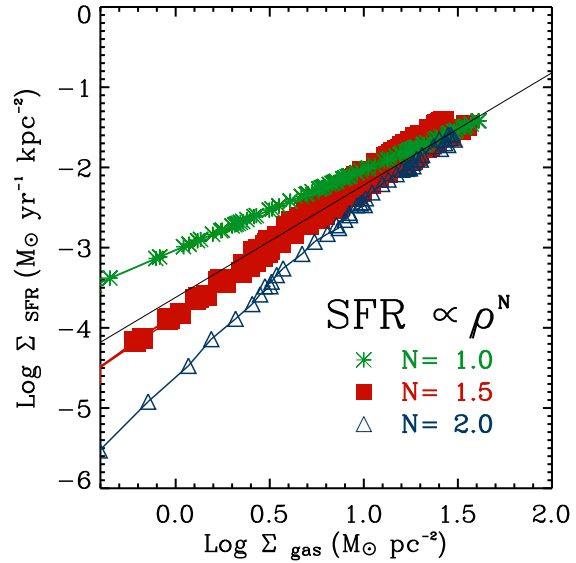}}\\
\caption{
This figure is identical to Figure~\ref{fig:isokenn} except
here we compare our fiducial $n2low$ model to variants that
have an altered star-formation law according to Equation~(\ref{sflawN}),
with $N=0,2$ (opposed to the fiducial $N=1.5$).  All simulations
are run with the threshold density for star-formation set to zero.
\label{fig:isokennsfrN}}
\end{center}
\end{figure}

The previous section demonstrated that the normalization of the
projected star-formation law, and hence the agreement with \citet{Kenn98},
is set by the free parameter $c_\star$.  However, we have still not 
demonstrated what sets the slope of the star-formation law.  Previous
efforts have argued that the observed star-formation law is a
natural consequence of the turbulent interstellar medium \citep{Kv03}
or that it results from gravitational instabilities in gaseous 
spiral disks \citep{Li04}.  In order to investigate the origin of
the slope of the star-formation law, we will follow a procedure
similar to \citet{MRB91} and modify our star-formation prescription,
i.e., Equation~(\ref{sflaw}).  To this end, we write a generalized
star-formation law as
\be
\frac{\dd \rho_{\star}}{\dd t} = C \rho_{\rm gas}^{N},
\label{sflawN}
\ee
where $C$ is a dimensionful free parameter and $N$ sets the
gas-density power-law dependence.  In our fiducial model, $C
=\sqrt{4 \pi G} c_\star$, and $N=1.5$.  As variants, we will try
$N=0$ and $2$, and fix $C$ such that the global star-formation
rate is identical at $\rt$.  To simplify the comparison, we only
show results from simulations where the star formation is not fixed
to be zero below $\rt$.  We also performed identical simulations 
that included the threshold density, however in this case the
star-formation surface density quickly plunged to zero below
$\sim 10$~\msunpc, as in Figure~\ref{fig:isokenn}, and thus there
was very little difference between the various star-formation
laws.

Figure~\ref{fig:isokennsfrN} shows the star-formation law for our
isolated Sbc disk galaxy for the three variants of the star formation
prescription.  We see a clear trend that the slope of the projected
star-formation law follows the power of $N$ in Equation~(\ref{sflawN}).
This correlation was also noted by \citet{MRB91}, who additionally
showed that the steeper powers of $N$ result in starbursts of 
greater intensity.  
From these considerations we conclude that the star-formation law, both in
normalization and slope, is fixed by the empircal Kennicutt law.

%
% -----------------------------
\subsubsection{Isothermal ISM}
\label{sssec:iism}

% ------------
%  Figure 18
% ------------
\begin{figure}
\begin{center}
\resizebox{8.0cm}{!}{\includegraphics{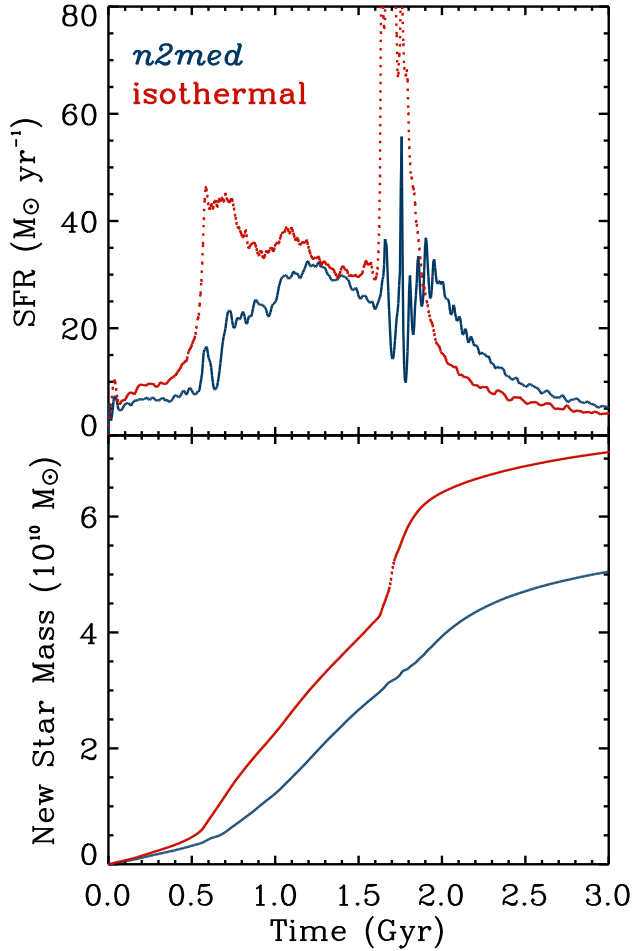}}\\
\caption{Star-formation rate for our \ntm\ model and for the identical
model except the temperature of non-star forming gas is fixed at 10$^4$ K.
\label{fig:sfr_isot}}
\end{center}
\end{figure}

Next we consider differences generated by assuming the ISM is an
isothermal gas with temperature $10^4$~K.  Because of the efficient
cooling in quiescent disk galaxies, it is reasonable to assume their
thermal energy is maintained near $10^4$~K.  Indeed, this is precisely
the case for our model galaxy when simulated in isolation, there is
no difference.

However,
galaxy mergers are a much more violent event in which there is ample
shock heating of low-density gas \citep{Cox04}.  This was demonstrated
in \S\ref{ssec:mmphase}, where phase diagrams showed large amounts of
hot ($\sim10^6$~K), low density ($10^{-4}-10^{-7}$\msunpcub) gas.  In
this state, the cooling time is much longer than the orbital time, and
the gas becomes locked in this phase, unable to cool.  If the
temperature of this gas were fixed at $10^4$~K, this gas would be much
more likely to fall back to the galactic center and could significantly
affect the star-formation rate.

To investigate this possibility, we have run our \ntm\ model with the
\textit{thermal} gas temperature (the $u$ reservoir) 
fixed to $10^4$~K.  (In order to
maintain a stable isolated disk, the feedback reservoir $q$ still
operates normally.)  Under these assumptions, the star-forming gas is
scarcely different than \ntm\ while all non-star forming gas is
effectively $10^4$~K, without gas which has recently formed stars and
hasn't thermalized its turbulent energy yet.  
The resulting star formation, in comparison to
\ntm, is shown in Figure~\ref{fig:sfr_isot}.  The isothermal
assumption results in a maximum star-formation rate that is
four times that of \ntm\ and
increases the gas consumption $e$ by 20\% over \ntm.  This 
gas fraction 
is nearly identical to the quantity of gas which ends up in the
hot, low-density phase in \ntm.  Further, the periods of maximum difference
between the two star-formation rates is precisely when large shock
heating occurs.  It thus appears that assuming the ISM
is isothermal at $10^4$~K grossly underpredicts the amount of 
shock heating due to
the merger and results in much more cold gas available for star formation.

%
% ------------------------
\subsubsection{SPH Version}
\label{sssec:sphv}

% -------------
%  Figure 19
% -------------
\begin{figure}
\begin{center}
\resizebox{8.0cm}{!}{\includegraphics{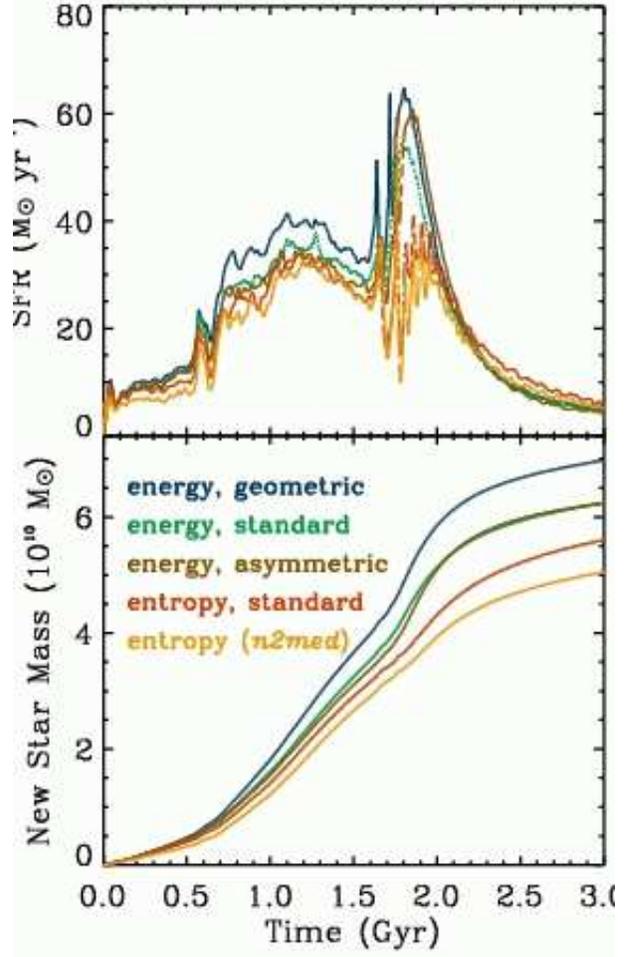}}\\
\caption{A comparison of our fiducial \ntm\ model run with different versions
of SPH commonly used in numerical simulations.
\label{fig:sfr_sph}}
\end{center}
\end{figure}

% -------------
%  Table 5
% -------------
\begin{table}
\bc
\caption{Star formation properties of major merger simulations
as a function of SPH formulation.  Quantities are defined in
Table~\ref{tab:sfprops}.}
\begin{tabular}{lccc}
\hline
SPH Version & $e$ & SFR$_{\rm max}$ & T$_{\rm LIRG}$  \\
 &  & (\msunyr) & (Gyr) \\
\hline
\hline
entropy (\ntm)  & 0.47 & 56 & 1.34 \\
energy, standard & 0.58 & 55 & 1.48 \\
energy, geometric & 0.65 & 65 & 1.52 \\
energy, asymmetric & 0.58 & 61 & 1.42 \\
entropy, standard & 0.52 & 60 & 1.38 \\
\hline
\end{tabular}
\label{tab:sfprops_sph}
\ec
\end{table}

All simulations presented up to this point have used a version of SPH which
integrates the entropy and was formulated to conserve both energy and entropy,
when appropriate \citep{SHEnt}.  In order to discern the
effects of our choice of SPH we resimulate the fiducial \ntm\ merger with
alternate versions.  Specifically, we use the energy integrating ``energy, geometric'',
``energy, standard'', and ``energy, asymmetric'' versions of SPH, as
well as the entropy integrating ``entropy, standard'' version,
where we follow the naming convention
of \citet{SHEnt} and refer the interested reader to their paper for further
details.  The galaxy-merger simulations of S00 and MH96
used the ``energy, standard'' version.

Figure~\ref{fig:sfr_sph} shows the star-formation rate during the SPH runs.
There are two notable features present in the star-formation histories owing to the 
version of SPH.  The first feature is the systematic enhancement of star formation
from $T=0.6$~to~$1.6$~Gyr in runs that do not use the fiducial \citet{SHEnt} version
of SPH.  The second feature is present after the final merger, from $T=1.8$~to~$2.0$~Gyr,
when the energy versions of SPH lead to a large peak of star formation while the
entropy versions (\ntm\ and ``entropy, standard'') of SPH have an oscillatory 
star-formation rate.  In general, the entropy versions lead to lower star-formation
rates and less gas consumption.  We speculate that these features arise because 
the entropy versions of SPH more accurately treat point-like energy injections and
adiabatic heating as gas falls into the central region of the halo, as demonstrated
in \citet{SHEnt}, and therefore may be more accurate than energy versions of SPH.
Additionally, while versions of SPH not tested here may produce still different
results, we feel that the tests performed here survey the plausible range of 
outcomes and accurately reflect the modeling uncertainties resulting from 
versions SPH.

While these tests show that the version of SPH can alter the star formation during a
galaxy interaction, we note that these differences are smaller than differences 
induced by other modeling assumptions.  For instance, Table~\ref{tab:sfprops_sph}
shows that the total amount of gas consumed and maximum star-formation rate for this
series of mergers varies by $<20$\%.  By comparison, changes in the star formation
efficiency ($c_\star$) or feedback parameters ($\tfb$ and $n$) can produce differences
of over a factor of 2 (see Tables~\ref{tab:sfprops} and \ref{tab:sfprops_cstar}).

%
% ----------------------------
\subsection{ISM Metallicity}
\label{ssec:cool}

% ------------
%  Figure 20
% ------------
\begin{figure} 
\begin{center}
\resizebox{8.0cm}{!}{\includegraphics{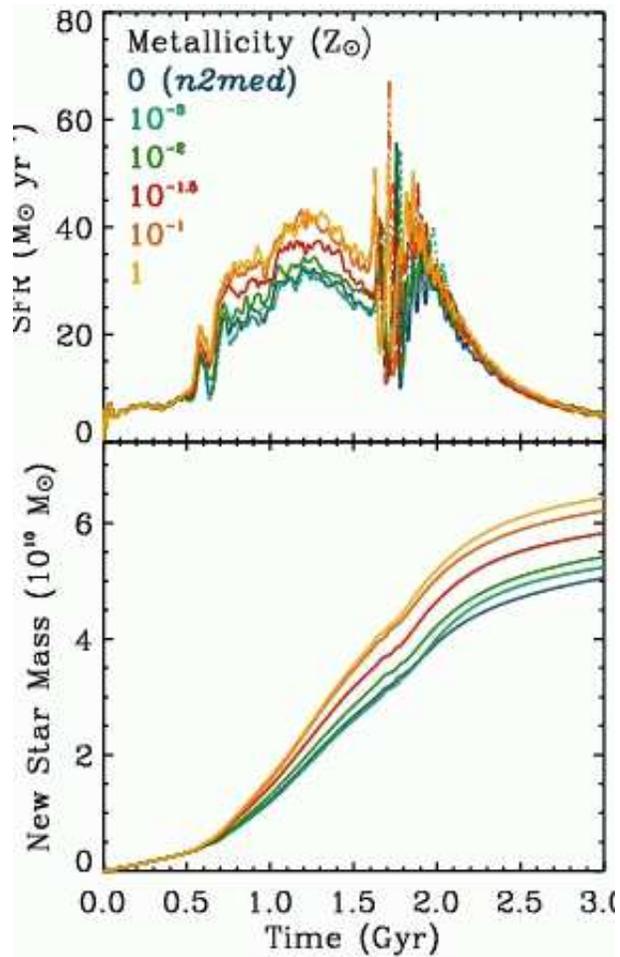}}
\caption{Star-formation rate for our fiducial major merger with 
various assumptions for the inter stellar medium metallicity.
The metallicities listed are in solar units.
\label{fig:sfr_cool}}
\end{center}
\end{figure}

% ------------
%  Table 7
% ------------
\begin{table}
\bc
\caption{Star formation properties of major merger simulations
as a function of ISM metallicity.  The metallicity is given
relative to solar and the remaining quantities are defined in
Table~\ref{tab:sfprops}.}
\begin{tabular}{lccc}
\hline
Metallicity & $e$ & SFR$_{\rm max}$ & T$_{\rm LIRG}$  \\
 (Z$_{\odot}$) &  & (\msunyr) & (Gyr) \\
\hline
\hline
zero (\ntm)  & 0.47 & 56 & 1.34 \\
$10^{-3}$ & 0.49 & 54 & 1.38 \\
$10^{-2}$ & 0.50 & 54 & 1.40 \\
$10^{-1.5}$ & 0.54 & 47 & 1.44 \\
$10^{-1}$ & 0.58 & 67 & 1.44 \\
1 & 0.60 & 53 & 1.52 \\
\hline 
\end{tabular}
\label{tab:sfprops_metals}
\ec
\end{table}

The ISM radiatively cools at a rate which is highly dependent upon the metallicity.
Up to now, cooling has been calculated assuming the ISM is a primordial
plasma with zero metallicity even though metals are tracked by the code, as
mentioned in \S~\ref{ssec:metals}.  In this section we resimulate the fiducial \ntm\
model with radiative cooling calculated from the tabulated cooling curves of
\citet{SD93} and assumed gas metallicities, in units of solar, of
$10^{-3}$, $10^{-2}$, $10^{-1.5}$, $10^{-1}$, and $1$.   As with our fiducial
model, we continue to ignore spatial and temporal changes in the metallicity.
The resulting star-formation rate is shown in Figure~\ref{fig:sfr_cool}.  As
expected, the increased metallicity leads to a higher fraction of the gas in the
cold phase, and increases the star-formation rate.  The more efficient metal-line
cooling is strongest for gas which is $\sim10^5$~K, and hence the star formation
differences are not apparent until gas begins to populate this temperature
range.  Gas at these temperatures is produced by the interaction itself, in shocks
that occur between gas in the progenitor disks.  While the maximum
star-formation rate does not dramatically change for simulations with increased
gas metallicity, the gas consumption $e$ steadily increases.  In fact, there is
a 13\% increase in $e$ during the merger with solar versus zero metallicity.  The 
star-formation properties are listed in Table~\ref{tab:sfprops_metals}.

The variations in gas consumption suggest that appreciable errors can be
generated when numerical simulations treat the ISM as a primordial plasma.
While these variations are smaller than those induced by altering the star 
formation and feedback parameters (see Tables~\ref{tab:sfprops} and 
\ref{tab:sfprops_cstar}) the cooling rate can be calculated fully self-consistently
because the simulation tracks metal enrichment (see \S\ref{ssec:metals}).  Toward
this end a number of groups have recently included metallicity-dependent cooling
based upon a model for the enrichment of metals from supernovae (SN) \citep[see, e.g.,][]
{cSca05,cSca06,Oka05,Tor04,Mos01} and have found similar trends to those reported here.
Unfortunately, these models necessarily
require a number of assumptions regarding the time delay of SNIa, the physical scale
that stellar particles distribute their metals, the diffusion of metals, and the
scale in which metals are smoothed when the cooling is calculated.  At 
the current moment it is unclear how best to implement the physics of metal
enrichment and metallicity-dependent cooling but these physical processes are
certain to garner a significant amount of attention in the near future.

% -------------------------
%  Conclusion: wrap it up
% -------------------------
\section{Conclusion}
\label{sec:concl}

We have carried out a series of isolated and merging disk galaxy
simulations which include simple recipes for implementing star formation
and feedback.  The free parameters of these models are selected based
upon the stability of our isolated disk galaxy model and excursions are
explored with increased and descreased stability.  These models
were then used during equal-mass prograde major mergers
to investigate the resulting star-formation and remnant properties.
The primary conclusions can be summarized as follows:

\begin{itemize}
\item The slope and normalization of the star-formation law are fixed
  by adopting a recipe in which stars are formed from high-density gas
  according to 
  $\dot{\rho}_{\star} \approx c_\star \rho_{\rm gas}^{1.5}$.
  The slope is a result of the 1.5 index, and the normalization is set
  by $c_\star$.  Star formation designed in this manner, independent of
  the feedback model, can naturally reproduce the empirical Kennicutt law.
  The above holds for both quiescent and interacting galaxies.
\item Provided that star formation is designed as in the first item
  above, the rate at which stars are formed is then determined by
  the feedback efficiency $\tfb$.  The star-formation rate in isolated
  disk galaxy simulations
  is largely insensitive to the equation of state at high
  densities (in our model this is controlled by $n$), because
  any feedback sufficient to stabilize the galactic disk is also
  efficient enough to confine star-forming gas to a narrow range of
  densities around the threshold density $\rt$.
\item During a merger, where gas can acheive much higher densities
  than in the isolated case, the star formation rate depends on both
  $\tfb $ and $n$.  However, the overall gas consumption is fixed
  primarily by $\tfb$, the feedback efficiency.  Because $\tfb$ determines
  the pressurization of
  star-forming regions around the threshold density, large values can
  prevent gas from ever reaching star-forming densities.
\item The maximum star-formation rate during the merger
  simulations depends on $\tfb $ in a way which is reasonably
  well approximated by a toy model of star-forming gas in a fixed
  potential (Equation~\ref {eq:sfrn}).
\end{itemize}

Our results extend, and are consistent with, previous numerical
  simulations of star formation in colliding disk galaxies.  A
  detailed comparison finds that previous versions of SPH predict
  an increased gas consumption by varying amounts.  We also 
  determined that an isothermal gas assumption at 10$^4$~K also
  predicts much more efficient star formation during a major merger.

The above items collectively imply that requiring free parameters
  to yield a stable isolated disk
  and star formation that is consistent with the Kennicutt law
  provides a fairly tight constraint on the star-formation recipe
  but does not uniquely determine the feedback parameters.
  The large amount of freedom in selecting the feedback parameters
  makes a significant difference to the maximum
  star-formation rate during a galaxy merger, while still fulfilling
  these basic requirements.  

Generally speaking, the integrated gas consumption during a merger is less
  sensitive to the details of the feedback parameters, due to a
  trade-off between star-formation rate and the duration of the burst.
  In models with highly efficient feedback, strong bursts of
  star-formation are suppressed. However, star formation then
  continues until all the gas at star-forming densities has been consumed,
  leading to a similar overall gas consumption.

While we argue that the formulation of star formation in the numerical
simulations is well defined, the determination of the feedback parameters
is not and thus we ask the question:
Is there any hope of determing a unique set of 
feedback parameters which may then be used for a large series of 
galaxy merger simulations?
In our (optimistic) opinion, the answer to this question is 
yes.  However, doing so may be a long, arduous, and 
ill-defined process.  As a first step we note that there 
is a strong observational link between
galaxy interactions and star formation\citep[e.g.,][]{Bor00,Lam03,Bar03,NCA04}.
Theoretically, models of galaxy formation with merger induced
star-formation naturally produce the Lyman-break galaxies \citep{SPF}
and possibly also submillimeter galaxies \citep{Bau05}.  Thus,
it seems reasonable to require that our merging gas-rich spiral galaxies
induce a significant burst of star formation.
In this sense, the ``high'' feedback models are disfavored.
Further constraints on burst strengths and ages may place tighter
bounds on feedback models.  For instance, the prolonged bursts present
in the $n=2$ models may be at odds with observations that suggest 
starburst have a shorter duration $\sim100$~Myr.  
We also note that it would be very
helpful to have a statistical
observational catalog of mergers and merger remnants, including
estimates of the orbits and galaxy masses, burst strengths and
ages, and any associated kinematics to compare with the simulations.

Ultimately, one of the most promising tests of star formation and 
feedback models may require detailed comparisons
between observations of quiescent or interacting galaxies and
simulations designed to reproduce these galactic systems. Modeling of
specific galaxy systems has already been done in several instances such
as ``The Mice'' \citep{B04}, M51 \citep{SL00}, Arp118 \citep{LHG98},
Arp119 \citep{HL01}, Arp220 \citep{McD03}, NGC 7714/15 \citep{Stck03,SSN05},
NGC 2442 \citep{MB97}, IC2163/NGC 2207 \citep{Stck05} 
and NGC 7252 \citep{HibM95}, although none of
these explored a range of feedback parameterizations.  It is evident
that much work must be done to determine what initial conditions are
plausible and which physical processes must be included to reproduce
the varied and extreme environments of galaxy mergers.

Even with the ambiguities present in our models it is encouraging
that under a wide range of assumptions we
are able to produce stable, gas-rich quiescent galaxies and large
prolonged bursts of star formation.  Furthermore, for the models that
produce large bursts of star formation, the insensitivity
of the global gas consumption to the details of the model leads us to
believe that we can apply these techniques to a large series of minor as well as major
merger simulations and achieve a reliable estimate of the ability
of galaxy interactions to drive star formation, a project which has
already been started.  While the same universality does not hold
for the maximum star-formation rate of each model, this is
something that might allow us to constrain feedback by matching to
observations, as we mentioned above.

In addition, we have presented evidence that the remnant properties,
such as half-mass radius and velocity dispersion, are highly dependent
upon the star formation and feedback assumptions.  Low feedback models
are more concentrated and have higher central velocity dispersions.
  Hence, it is likely
that performing a large series of runs, and comparing them to a
statistical sample of elliptical galaxies, will significantly
discriminate between the models.  It is also possible that some leverage
can be garnered from the significant differences in gas properties
predicted by the different feedback models.  By comparing phase
diagrams such as those in \S\ref{ssec:mmphase} to mass-weighted phase
diagrams of observed galaxies there may be some models that could be
ruled out.  Another avenue for comparison between the models and
observations is the gas morphology.  For instance, the feedback
parameters affect the vertical structure of the gas disk, and hence
quantities like the inclination dependence of dust attenuation
\citep[][]{Patrik}.

\section*{Acknowledgments}

We thank Volker Springel for very useful comments and for making
GADGET and his initial conditions generator available to us.  We also
thank Anthony Aguirre, Avishai Dekel, and Jennifer Lotz for a careful
reading of a prior version of this paper.  This research used
computational resources of the National Energy Research Scientific
Computing Center (NERSC), which is supported by the Office of Science
of the US Department of Energy and also UpsAnd, a Beowulf at UCSC 
supported by NSF.  T.J.C. and J.R.P. were supported by grants from
NASA and NSF, and P.J. is supported by HST-AR-10678.01-A, provided by 
NASA through a grant from the Space Telescope Science Institute, which
is operated by the Association of Universities for Research in
Astronomy, Incorporated, nder NASA contract NAS5-26555.  P.J. also
acknowledges support by grants from IGPP/LLNL.

\bibliographystyle{mn2e}
\bibliography{paper}

\end{document}